\documentclass{emulateapj}
\usepackage{natbib}
\usepackage{apjfonts}
\def\ltsim {\lower .1ex\hbox{\rlap{\raise .6ex\hbox{\hskip .3ex
        {\ifmmode{\scriptscriptstyle <}\else
          {$\scriptscriptstyle <$}\fi}}}
    \kern -.4ex{\ifmmode{\scriptscriptstyle \sim}\else
      {$\scriptscriptstyle\sim$}\fi}}}

\def\gtsim {\lower .1ex\hbox{\rlap{\raise .6ex\hbox{\hskip .3ex
        {\ifmmode{\scriptscriptstyle >}\else
          {$\scriptscriptstyle >$}\fi}}}
    \kern -.4ex{\ifmmode{\scriptscriptstyle \sim}\else
      {$\scriptscriptstyle\sim$}\fi}}}

\begin{document}

\title{\emph{Hubble Space Telescope} Spectroscopic Observations of the 
Narrow-Line Region in Nearby Low-Luminosity Active Galactic 
Nuclei\footnotemark[1]}

\author{Jonelle L. Walsh \footnotemark[2], Aaron J. Barth
\footnotemark[2], Luis C. Ho \footnotemark[3], Alexei V. Filippenko
\footnotemark[4], Hans-Walter Rix \footnotemark[5], \\Joseph
C. Shields \footnotemark[6], Marc Sarzi \footnotemark[7], and Wallace
L. W. Sargent \footnotemark[8]}

\footnotetext[1]{Based on observations made with the NASA/ESA
\emph{Hubble Space Telescope}, obtained from the Data Archive at the
Space Telescope Science Institute, which is operated by the
Association of Universities for Research in Astronomy, Inc., under
NASA contract NAS 5-26555. These observations are associated with
programs GO-7403 and GO-7354.}

\footnotetext[2]{Department of Physics and Astronomy, University of
California at Irvine, 4129 Frederick Reines Hall, Irvine, CA
92697-4574; jlwalsh@uci.edu, barth@uci.edu .}

\footnotetext[3]{The Observatories of the Carnegie Institution of
Washington, 813 Santa Barbara Street, Pasadena, CA 91101; lho@ociw.edu .}

\footnotetext[4]{Department of Astronomy, University of California,
Berkeley, CA 94720-3411; alex@astron.berkeley.edu .}

\footnotetext[5]{Max-Planck Institut f\"{u}r Astronomie,
K\"{o}nigstuhl 17, 69117 Heidelberg, Germany; rix@mpia-hd.mpg.de .}

\footnotetext[6]{Department of Physics and Astronomy, Ohio University,
Athens, OH 45701; shields@phy.ohiou.edu .}

\footnotetext[7]{Centre for Astrophysics Research, University of
Hertfordshire, AL10 9AB Hatfield, UK; sarzi@star.herts.ac.uk .}

\footnotetext[8]{Palomar Observatory, California Institute of
Technology, Pasadena, CA 91125; wws@astro.caltech.edu .}

\begin{abstract}

We present Space Telescope Imaging Spectrograph observations of 14
nearby low-luminosity active galactic nuclei, including 13 LINERs and
1 Seyfert, taken at multiple parallel slit positions centered on the
galaxy nuclei and covering the H$\alpha$ spectral region. For each
galaxy, we measure the emission-line velocities, line widths, and
strengths, to map out the inner narrow-line region structure,
typically within $\sim$100 pc from the galaxy nucleus. There is a wide
diversity among the velocity fields: in a few galaxies the gas is
clearly in disk-like rotation, while in other galaxies the gas
kinematics appear chaotic or are dominated by radial flows with
multiple velocity components. In most objects, the emission-line
surface brightness distribution is very centrally peaked. The
[\ion{S}{2}] line ratio indicates a radial stratification in gas
density, with a sharp increase within the inner 10--20 pc, in the
majority of the Type 1 (broad-lined) objects. The electron-density
gradients of the Type 1 objects exhibit a similar shape that is well
fit by a power law of the form $\emph{n}_{\mathrm{e}} =
\emph{n}_0(r/1~ \mathrm{pc})^{\alpha}$, where $\alpha = -0.60
\pm{0.13}$. We examine how the [\ion{N}{2}] $\lambda 6583$ line width
varies as a function of aperture size over a range of spatial scales,
extending from scales comparable to the black hole's sphere of
influence to scales dominated by the host galaxy's bulge. For most
galaxies in the sample, we find that the emission-line velocity
dispersion is largest within the black hole's gravitational sphere of
influence, and decreases with increasing aperture size toward values
similar to the bulge stellar velocity dispersion measured within
ground-based apertures. We construct models of gas disks in circular
rotation and show that this behavior can be consistent with virial
motion, although for some combinations of disk parameters we show that
the line width can increase as a function of aperture size, as
observed in NGC 3245. Future dynamical modeling in order to determine
black hole masses for a few objects in this sample may be worthwhile,
although disorganized motion will limit the accuracy of the mass
measurements.

\end{abstract}


\section{Introduction}
\label{sec:intro}

With its $\sim$ 0\farcs1 resolution, the Space Telescope Imaging
Spectrograph (STIS) aboard the \emph{Hubble Space Telescope (HST)} has
provided the exceptional ability to spatially resolve gas kinematics
within the gravitational sphere of influence of the putative central
black hole in nearby galaxies. Resolving the gas kinematics very close
to the black hole allows for the determination of black hole mass in a
relatively straightforward manner, provided that the gas is in
Keplerian rotation in a thin, disk-like structure. STIS has thus had a
major impact on the field of supermassive black hole detection in
galactic nuclei, and data from STIS have contributed greatly toward
establishing the correlations between black hole mass and host-galaxy
properties, such as those with the stellar velocity dispersion
\citep{Ferrarese_2000,Gebhardt_2000,Tremaine_2002} and bulge
luminosity \citep{Kormendy_Gebhardt_2001}.

More recently, ground-based integral field unit (IFU) observations
have provided two-dimensional (2D) kinematic information on the
ionized gas in galaxy centers \citep{Sarzi_2006,
Dumas_2007}. Additionally, ground-based IFU observations with the
assistance of adaptive optics have produced gas dynamical black hole
mass measurements \citep{Hicks_Malkan_2007, Neumayer_2007}. Without
the use of adaptive optics, however, ground-based observations are
usually unable to resolve scales comparable to the black hole radius
of influence in nearby galaxies. Although STIS is a long-slit
instrument, 2D information similar to IFU data can be obtained with
STIS by using multiple slit positions.

Even with the past decade of high-resolution observations, only about
30 dynamical black hole mass measurements have been made
\citep{Ferrarese_Ford_2005}. Consequently, searching for regular gas
velocity fields dominated by gravitational motion is essential for
extending the number of black hole mass measurements and further
establishing the connections between black holes and host
galaxies. While searching for regular gas velocity fields is clearly
important, it appears, however, that they are fairly rare. Instead,
complicated and chaotic emission-line velocity fields are more
commonly found near the centers of galaxies \citep{Ho_2002,
Atkinson_2005}. Although only upper limits on the black hole mass can
be estimated using gas dynamical modeling in these situations
\citep{Sarzi_2002}, subarcsecond-resolution observations can still be
useful for studying the dynamical state of the gas and searching for
radial motions that might be related to active galactic nucleus (AGN)
fueling or outflows. Modeling AGN outflows on subarcsecond scales
\citep{Capetti_1996, Crenshaw_Kraemer_2000} is important for
furthering our understanding of the connection between AGN feedback
and host-galaxy properties. Of particular interest are the connection
to star formation, the growth of the supermassive black hole, and how
gas may be driven out from the host galaxy as predicted in some
feedback scenarios \citep{Springel_2005, Hopkins_2006, Sijacki_2007}.

Moreover, the narrow-line region (NLR) velocity dispersion
($\sigma_{g}$), often measured from the width of the [\ion{O}{3}]
$\lambda 5007$ line, may provide a reasonable estimate of the
host-galaxy stellar velocity dispersion ($\sigma_\star$), albeit with
substantial scatter \citep{Nelson_Whittle_1996, Greene_Ho_2005}.
Detailed NLR studies are essential in understanding the origin of the
scatter in the $\sigma_{g}$--$\sigma_\star$ relationship. This
relationship can be useful for distant and luminous AGNs where
measurement of the stellar velocity dispersion is not feasible
\citep{Salviander_2007, Shields_2003}. Using the narrow emission-line
widths as a proxy for stellar velocity dispersion is only meaningful
if the NLR gas is dominated by virial motion in the host-galaxy bulge
and not by the gravitational influence of the black hole or
nongravitational motions. Therefore, outflows and similar
nongravitational motions, which can disrupt viral motion, are one
potential cause of the scatter in the $\sigma_{g}$--$\sigma_\star$
relationship.

While nongravitational motions are clearly important in setting the
narrow-line widths in AGNs with strong outflows, there can be an
important contribution to the line widths from nongravitational motion
even in objects that are not outflow-dominated. From \emph{HST} STIS
observations, \cite{VerdoesKleijn_2006} found that circumnuclear gas
in Fanaroff-Riley 1 radio galaxies typically has an intrinsic velocity
dispersion in excess of that expected from purely gravitational motion
in the host-galaxy potential. They concluded that this excess line
width is driven by the injection of energy by the AGN, although there
was no obvious relationship between the AGN radiative luminosity and
the magnitude of the nongravitational line dispersion.

Other origins for the scatter were studied by \cite{Greene_Ho_2005}
using a large sample of Sloan Digital Sky Survey (SDSS) type 2
AGNs. They looked for correlations of nuclear properties (radio power,
AGN luminosity, Eddington ratio), as well as global properties
(host-galaxy morphology, local environment, star-formation rate), with
the excess line width, defined to be the difference between the
dispersion of the [\ion{O}{3}] $\lambda 5007$ line and the stellar
velocity dispersion. They found a strong correlation between the
excess line width and Eddington ratio, where larger excess line
widths, indicative of outflowing gas, were seen in the higher
Eddington ratio objects.

Another potential explanation for the scatter in the
$\sigma_{g}$--$\sigma_\star$ relationship is the effect of
observational aperture size. \cite{Rice_2006} investigated the
relationship between line width and aperture size in a sample of
nearby Seyferts using single-slit STIS observations. They found a wide
range of behavior in how the widths of the [\ion{O}{3}] $\lambda 5007$
and [\ion{S}{2}] $\lambda\lambda$6716, 6731 lines varied as a function
of radius, but the most common trends were either a roughly constant
width or an increase in width with increasing aperture size. Their
analysis showed that the line widths measured within the largest STIS
aperture were systematically smaller by 10\%--20\% than both the
stellar velocity dispersion and the line widths measured within a
ground-based sized aperture. They concluded that the smaller line
width measured within the randomly oriented STIS slit is the result of
sampling a smaller portion of the velocity field than the larger
ground-based aperture.

Beyond the $\sigma_{g}$--$\sigma_\star$ relationship, we can address
other open questions concerning low-luminosity AGNs. \cite{Laor_2003}
pointed out that while the NLR in luminous AGNs occurs on scales of
tens of parsecs, where its dynamics are dominated by the bulge, the
NLR in low-luminosity AGNs is more compact, and its dynamics could be
dominated by the black hole. It is therefore interesting to ask
whether the narrow-line widths in low-luminosity AGNs are set by the
black hole or by the bulge potential.

The NLR emission lines also provide information on ionization
mechanisms. Spectra of low-ionization nuclear emission-line regions
\citep[LINERs;][]{Heckman_1980} can be reproduced with photoionization
models involving a nonstellar continuum and low ionization parameters
\citep{Ferland_Netzer_1983, Halpern_Steiner_1983}, but high electron
densities, as well as a large range of densities, are needed in order
to reproduce the strength of the [\ion{O}{3}] $\lambda 4363$ line
\citep{Filippenko_Halpern_1984, Filippenko_1985,
Ho_1993}. \cite{Filippenko_Halpern_1984} demonstrated that such
conditions exist in the LINER NGC 7213, and that there is a
correlation between line width and the critical density for
collisional deexcitation. A similar relationship between line width
and ionization potential was also previously seen in Seyferts
\citep{Pelat_1981}. This implied that the NLR is not a homogeneous
environment, but must have radial gradients in velocity and
density. In ground-based spectra these spatial gradients were
generally unresolved, but using STIS it is possible to directly
resolve the density gradients in the inner NLR \citep{Barth_2001a,
Shields_2007}. It is therefore of interest to determine what fraction
of LINERs show spatially resolved density gradients, and whether the
type 2 LINERs, where a central photoionizing source is not always
seen, exhibit density gradients.

In this paper, we present multislit \emph{HST} STIS observations of 14
low-luminosity AGNs, including 13 LINERs and 1 Seyfert. We use the
high spatial resolution of STIS to map out the 2D kinematic structure
of the NLR within the inner $\sim 100$ pc, and search for velocity
fields that show signs of regular rotation for which gas dynamical
modeling may be performed. The subarcsecond resolution STIS data also
allow us to detect electron-density gradients through the [\ion{S}{2}]
$\lambda 6716/\lambda 6731$ line ratio diagnostic. Finally, we examine
how the [\ion{N}{2}] line width varies as a function of aperture size
over spatial scales ranging from those comparable to the black hole
sphere of influence to scales dominated by the bulge.

\section{Observations and Data Reduction}
\label{sec:obs}

Our goal is to examine the NLR of low-luminosity AGNs through STIS
spectra of the H$\alpha$ spectral region at multiple slit
positions. Eight galaxies observed under program GO-7354 (PI Dressel)
fit this description, and we retrieved those from the \emph{HST}
archive. We combined these eight galaxies with six galaxies that were
observed as part of our own program, GO-7403, to arrive at a sample of
14 galaxies. Program GO-7354 was originally designed to study LINER
nuclei with compact, flat-spectrum radio cores, while the observations
under program GO-7403 focused on obtaining black hole masses in a
sample of low-luminosity broad-lined AGNs. All of the galaxies are
optically classified as LINERs except NGC 3227, which contains a
Seyfert 1 nucleus, and NGC 3245, which is a transition
object. Properties of all 14 galaxies in the sample are given in Table
\ref{tab:galprop} and details of the observations are presented in
Table \ref{tab:obsparams}.

With the exception of NGC 3998, NGC 4594, and NGC 6500, all of the
galaxies from program GO-7354 were observed with the STIS
\texttt{52x0.1} aperture with no gap between the three adjacent slit
positions. NGC 3998, NGC 4594, and NGC 6500 were observed with the
same \texttt{52x0.1} aperture, but consist of five slit positions,
again with no gap between neighboring slit positions. The CCD was read
out with binning by two along the dispersion axis, creating a scale of
$0$\farcs$0507$ pixel$^{-1}$ along the spatial axis and $1.108$ \AA\
pixel$^{-1}$ along the dispersion axis. The slit position angle was
constrained to be within 15$^\circ$ of the major axis of the gaseous
disk. The G750M grating provided coverage of 6490--7050 \AA, which
included the [\ion{N}{2}] $\lambda\lambda$6548, 6583, H$\alpha$, and
[\ion{S}{2}] $\lambda\lambda$6716, 6731 lines. For this program, a
typical exposure time per position was $\sim 400$ s.

Observations of the six galaxies under program GO-7403 were obtained
with the STIS \texttt{52x0.2} and a spacing of $0$\farcs$25$ between
adjacent slits (i.e., a gap of $0$\farcs$05$ between slits). Two of
these galaxies, NGC 1052 and NGC 3227, were observed with seven
parallel slit positions, while NGC 3245, NGC 4036, NGC 4278, and NGC
4579 were observed with five slit positions. For these galaxies, the
CCD was read out in unbinned mode, and the slit position angle was
left unconstrained. With this setting the 2D spectral image has a
scale of $0$\farcs$0507$ pixel$^{-1}$ along the slit and $0.554$ \AA\
pixel$^{-1}$ along the dispersion direction. The G750M grating was
used to cover 6300--6860~\AA, which included the H$\alpha +
$[\ion{N}{2}], [\ion{S}{2}] $\lambda\lambda$6716, 6731, and
[\ion{O}{1}] $\lambda\lambda$6300, 6364 emission lines. A typical
exposure time per position was $\sim 3000$ s. The combination of
longer exposure times and the use of a wider slit made it possible to
map out the NLR to larger radii and at higher signal-to-noise ratio
(S/N) than for program GO-7354.

The data for all 14 galaxies were reduced using the standard STScI
pipeline. Dark and bias subtraction, as well as flat-field
corrections, were performed, along with the rejection of cosmic-ray
events through the combination of multiple subexposures. The data were
wavelength and flux calibrated and rectified for geometric
distortions. Occasionally a number of hot pixels remained even after
the data were run through the pipeline. In those cases, we performed
an additional cleaning step by replacing the bad pixel with a local
mean value in the flattened data, and then applied the geometric
distortion correction to the cleaned images.

In addition to the STIS observations, we obtained Wide Field Planetary
Camera 2 (WFPC2) images from the archives in order to examine the
nuclear morphology. When available, we retrieved the images from the
WFPC2 Associations web site
\footnote{http://archive.stsci.edu/hst/wfpc2/search.html
.}. Otherwise, they were retrieved from the \emph{HST} archive. In
order to remove cosmic rays, ``CR-split'' observations were combined
using the IRAF {\tt combine} task. If only single exposures were
available, they were cleaned using the LA-COSMIC task
\citep{vanDokkum}.

Black hole mass measurements have been previously made for three of
the galaxies in the sample: NGC 3245 \citep{Barth_2001}, NGC 3998
\citep{DeFrancesco_2006}, and NGC 5077 \citep{DeFrancesco_2008}. In
addition, \cite{DeFrancesco_2008} presented velocity curves for NGC
6500 and IC 989; however, both velocity fields were too complex to
model the gas as a thin circularly rotating disk, so no black hole
masses were determined for these two galaxies.

\section{Measurement of Emission Lines}
\label{sec:measurement}

Spectra were extracted from individual rows of the geometrically
rectified 2D images, and the rows were extracted as far out as
emission lines were detectable. For the eight galaxies from program
GO-7354, the emission lines were typically detectable out to about
$0$\farcs$5$ from the slit center, and for the six galaxies from
program GO-7403, emission lines were measured out to about
$1$\farcs$5$. Before any fitting was performed, the continuum was
removed from the spectrum on a row-by-row basis by fitting a straight
line to the continuum regions covering $6280$--$6380$ \AA,
$6500$--$6650$ \AA, and $6700$--$6750$ \AA. This continuum fit was
then subtracted from the spectrum. In general, a straight line is not
a very good continuum model near H$\alpha$, but given the low S/N in
the continuum at most positions, the small wavelength range, and the
AGN dominance in the nuclei of some objects, a proper starlight
subtraction could not be performed. During the emission-line fitting
process, multiple rows that contained weak emission lines far from the
slit center were binned together to improve the S/N.

In order to obtain the emission-line velocity, velocity dispersion,
and flux, we fit each spectrum with a set of Gaussians, where the fit
was carried out row-by-row with a Levenberg-Marquardt least-squares
minimization technique using C. Markwardt's ``MPFIT'' library in
IDL\footnote{http://cow.physics.wisc.edu/$\sim$craigm/idl/idl.html .}.
Single Gaussians were simultaneously fit to all the emission lines
present in the spectra that were strong enough to be fit
accurately. The fluxes of the [\ion{N}{2}] $\lambda\lambda$6548, 6583
lines were held at a 3:1 ratio, and all lines were required to have a
common velocity. The [\ion{N}{2}] lines were constrained to have equal
velocity widths. The widths of the [\ion{S}{2}] lines were also
required to be the same, which was necessary due to the lower S/N in
the off-nuclear rows. Although the widths of the two [\ion{S}{2}]
lines can differ, the effect is small \citep{Filippenko_Halpern_1984,
Filippenko_Sargent_1988}. If present, the [\ion{O}{1}] line widths
were constrained to be equal.

Whenever this basic model did not return adequate fits due to either
low S/N or to severe blending between the [\ion{N}{2}] and H$\alpha$
lines, we applied further constraints, such as tying together the
H$\alpha$ and [\ion{N}{2}] widths, or linking the H$\alpha$,
[\ion{N}{2}], and [\ion{S}{2}] widths. Additionally, some spectra
appeared to contain two kinematically separate components at some
locations. In these instances, we fit two Gaussians to each of the
H$\alpha$, [\ion{N}{2}] $\lambda\lambda$6548, 6583, [\ion{S}{2}]
$\lambda\lambda$6716, 6731, and the [\ion{O}{1}] $\lambda 6300$
lines. We fit a single Gaussian to the [\ion{O}{1}] $\lambda 6364$
line, and sometimes to the [\ion{O}{1}] $\lambda 6300$ line, in place
of two Gaussians since often these lines were too weak to constrain a
multiple-component model. Each set of Gaussians in the
multiple-velocity component model was constrained in the manner
previously described above for the basic model.

\begin{figure}
\begin{center}
\epsscale{0.5}
\plotone{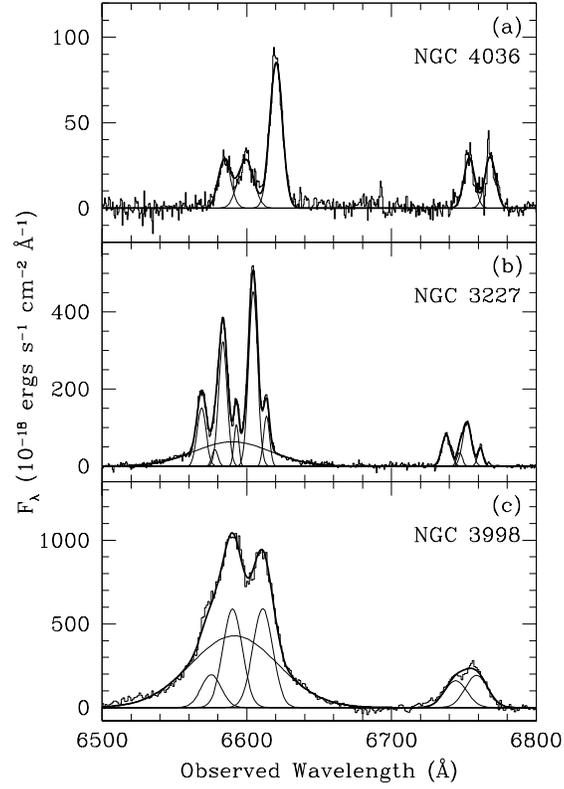}
\caption{Examples of various models used to fit the spectra: (a) a
spectrum from NGC 4036 with a single-Gaussian fit to the narrow lines,
(b) a spectrum from NGC 3227 having multiple velocity components with
double-peaked profiles and a broad H$\alpha$ component, and (c) a
central spectrum from NGC 3998 fit with single Gaussians to the narrow
lines and a broad H$\alpha$ component. Thin solid lines represent the
individual Gaussian components, and the sum of the model components is
given by the thick solid line. \label{fig:fitex}}
\end{center}
\end{figure}

Near the nuclear regions, several galaxies in the sample exhibited a
broad H$\alpha$ component in addition to the narrow lines of H$\alpha$
and [\ion{N}{2}] $\lambda\lambda$6548, 6583. For these regions, an
additional component in the form of either a single or a double
Gaussian was added to the fit. If the broad component could not be
adequately fit by two Gaussians, we selected points along the broad
base of the H$\alpha$ line and fit a cubic spline to trace the shape
of the broad component. This was then subtracted from the spectrum,
leaving just the narrow emission lines. A cubic-spline model was
needed for several rows while fitting the broad component in NGC 3227,
NGC 4278, and NGC 4579. We judged the effectiveness of an additional
H$\alpha$ broad component both by eye and by comparing the reduced
$\chi^2$, where if the reduced $\chi^2$ was improved by $15\%$ or
more, we accepted the broad-component fit. Figure \ref{fig:fitex}
shows examples of fits with single or double narrow-line components
and with a broad H$\alpha$ component.

From the model fits, we obtained the velocity, velocity dispersion,
and flux for the [\ion{N}{2}] $\lambda 6583$ emission line. In
calculating the velocity dispersion, we accounted for the instrumental
line width by using observations of the STIS wavelength calibration
lamp. Gaussian models were fit to the emission-line profiles of
extracted calibration-lamp spectra. The variation in line width as a
function of wavelength was insignificant across the spectrum and as a
result, we used the average value from measurements of several
lines. The instrumental velocity dispersion was $\sigma_i \approx 31$
km s$^{-1}$ for program GO-7403 and $\sigma_i \approx 17$ km s$^{-1}$
for program GO-7354. These instrumental line widths were then
subtracted in quadrature from the observed line widths.

\section{Results}
\label{sec:resuls}

\begin{figure}
\begin{center}
\figurenum{2a}
\epsscale{0.9}
\plottwo{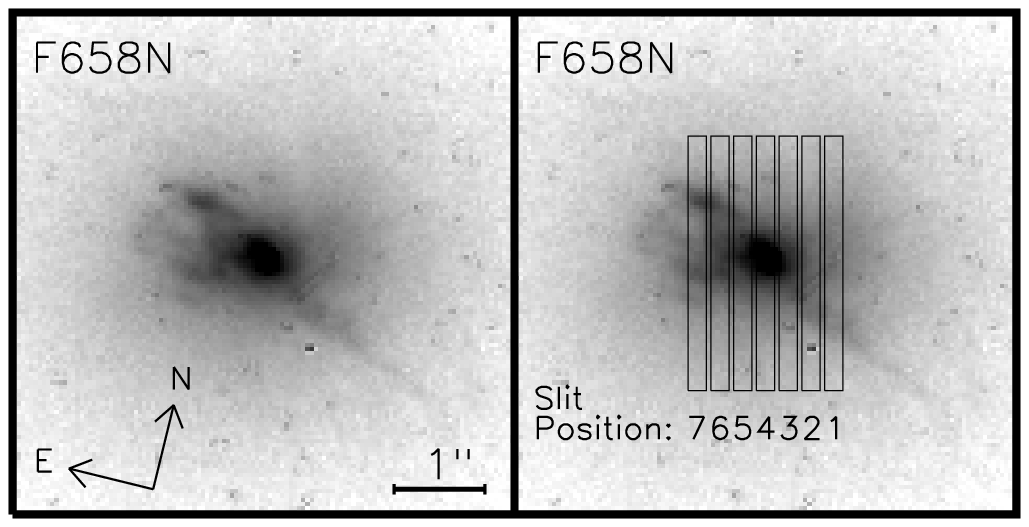}{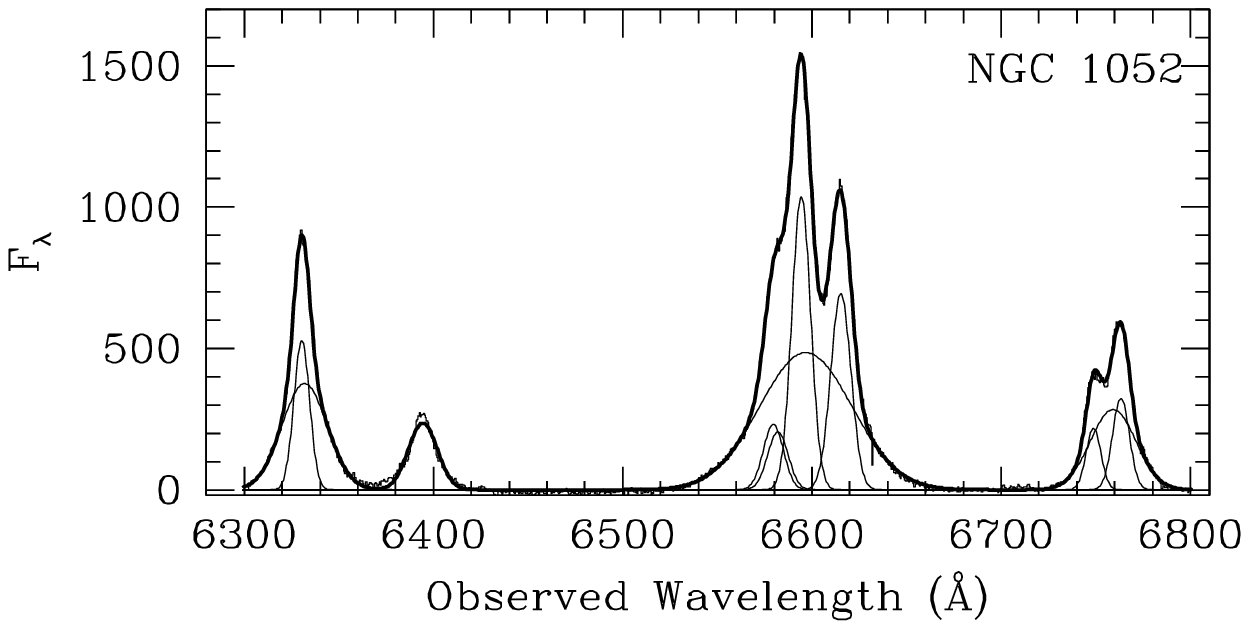}
\epsscale{0.9} 
\plotone{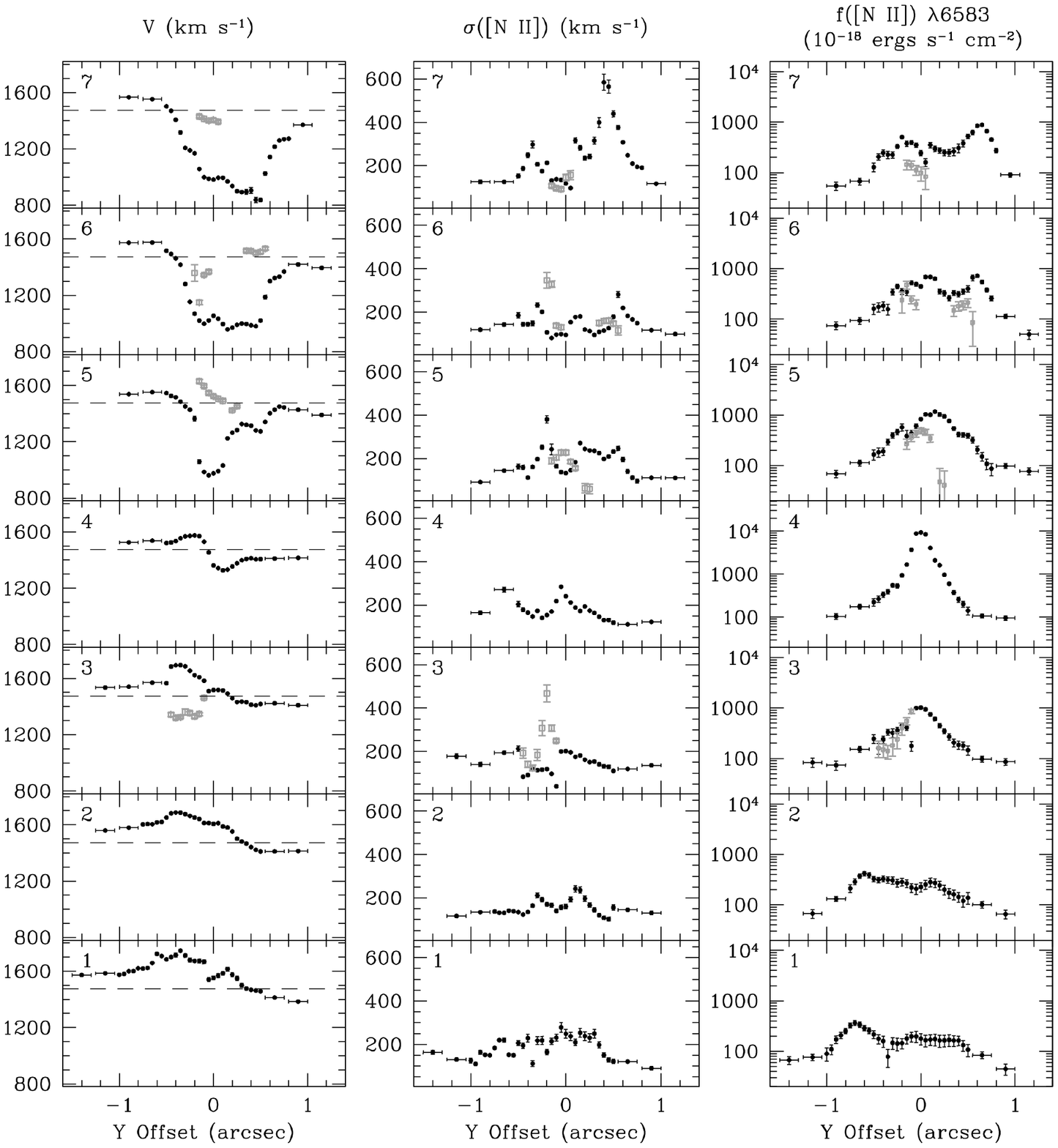}
\caption{The velocity, velocity dispersion, and flux as a function of
position along the slit for each of the slit positions in NGC
1052. Spectral rows that were binned together in order to improve the
S/N are shown with error bars in the $x$-direction, where the error
bars denote the radial range over which the spectra were
extracted. The grey points represent separate velocity components. The
horizontal dashed line shows the galaxy's systemic velocity that was
taken from optical line measurements in RC3. The positive and negative
$y$-offsets correspond to the spectra extracted from the rows above
and below the central row, respectively, when the slits are oriented
such that the STIS instrumental $y$-axis points directly upward. Thus,
the bottom half of the slit drawn over the F658N WFPC2 image
corresponds to the left side of the velocity, velocity dispersion, and
flux plots. The length of the slits in the same diagram corresponds to
the length over which we could measure the emission lines. The
upper-right panel shows the spectrum extracted from the nuclear region
of the central slit, where the flux density is in units of $10^{-18}$
ergs s$^{-1}$ cm$^{-2}$ \AA$^{-1}$. No WFPC2 continuum images were
available. \label{fig:ngc1052}}
\end{center}
\end{figure}

\begin{figure}
\begin{center}
\figurenum{2b}
\epsscale{1.0} 
\plottwo{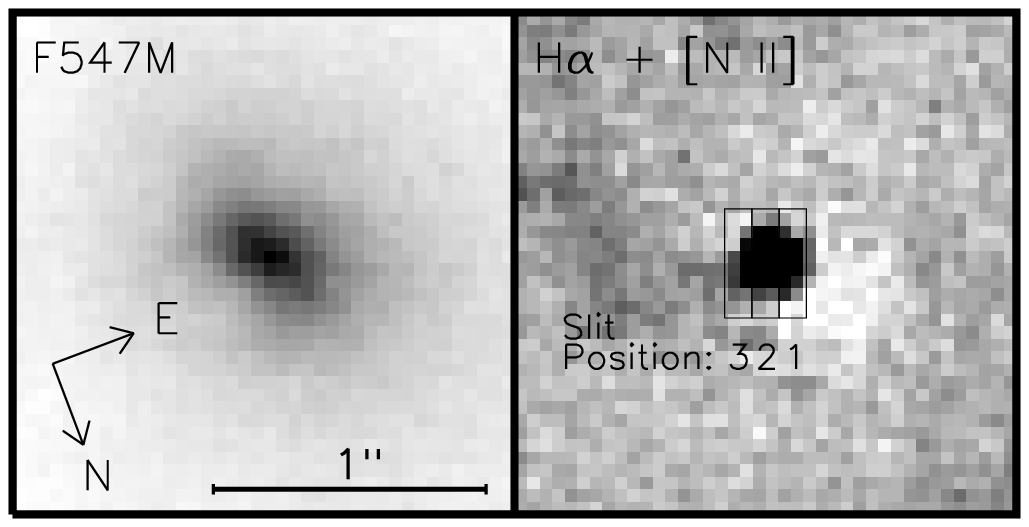}{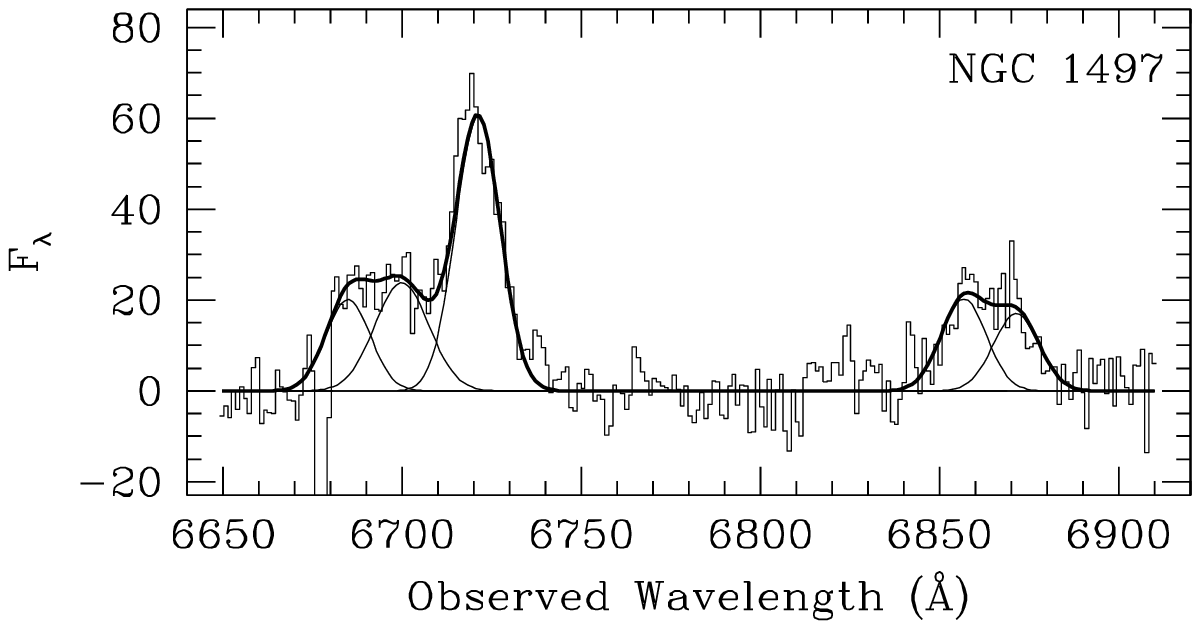}
\epsscale{1.0} 
\plotone{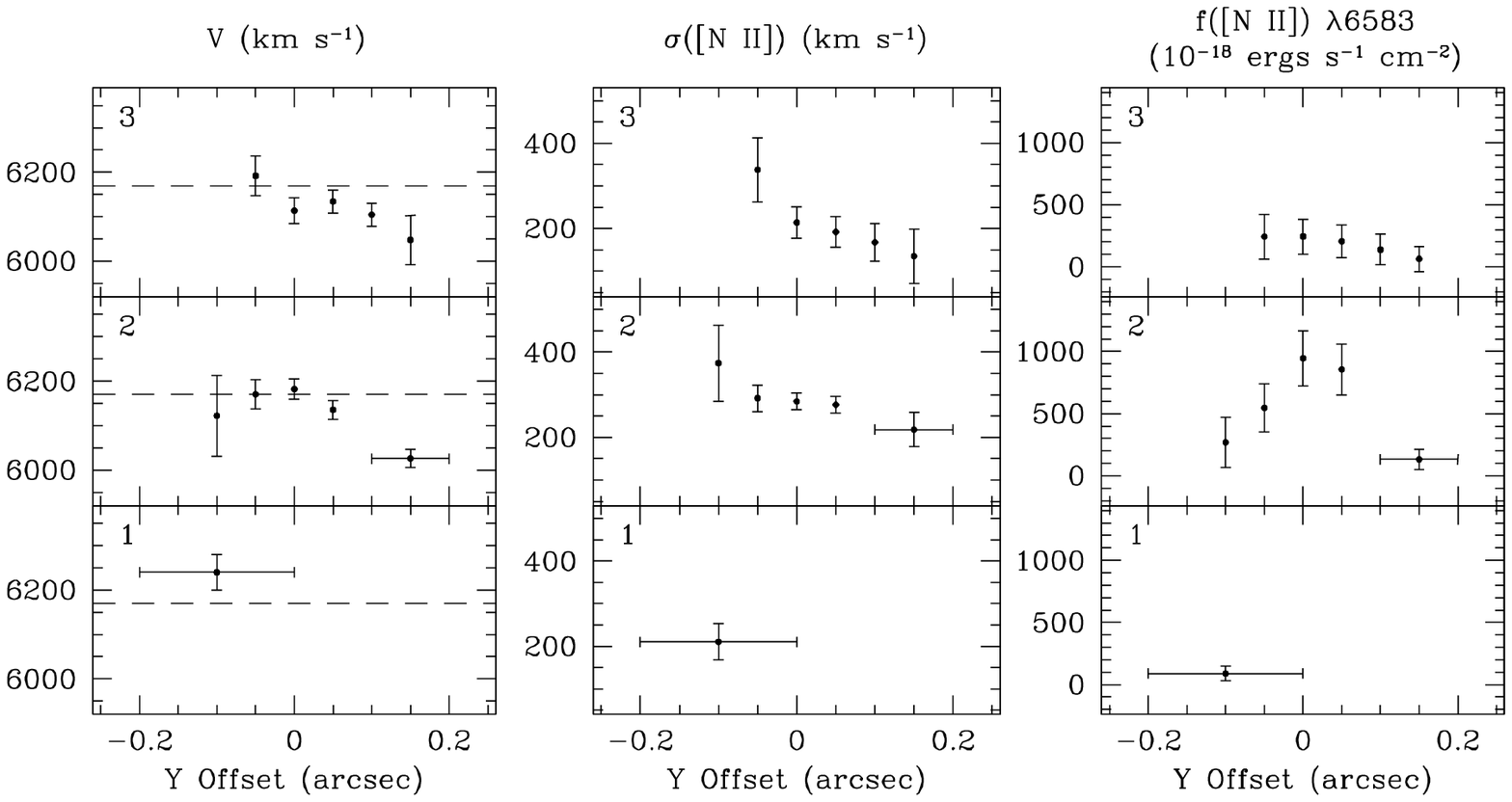}
\caption{NGC 1497; see Figure \ref{fig:ngc1052} for
description. \label{fig:ngc1497}}
\end{center}
\end{figure}

\begin{figure}
\begin{center}
\figurenum{2c}
\epsscale{1.0} 
\plottwo{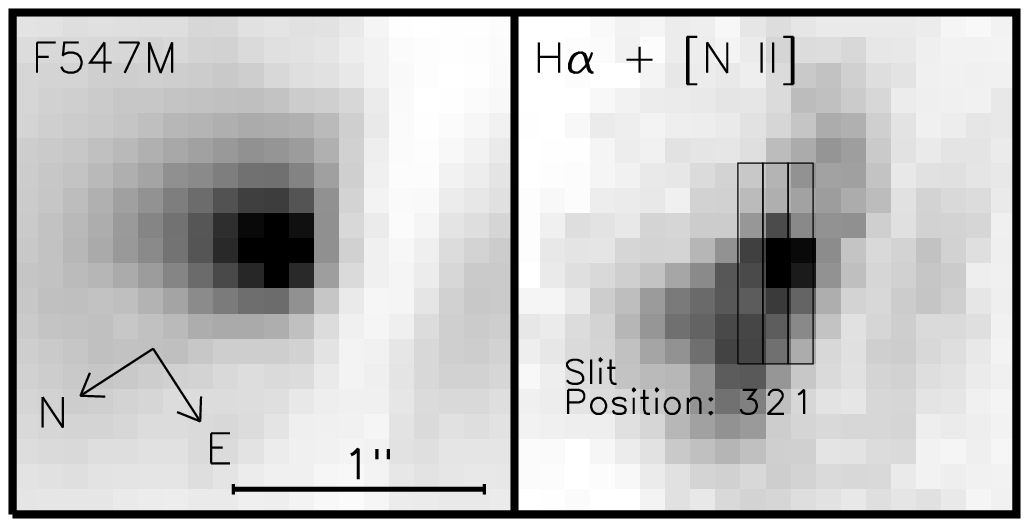}{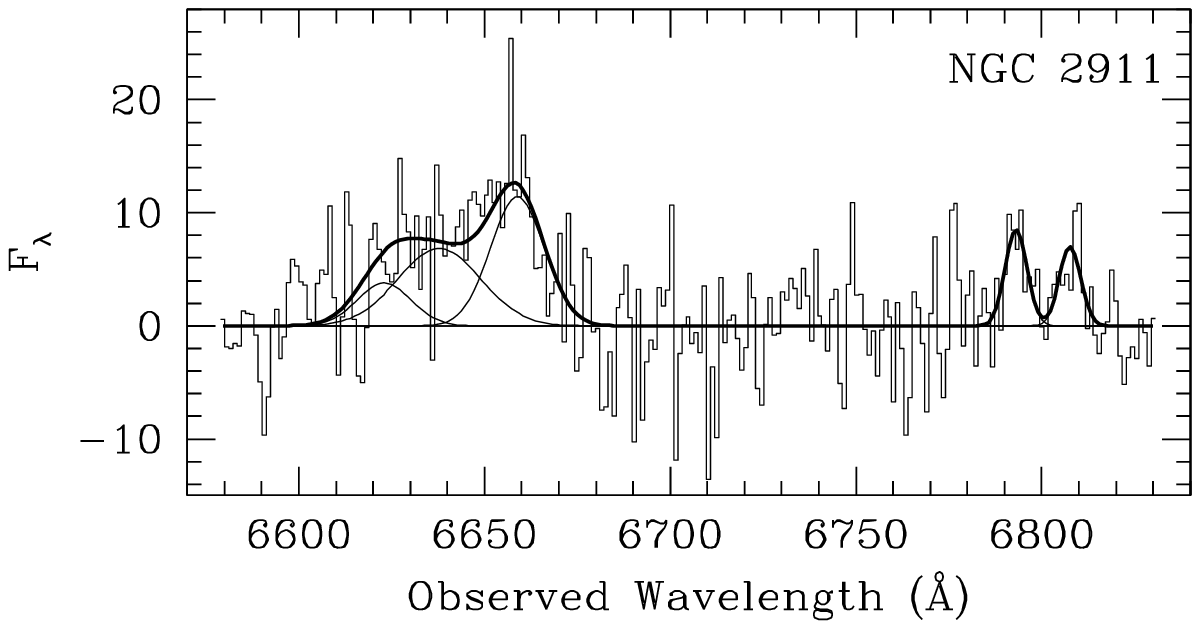}
\epsscale{1.0} 
\plotone{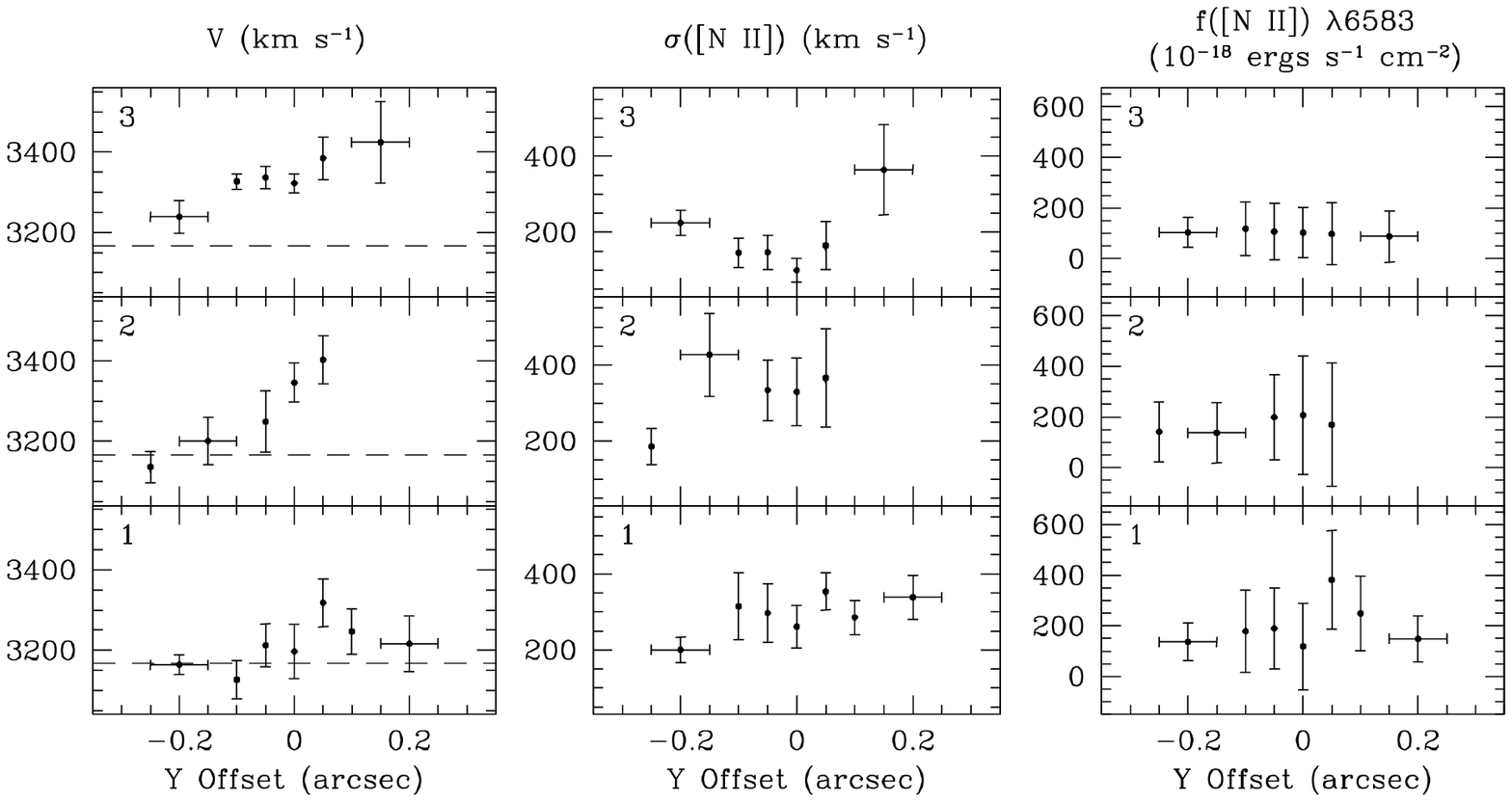}
\caption{NGC 2911; see Figure \ref{fig:ngc1052} for
description. \label{fig:ngc2911}}
\end{center}
\end{figure}

\begin{figure}
\begin{center}
\figurenum{2d}
\epsscale{0.9}
\plottwo{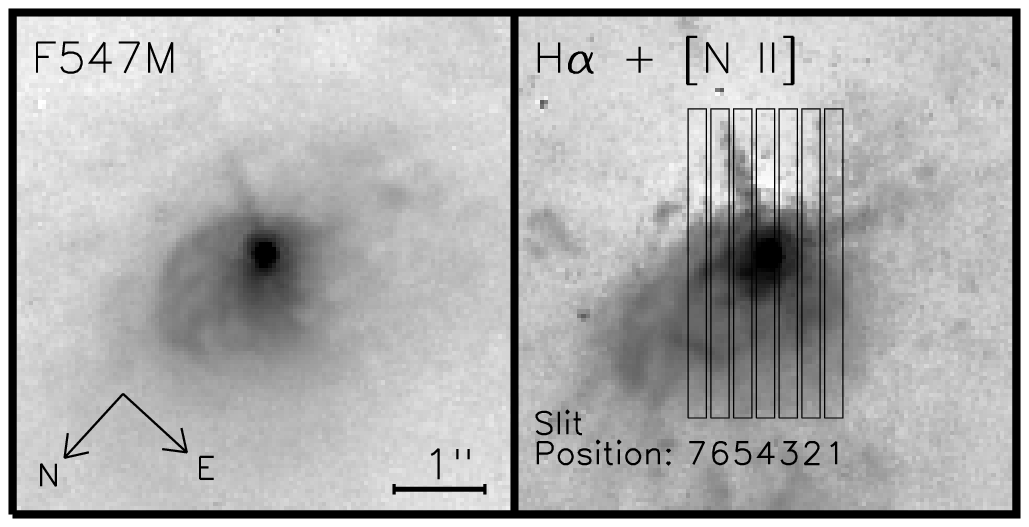}{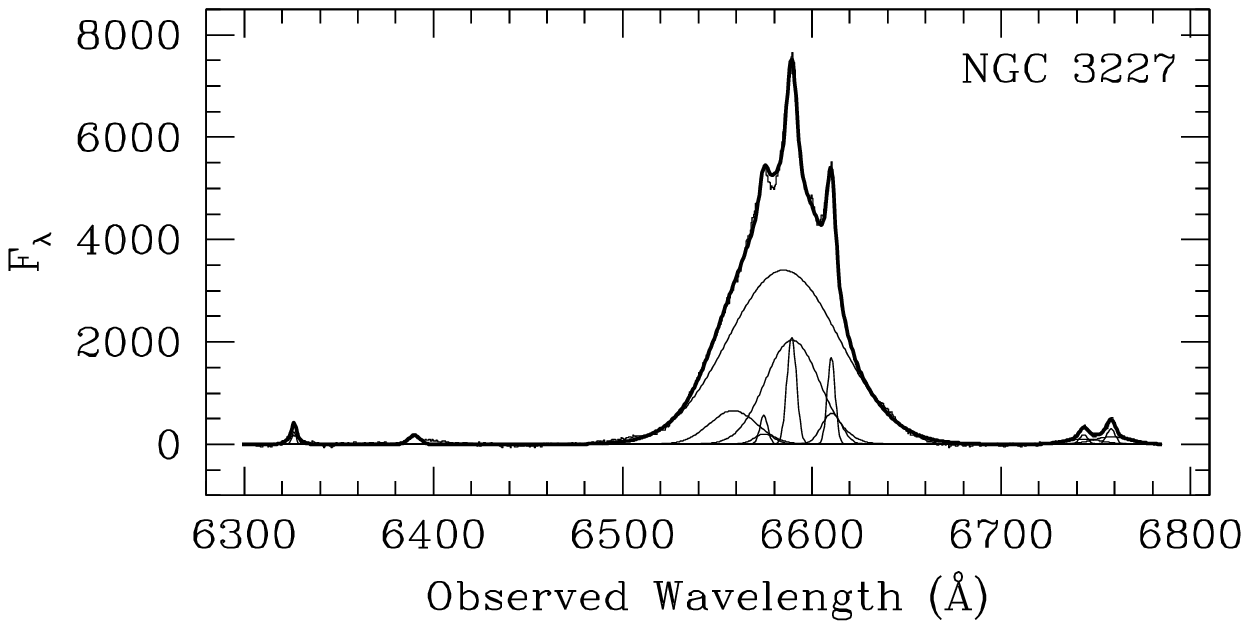}
\epsscale{0.9} 
\plotone{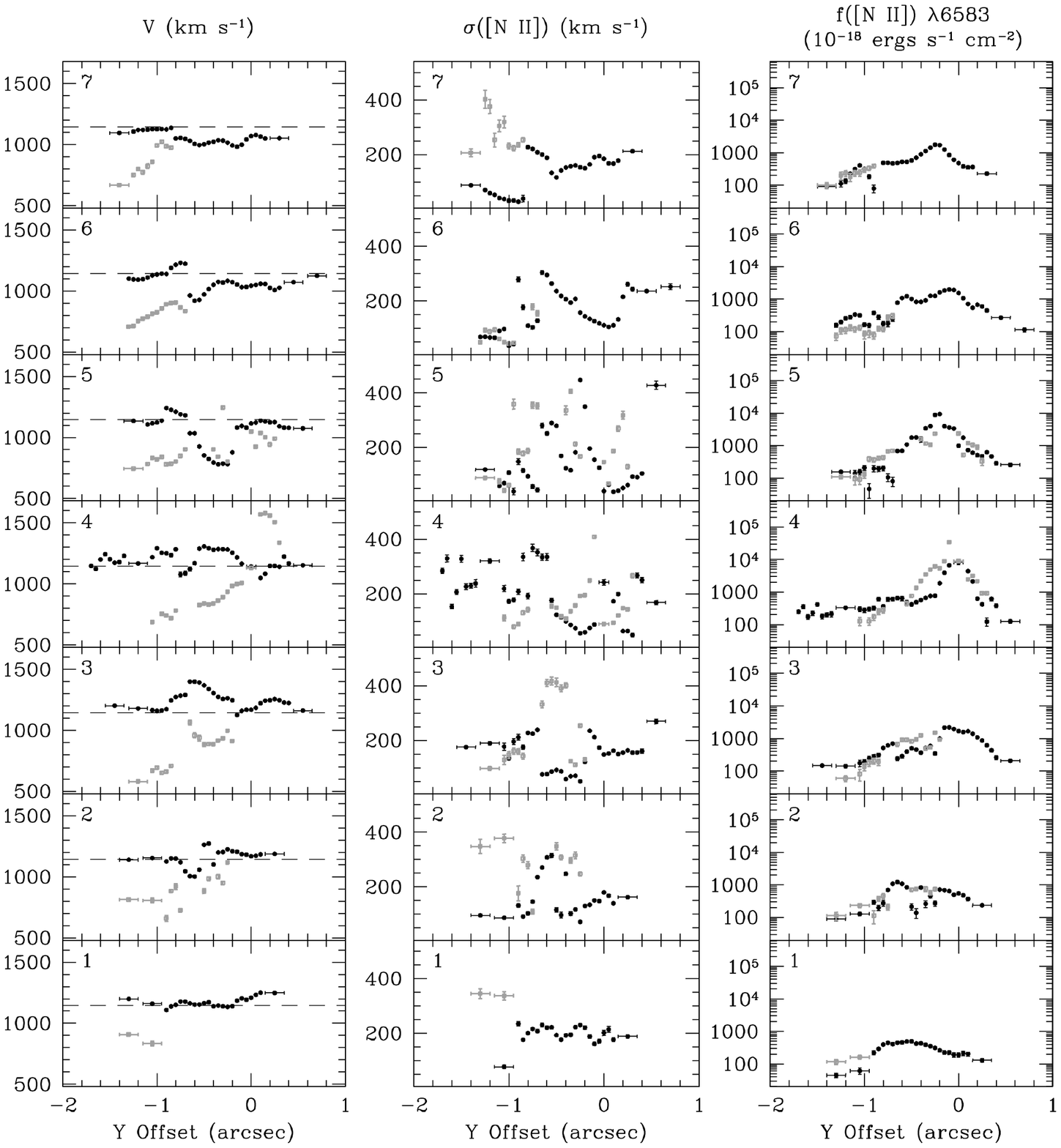}
\caption{NGC 3227; see Figure \ref{fig:ngc1052} for
description. \label{fig:ngc3227}}
\end{center}
\end{figure}

\begin{figure}
\begin{center}
\figurenum{2e}
\epsscale{1.0}
\plottwo{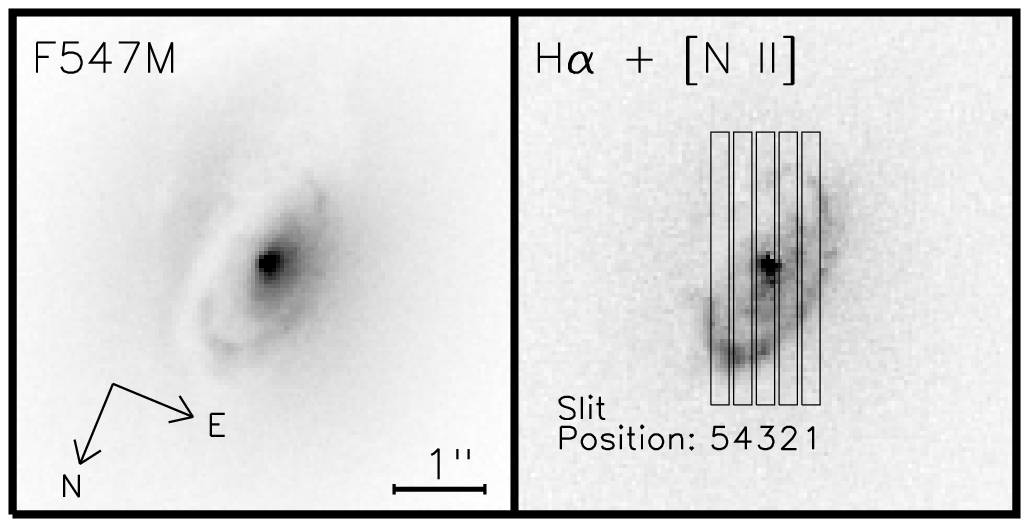}{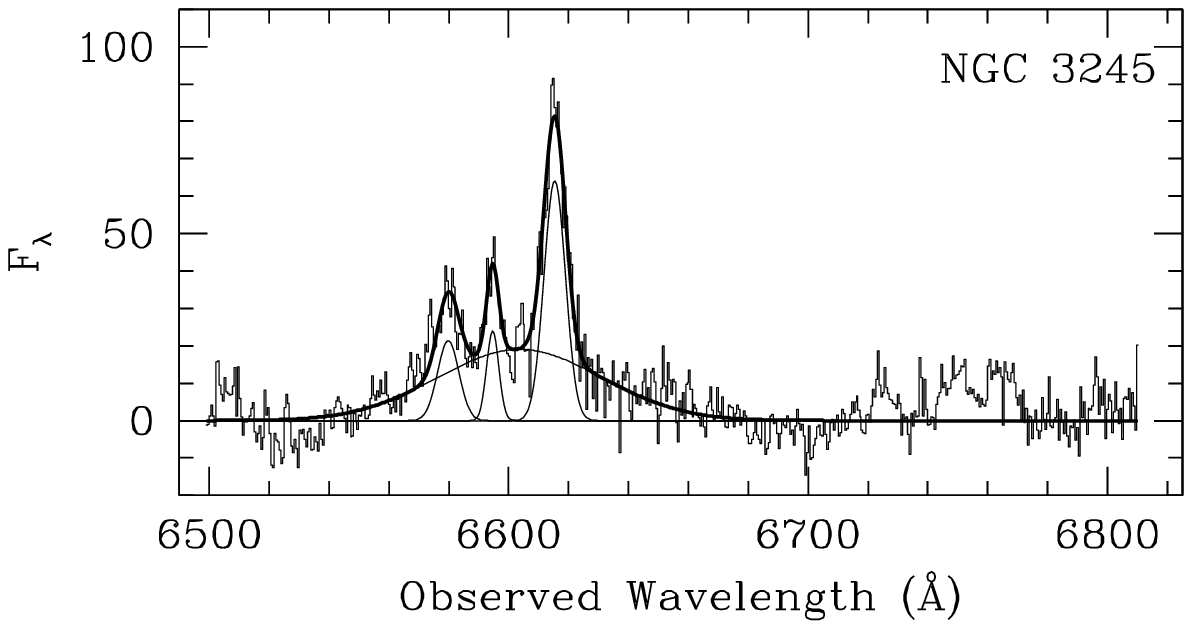}
\epsscale{1.0} 
\plotone{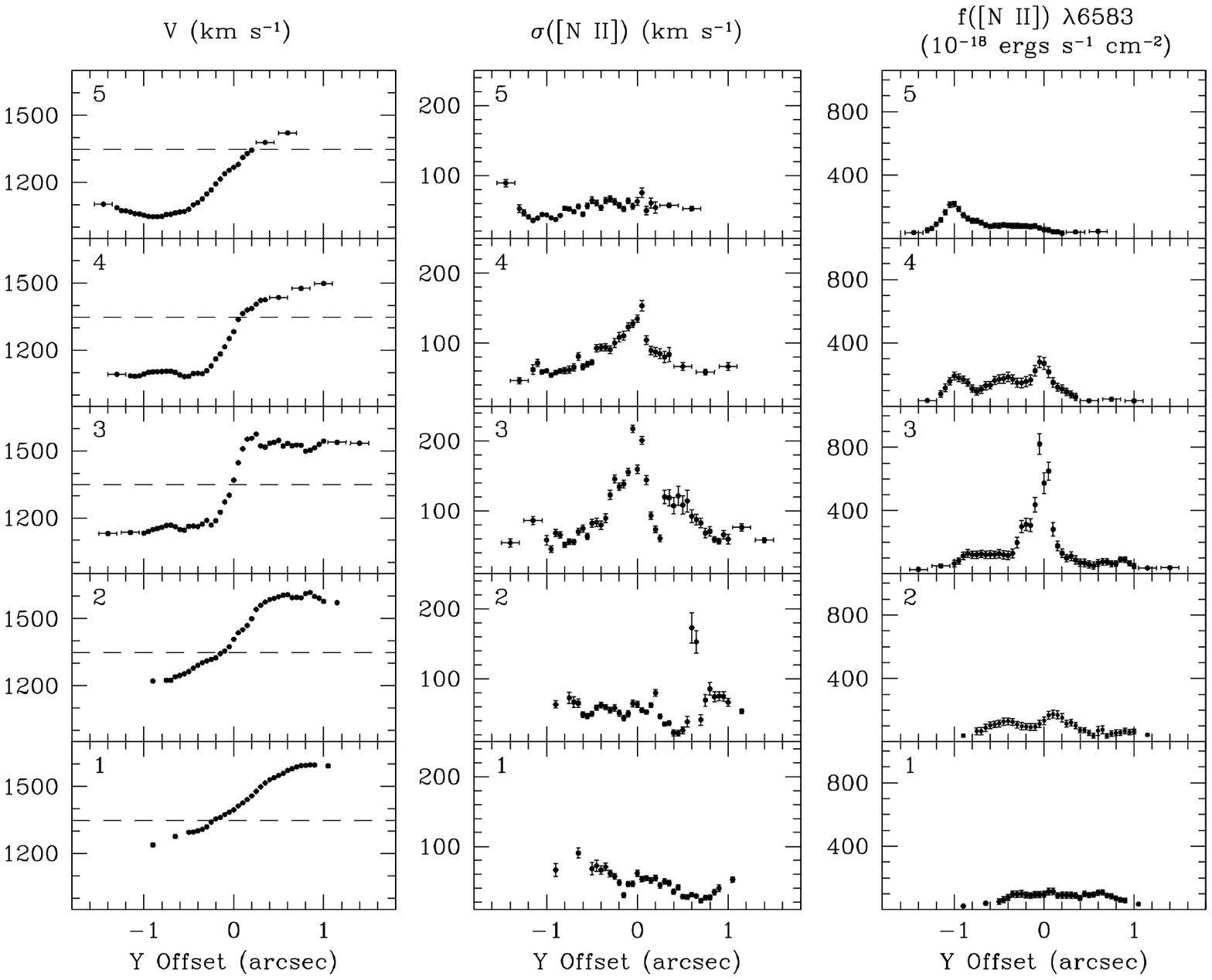}
\caption{NGC 3245; see Figure \ref{fig:ngc1052} for description. The
S/N was too low to accurately fit the [S II] lines. See also
\cite{Barth_2001}. \label{fig:ngc3245}}
\end{center}
\end{figure}

\begin{figure}
\begin{center}
\figurenum{2f}
\epsscale{1.0}
\plottwo{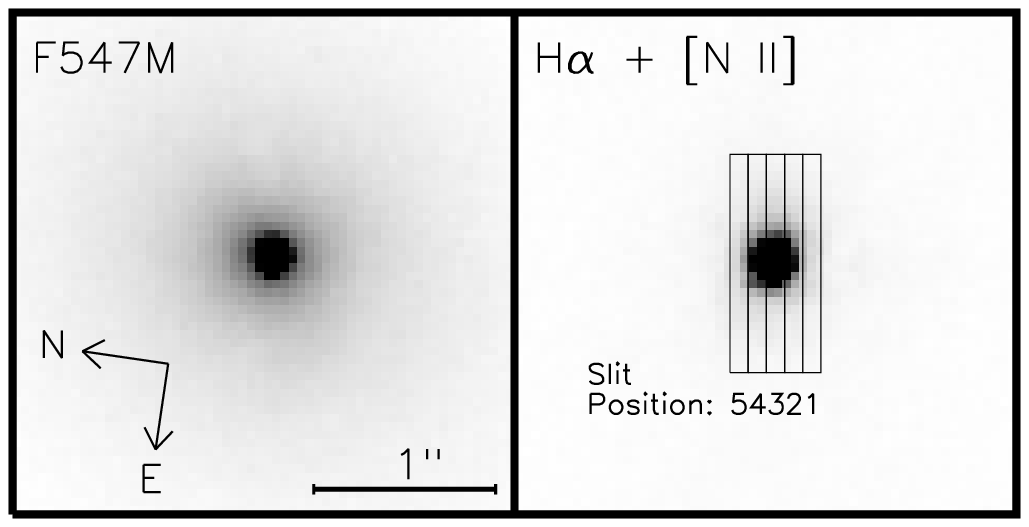}{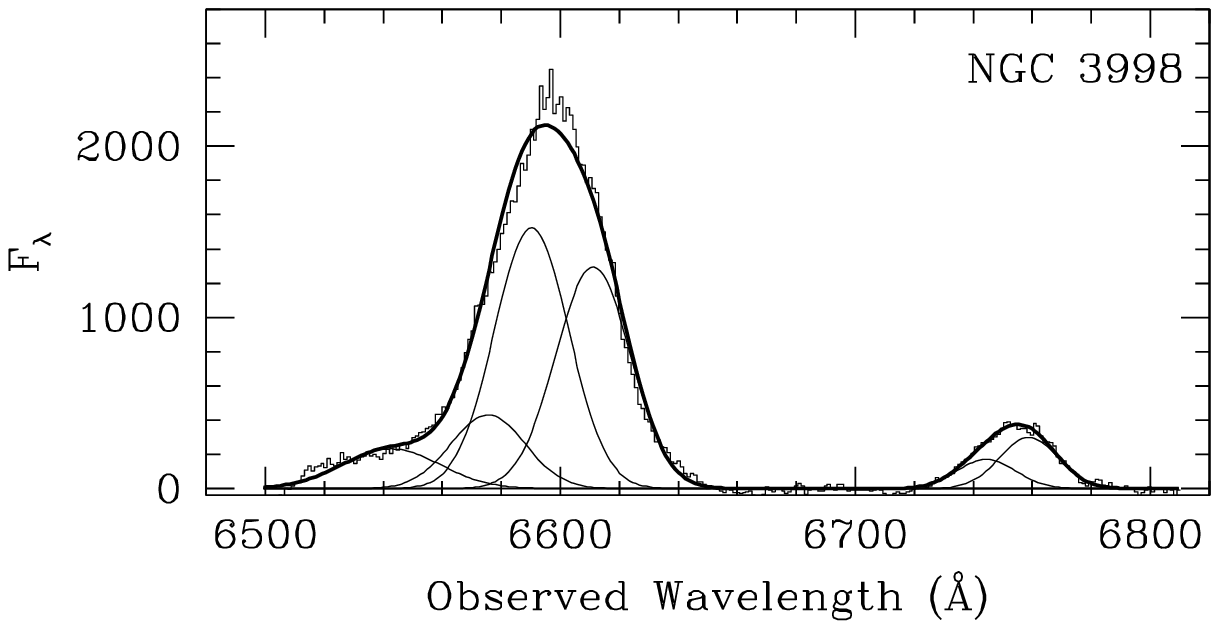}
\epsscale{1.0} 
\plotone{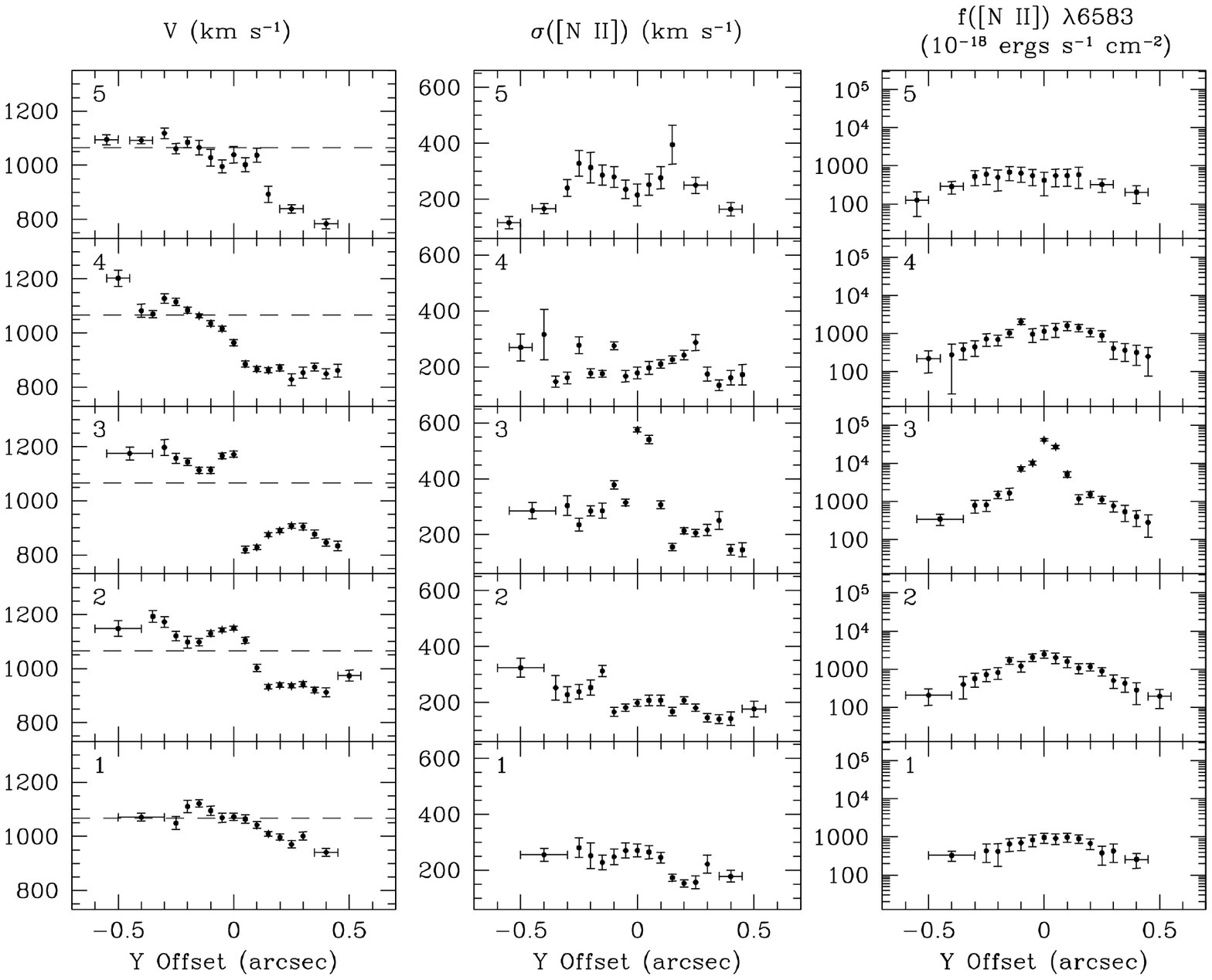}
\caption{NGC 3998; see Figure \ref{fig:ngc1052} for description. See
also \cite{DeFrancesco_2006}. \label{fig:ngc3998}}
\end{center}
\end{figure}

\begin{figure}
\begin{center}
\figurenum{2g}
\epsscale{1.0}
\plottwo{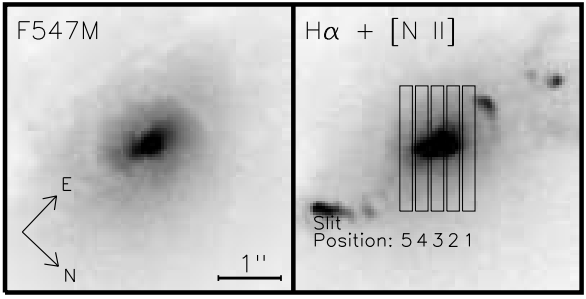}{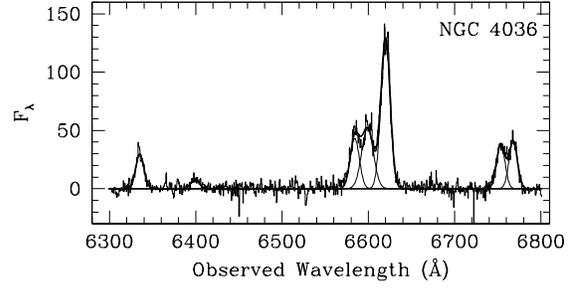}
\epsscale{1.0} 
\plotone{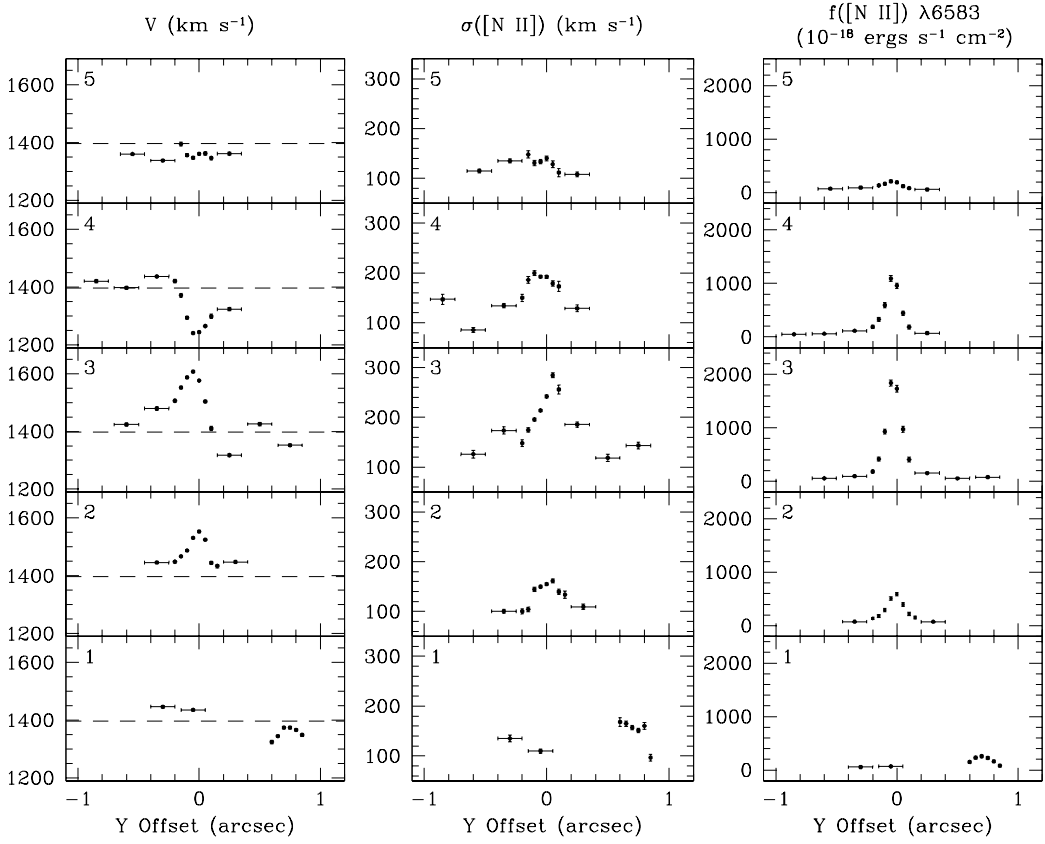}
\caption{NGC 4036; see Figure \ref{fig:ngc1052} for
description. \label{fig:ngc4036}}
\end{center}
\end{figure}

\begin{figure}
\begin{center}
\figurenum{2h}
\epsscale{1.0}
\plottwo{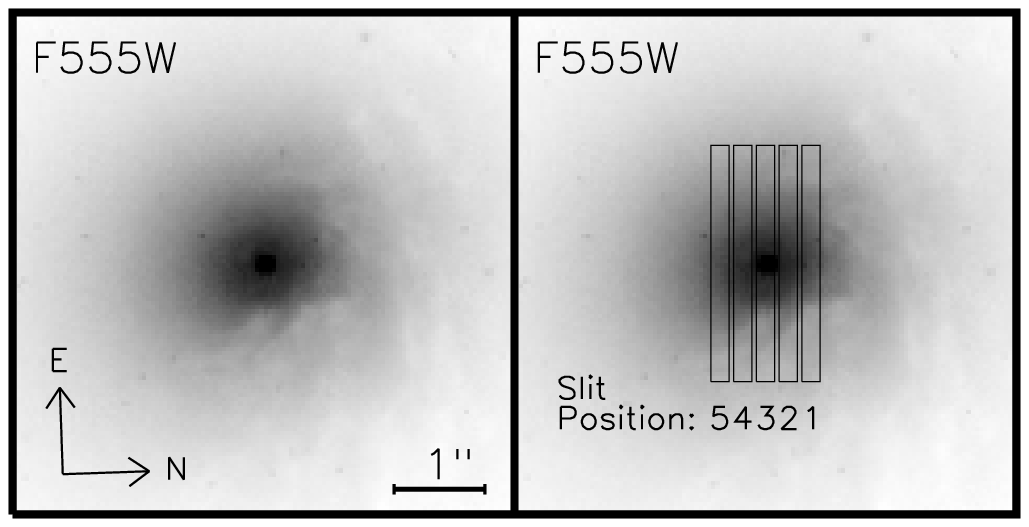}{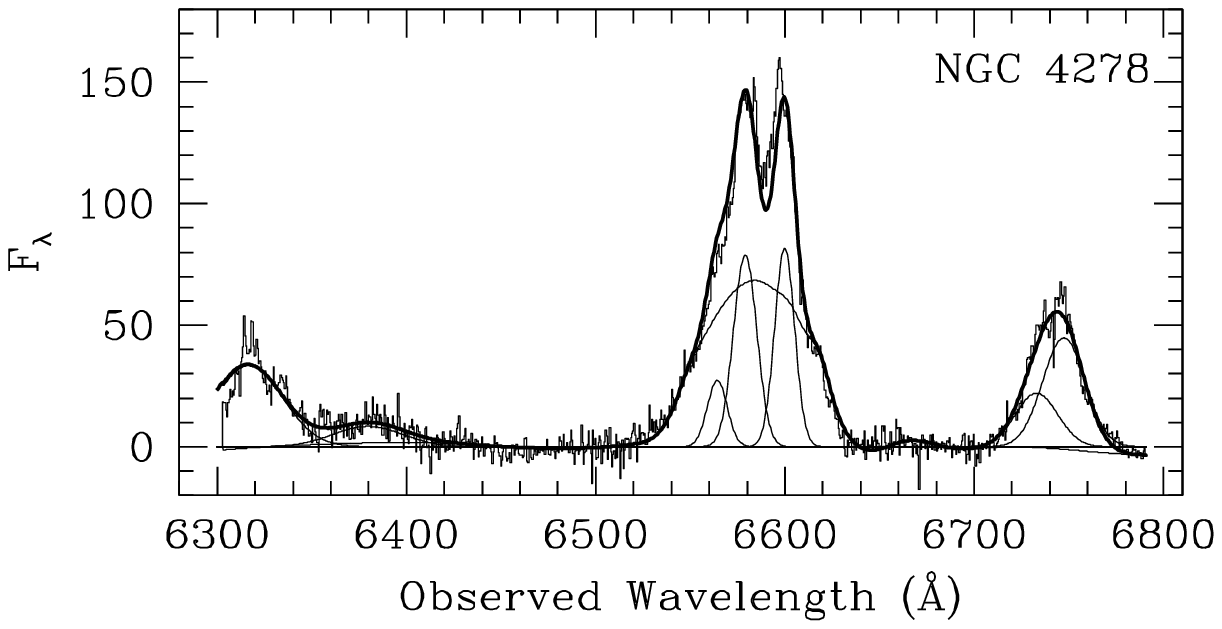}
\epsscale{1.0} 
\plotone{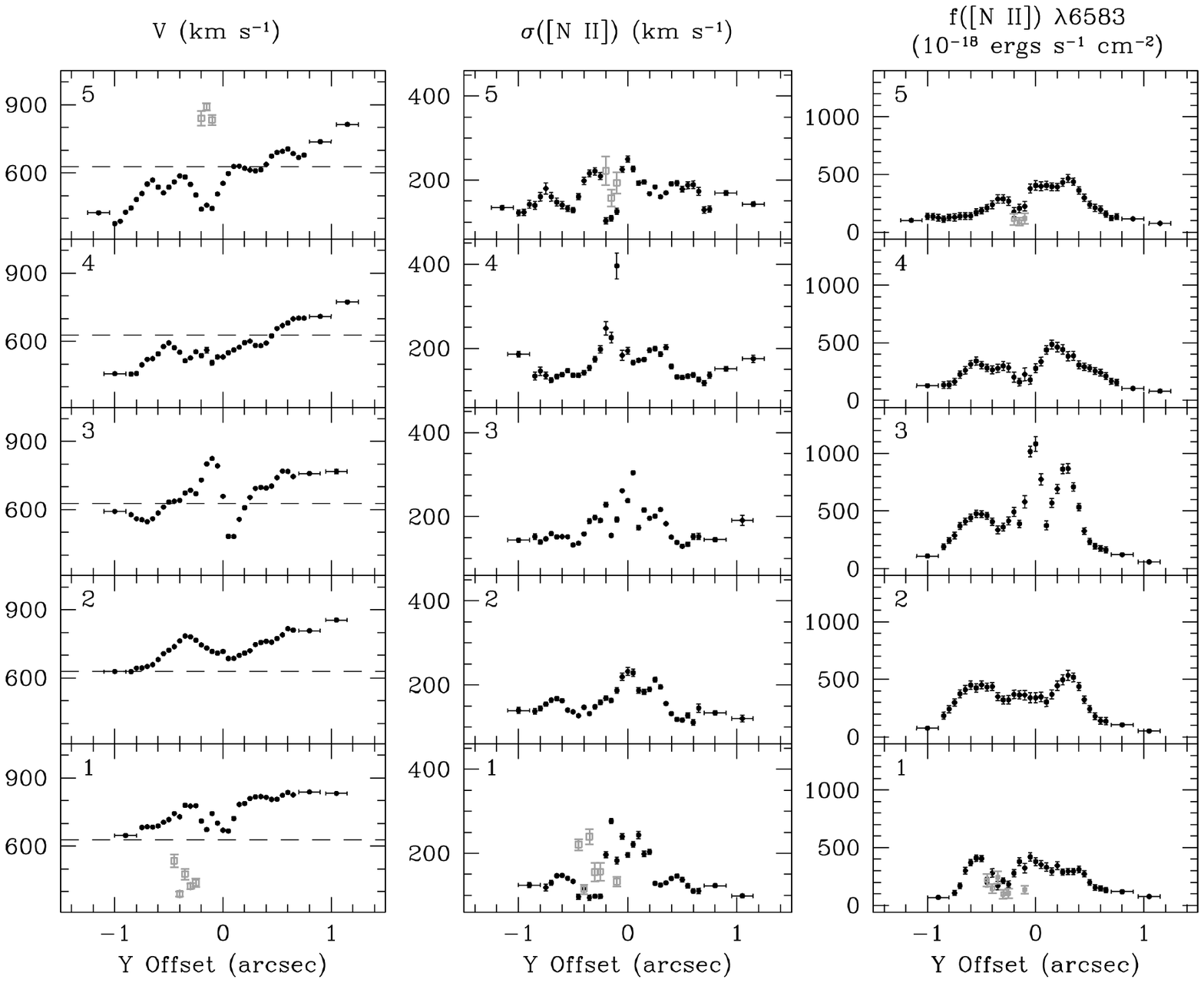}
\caption{NGC 4278; see Figure \ref{fig:ngc1052} for
description. \label{fig:ngc4278}}
\end{center}
\end{figure}

\begin{figure}
\begin{center}
\figurenum{2i}
\epsscale{1.0} 
\plottwo{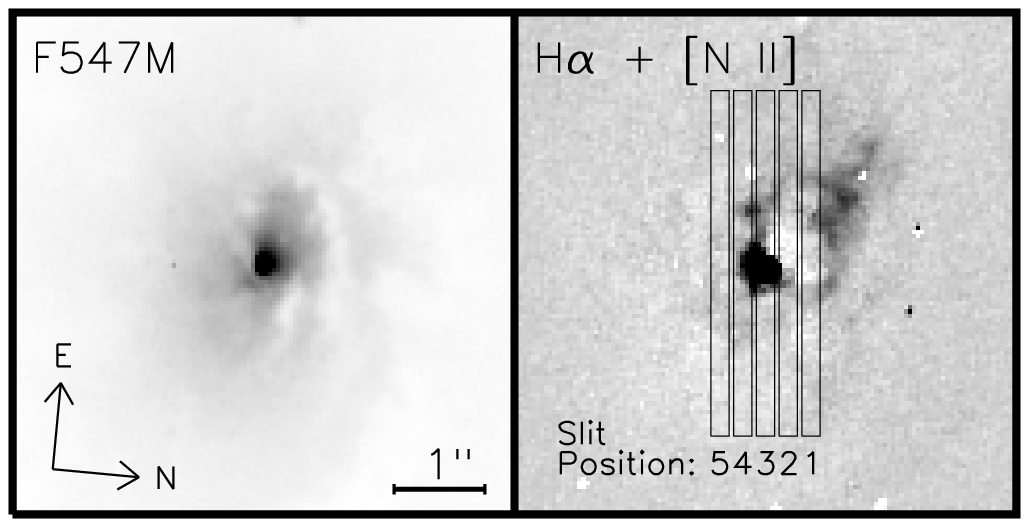}{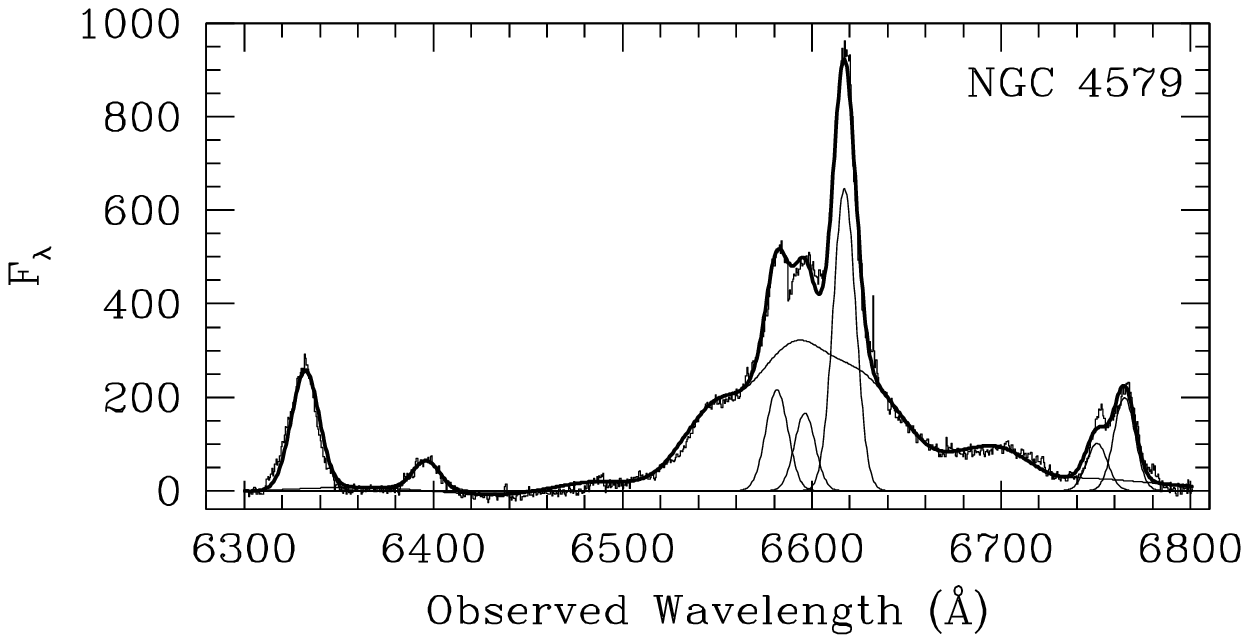}
\epsscale{1.0} 
\plotone{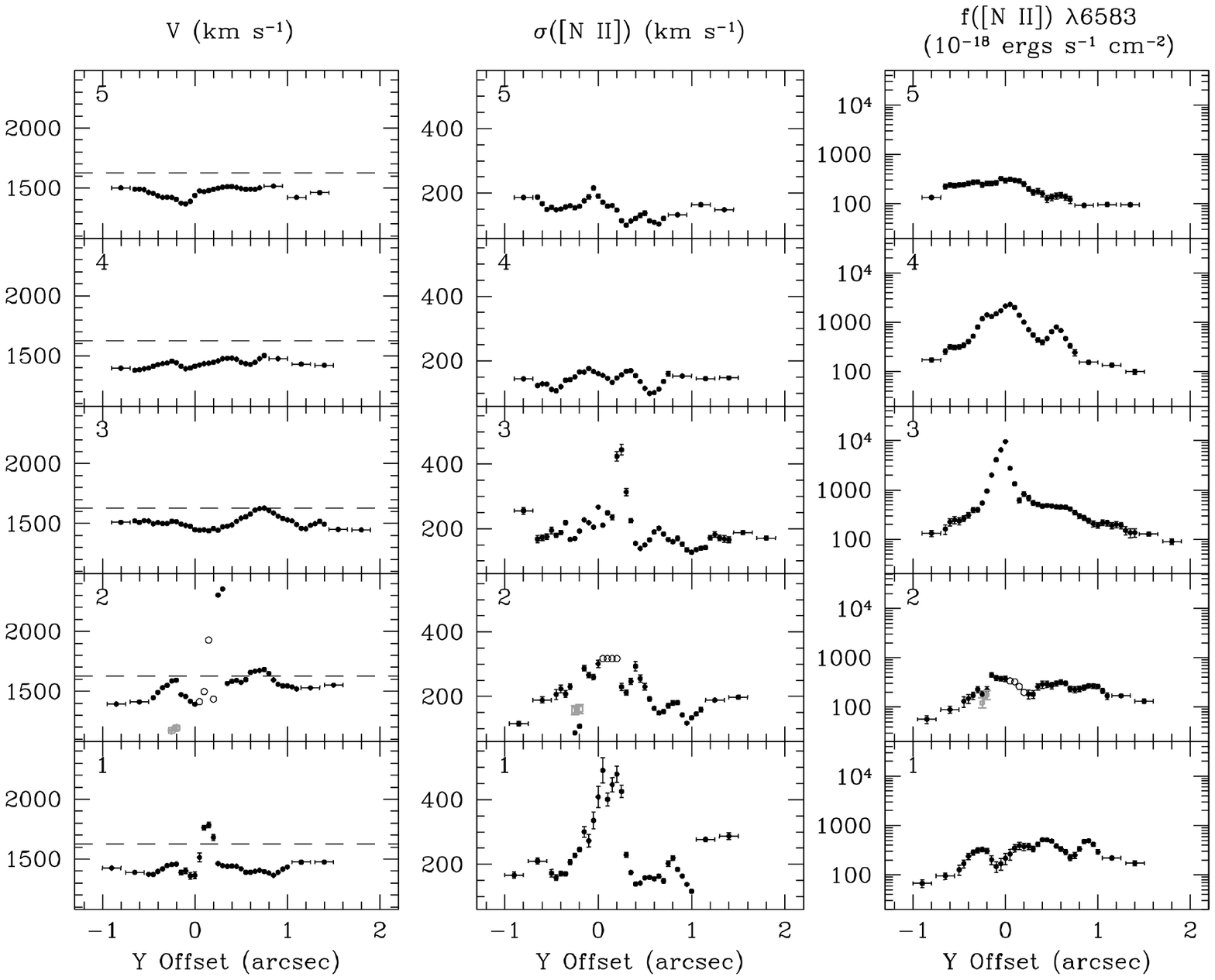}
\caption{NGC 4579; see Figure \ref{fig:ngc1052} for description. The
spectra near the central regions in the second slit position contained
severely blended H$\alpha$ and [\ion{N}{2}] lines, where the widths of
the three Gaussians were held fixed at a specific value in order to
achieve the most reasonable fit. The open circles correspond to rows
where this width restriction was necessary. \label{fig:ngc4579}}
\end{center}
\end{figure}

\begin{figure}
\begin{center}
\figurenum{2j}
\epsscale{1.0}
\plottwo{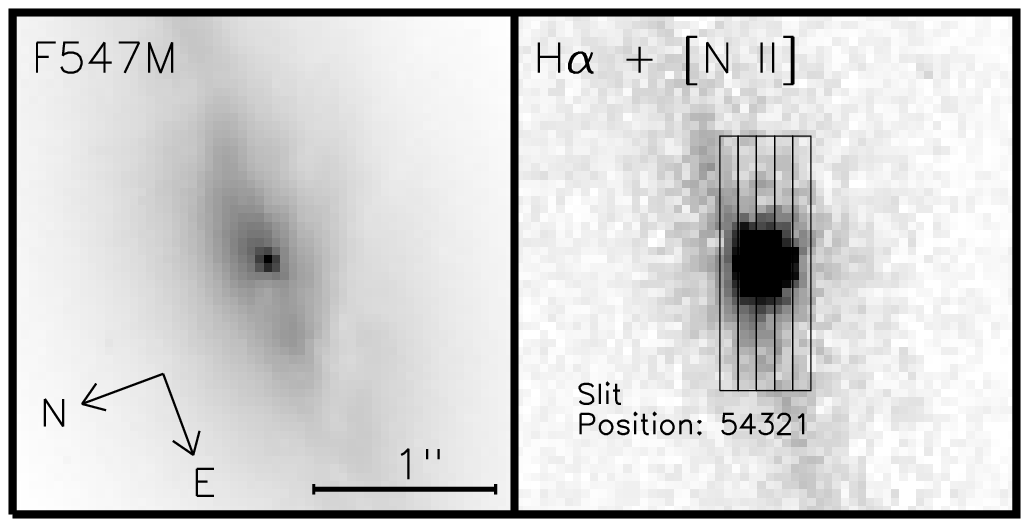}{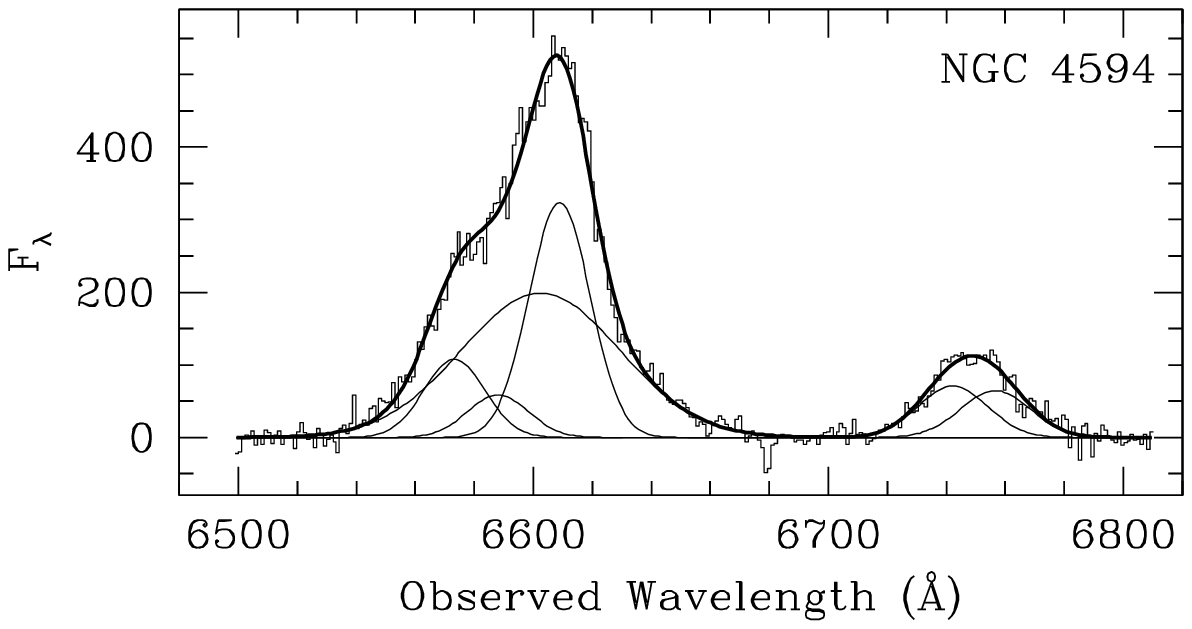}
\epsscale{1.0} 
\plotone{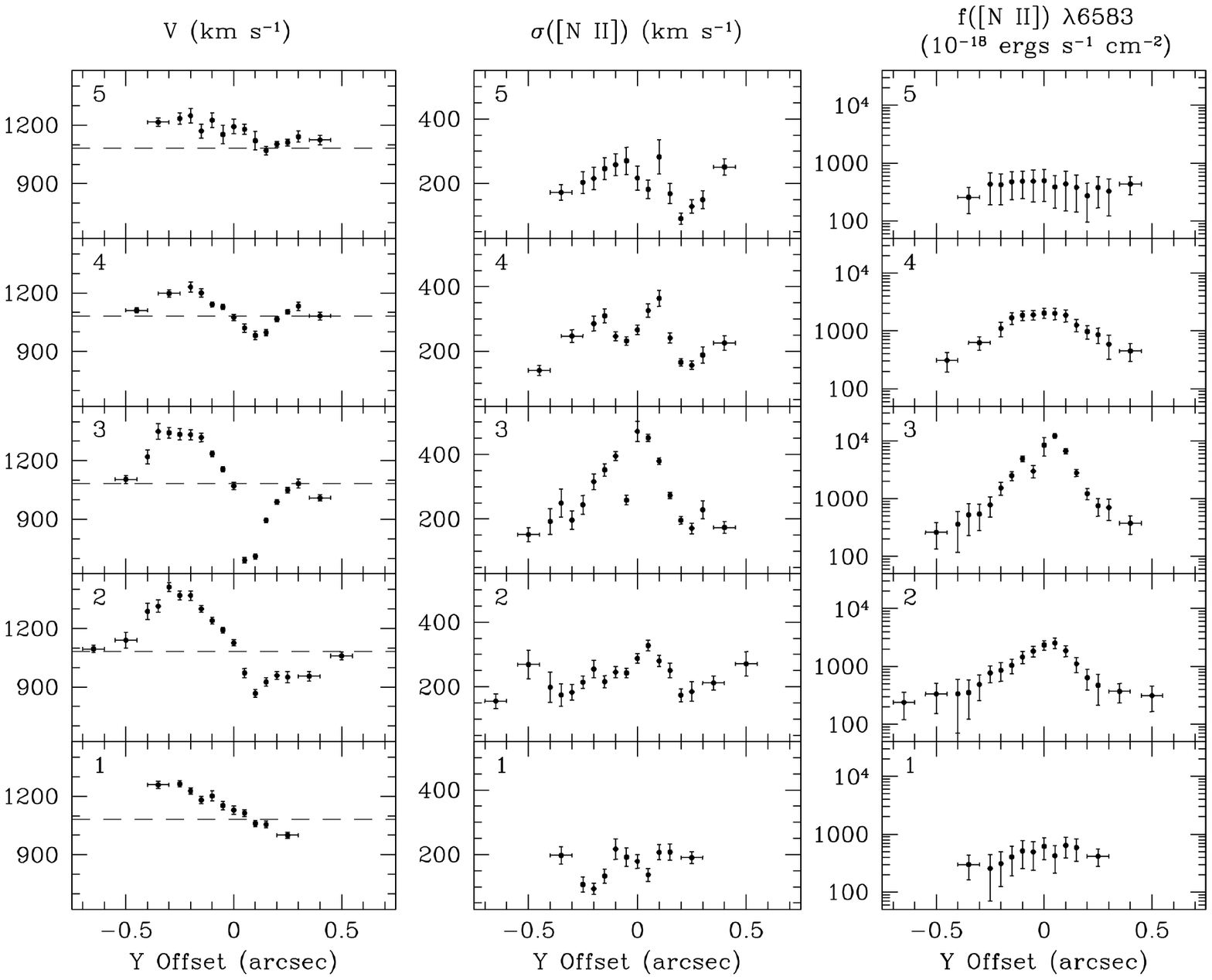}
\caption{NGC 4594; see Figure \ref{fig:ngc1052} for
description. \label{fig:ngc4594}}
\end{center}
\end{figure}

\begin{figure}
\begin{center}
\figurenum{2k}
\epsscale{1.0} 
\plottwo{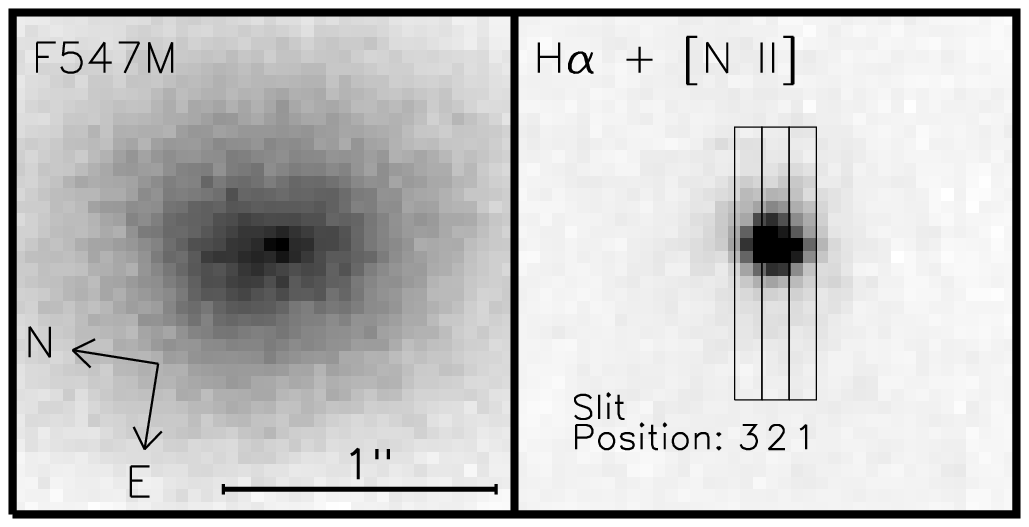}{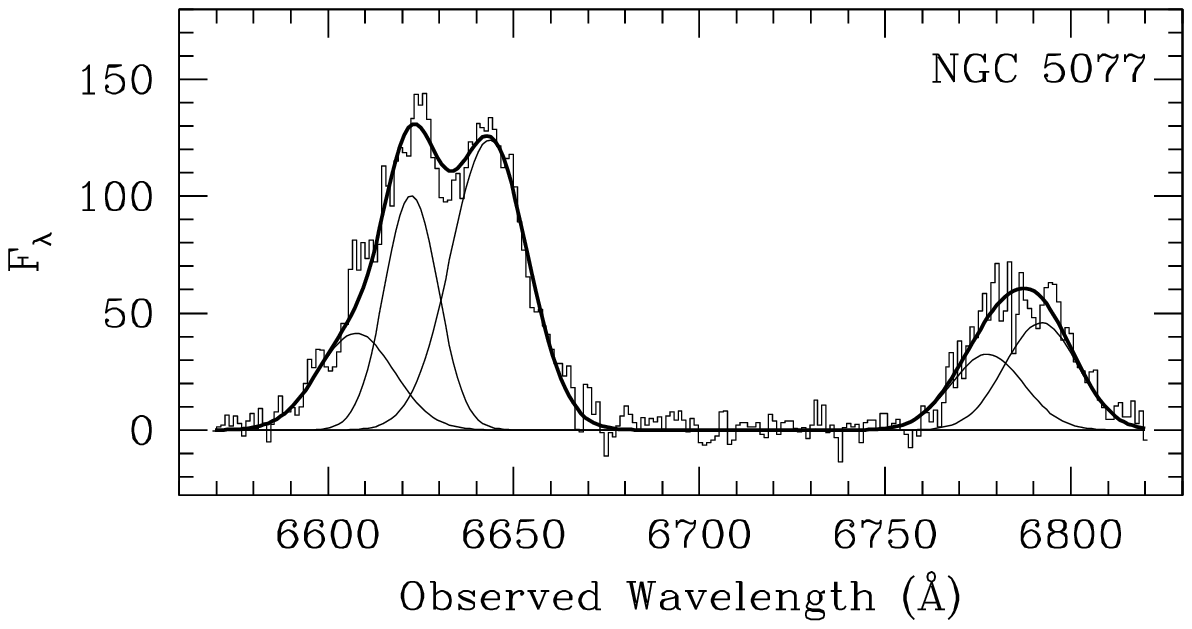}
\epsscale{1.0} 
\plotone{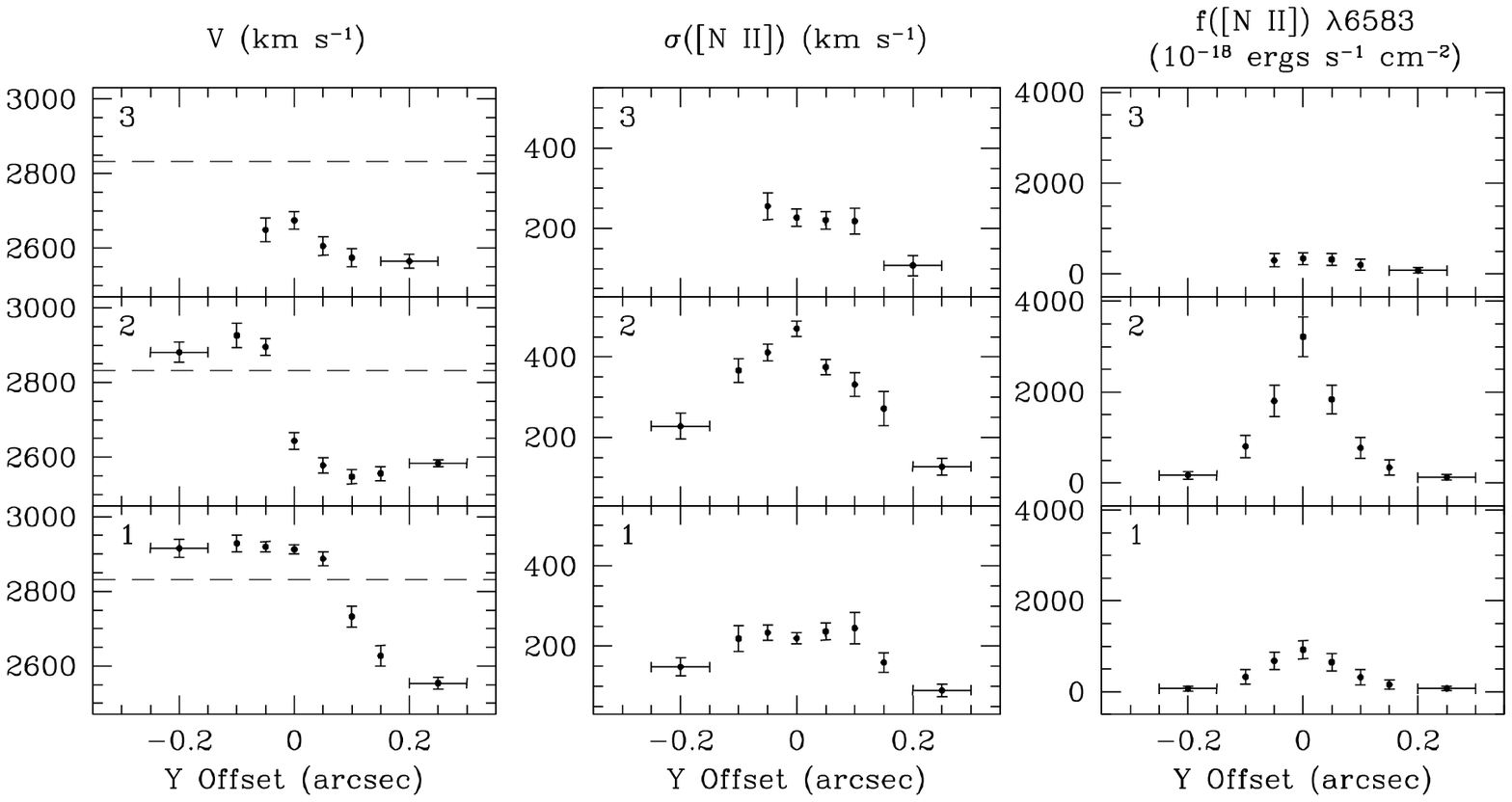}
\caption{NGC 5077; see Figure \ref{fig:ngc1052} for description. See
also \cite{DeFrancesco_2008}. \label{fig:ngc5077}}
\end{center}
\end{figure}

\begin{figure}
\begin{center}
\figurenum{2l}
\epsscale{1.0} 
\plottwo{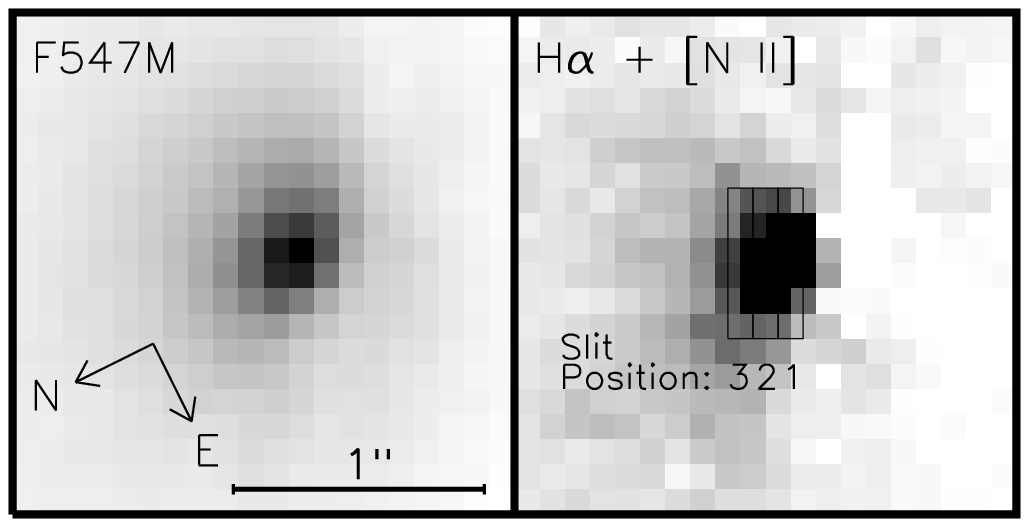}{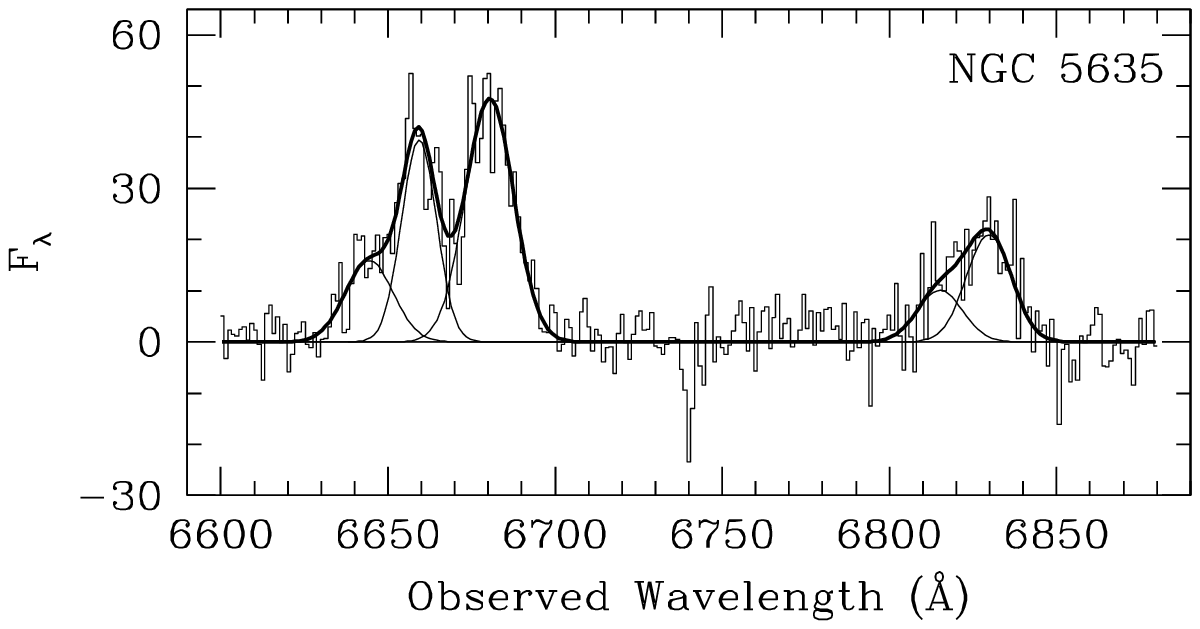}
\epsscale{1.0} 
\plotone{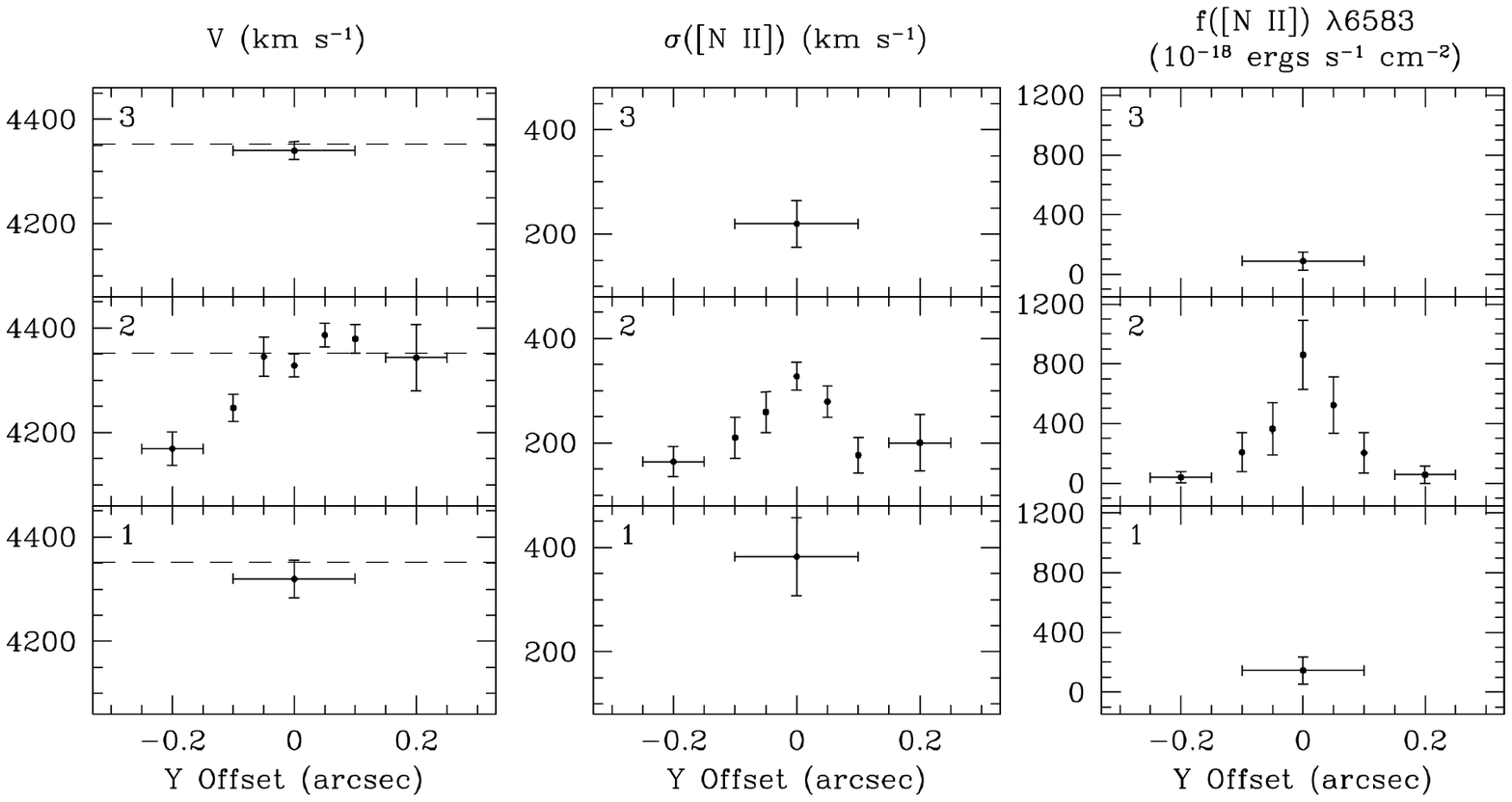}
\caption{NGC 5635; see Figure \ref{fig:ngc1052} for
description. \label{fig:ngc5635}}
\end{center}
\end{figure}

\begin{figure}
\begin{center}
\figurenum{2m}
\epsscale{1.0} 
\plottwo{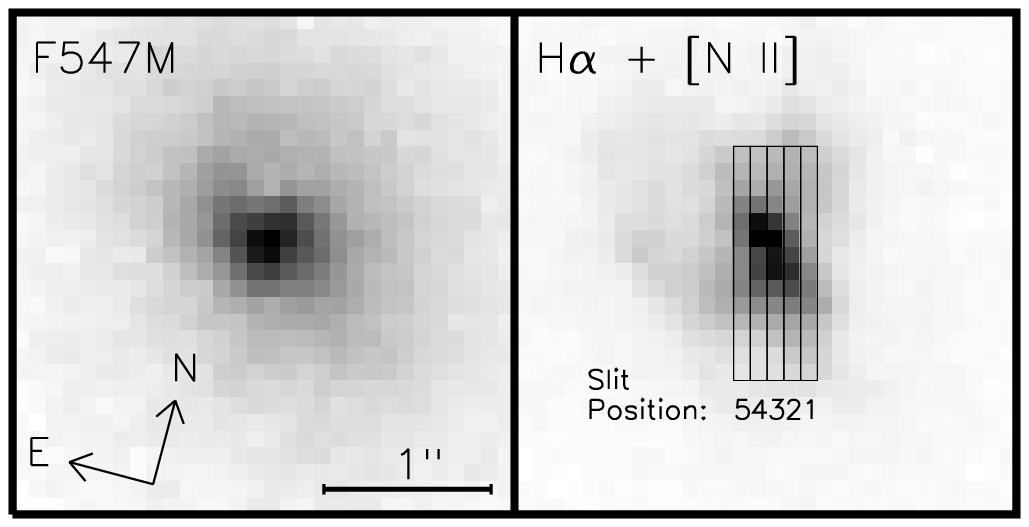}{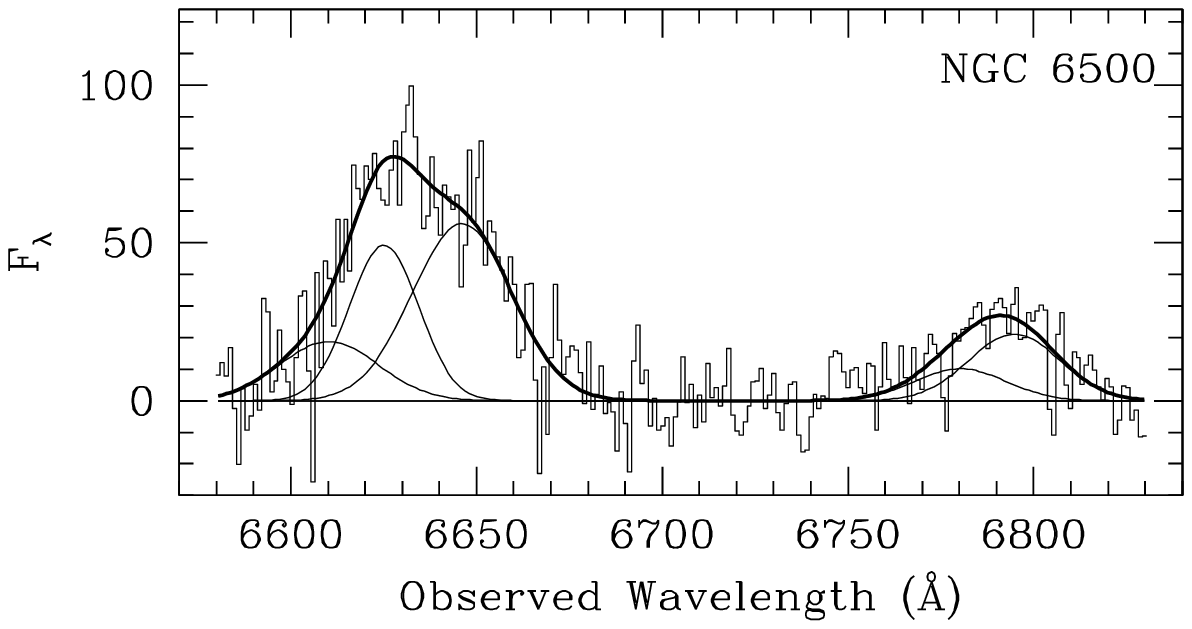}
\epsscale{1.0} 
\plotone{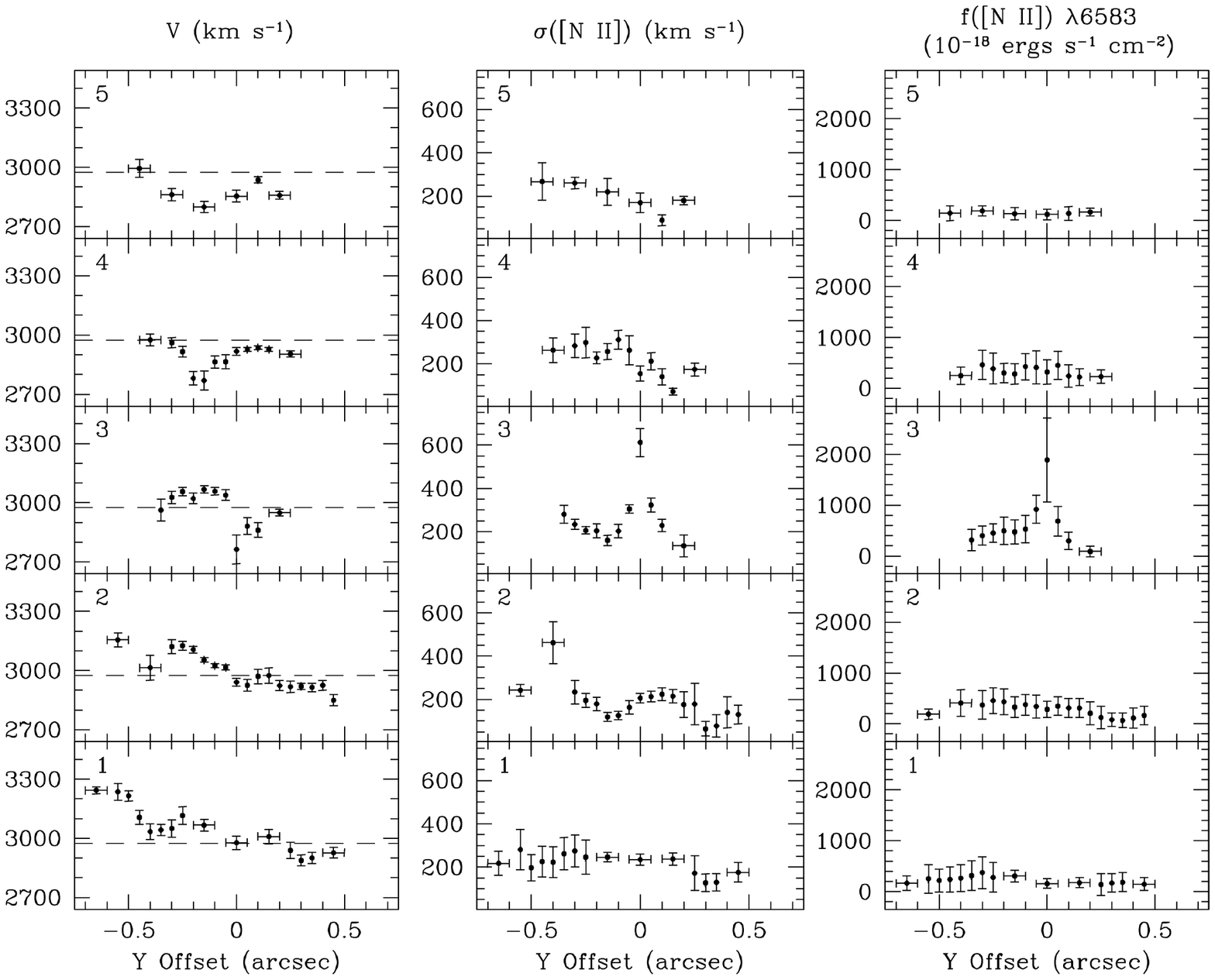}
\caption{NGC 6500; see Figure \ref{fig:ngc1052} for description. See
also \cite{DeFrancesco_2008}. \label{fig:ngc6500}}
\end{center}
\end{figure}

\begin{figure}
\begin{center}
\figurenum{2n}
\epsscale{1.0} 
\plottwo{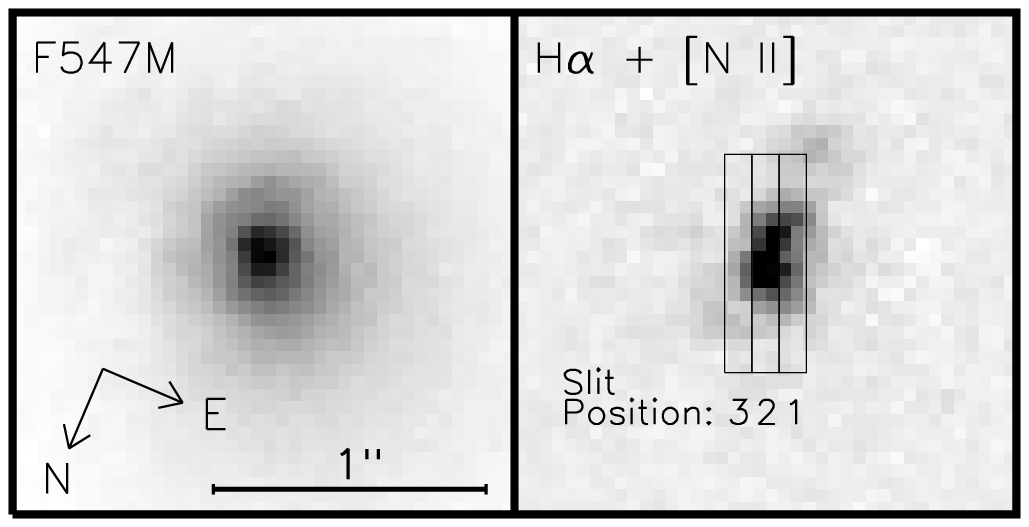}{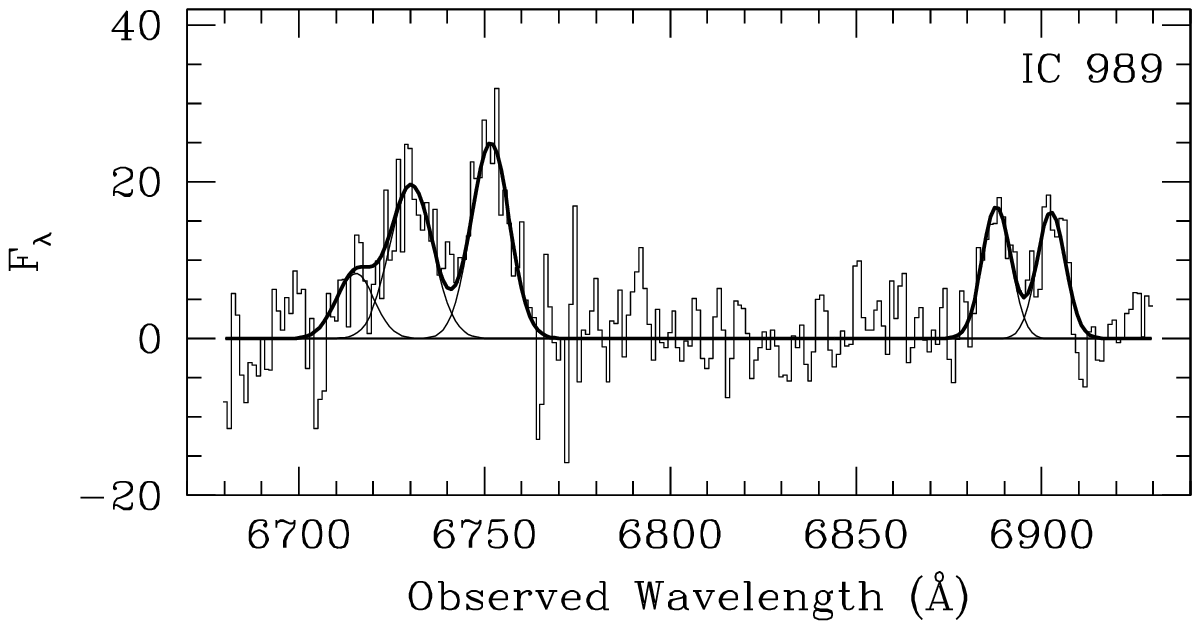}
\epsscale{1.0} 
\plotone{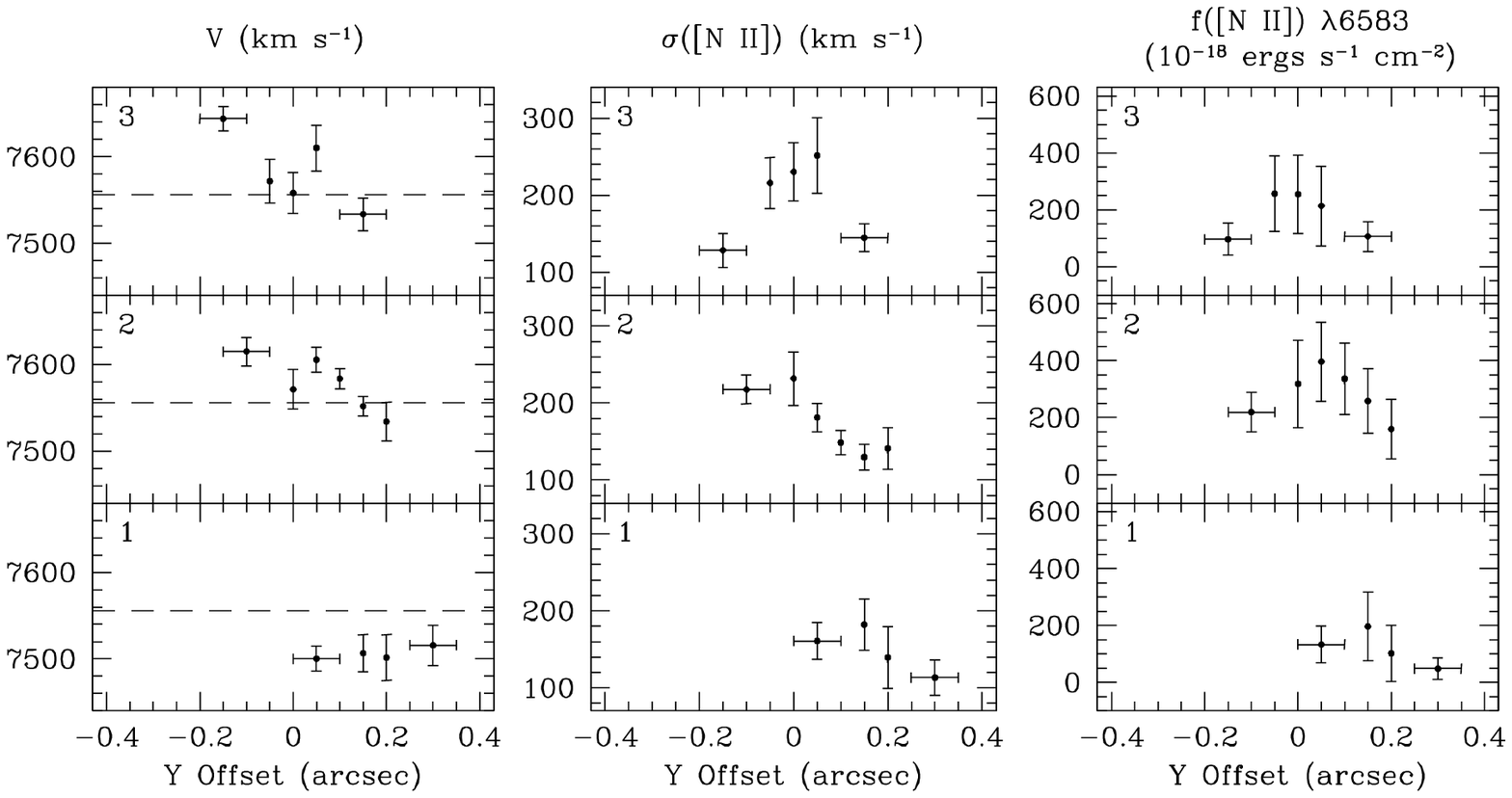}
\caption{IC 989; see Figure \ref{fig:ngc1052} for description. See
also \cite{DeFrancesco_2008}. \label{fig:ic989}}
\end{center}
\end{figure}

The velocity, velocity dispersion, and flux measured from the
[\ion{N}{2}] $\lambda 6583$ emission line for 12 of the 14 galaxies is
plotted as a function of the position along a single slit in Figures
\ref{fig:ngc1052}--\ref{fig:ic989}. We show the galaxy's systemic
velocity, taken from optical line measurements in the Third Reference
Catalogue of Bright Galaxies (RC3) \citep{RC3_catalog}, as the dashed
line in the radial-velocity plots. For each galaxy, we also calculated
the flux-weighted average velocity measured from the largest square
aperture possible given the number of STIS slits for the galaxy and
the number of rows from which spectra could be extracted. For five
galaxies in the sample, the flux-weighted mean velocity differs from
the RC3 velocity by $> 60$ km s$^{-1}$.

We define the galaxy center ($y$-offset = 0) to be the row with the
largest continuum flux. The peak of the continuum flux coincided with
the peak of the H$\alpha$ broad-component flux in five of the seven
broad-line galaxies, and in the cases of NGC 4579 and NGC 3998, the
peak broad-component flux was offset from the peak of the continuum by
one CCD row, corresponding to 4 and 8 pc, respectively. One possible
explanation for the small offset is differential extinction between
the continuum and the emission-line flux. Similarly, the peak of the
narrow-line emission coincided with the peak of the continuum in five
of the seven narrow-line galaxies, but for NGC 4036 and IC 989, the
peak of the [\ion{N}{2}] $\lambda 6583$ flux was offset from the
continuum peak by one CCD row, corresponding to 5 and 50 pc,
respectively. The H$\alpha$ broad-component flux peak coincided with
the narrow-line flux maximum in only two (NGC 1052 and NGC 4278) of
the seven type 1 AGNs. In the other type 1 objects, the offset between
broad and narrow-flux peaks was just one CCD row.

For completeness we include the velocity fields from NGC 3245, NGC
3998, and NGC 5077, although the data have been previously presented
by \cite{Barth_2001}, \cite{DeFrancesco_2006}, and
\cite{DeFrancesco_2008}, respectively. We also include the velocity
fields for NGC 6500 and IC 989 since \cite{DeFrancesco_2008} present
the velocity curves from the central slit positions, while we display
the velocity fields from every slit position. Our measurements of the
velocity fields from all five of these galaxies are consistent with
those results previously published.

For all but two galaxies, a continuum-subtracted H$\alpha +
$[\ion{N}{2}] image was created by subtracting a scaled WFPC2 702W
image, or a scaled combination of F547M and F791W images, from a F658N
image. This is shown along with the positions of the STIS slits in the
second panel of the set of WFPC2 images at the top of the figures. We
verified that the central slit position was placed on the nucleus by
examining the peak-up image, which is an undispersed image taken
through the slit, and the acquisition image, which is an initial image
of the object taken with a larger aperture. We compared the spatial
profile of the peak-up image to the spatial profile produced by
vertical cuts through the nucleus of the acquisition image, and found
similar profiles, indicating that the slit was centered on the
nucleus.

The 14 galaxies in the sample show a large diversity in their observed
velocity fields. Some velocity fields are consistent with rotation but
perhaps at a low S/N, some galaxies seem to be in overall rotation
with significant random components, some appear to be dominated by
irregular motions, and others demonstrate clear evidence for
outflows. As mentioned earlier, three galaxies (NGC 3245, NGC 3998,
NGC 5077) have velocity fields previously shown to be consistent with
circularly rotating disks. While there is no foolproof way to predict
which galaxies will contain regularly rotating gas, \cite{Ho_2002}
determine that the morphology of dust lanes can be a reliable
indicator of organized velocities fields. They find that galaxies with
smooth, circularly symmetric dust rings often have velocity fields
suitable for gas dynamical modeling, while those objects with clearly
disorderly dust lanes almost never contain regular velocity fields.

\begin{figure}
\begin{center}
\figurenum{3}
\epsscale{1.0}
\plotone{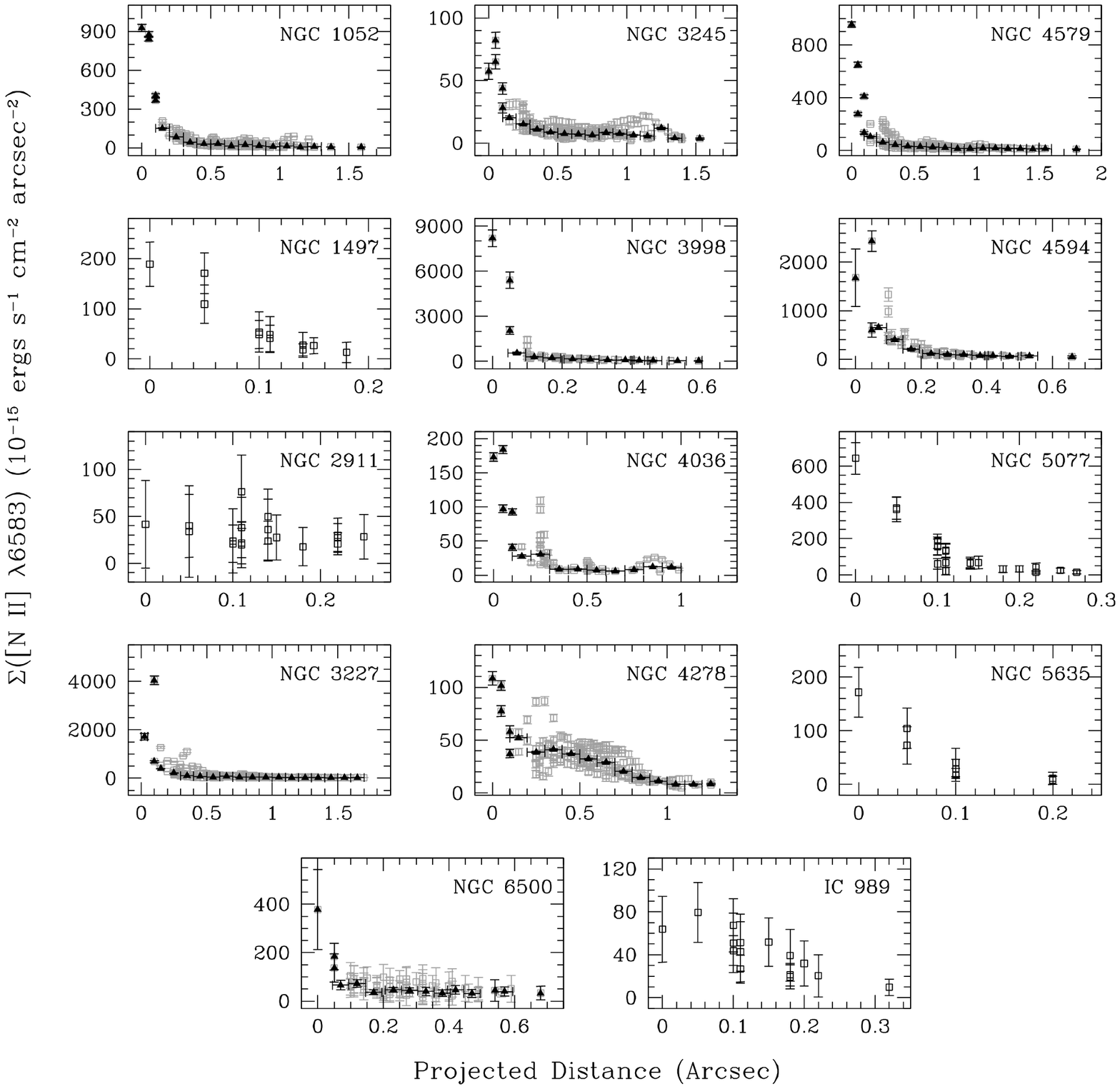}
\caption{[N II] $\lambda 6583$ surface brightness as a function of
projected angular distance from the nucleus. In the plots of the
galaxies from program GO-7403, the black triangles are averages of the
measurements in 0\farcs1 bins, and the grey squares represent the
entire, unbinned set of measurements. In the plots of NGC 3998, NGC
4594, and NGC 6500, the black triangles are averages of the surface
brightness within a 0\farcs05 bin. \label{fig:nii2sb_radial}}
\end{center}
\end{figure}

Figure \ref{fig:nii2sb_radial} shows the [\ion{N}{2}] $\lambda 6583$
surface brightness as a function of projected angular distance from
the nucleus, where each point corresponds to a single row
measurement. In order to emphasize the trends, we averaged together
points within 0\farcs1-wide annuli for the galaxies observed under
program GO-7403. For NGC 3998, NGC 4594, and NGC 6500, we averaged
together points that fell within 0\farcs05-wide bins. These averages
are plotted as black triangles. The shape of the surface brightness
profile varies between galaxies, where some galaxies show a dramatic
increase in surface brightness within the innermost regions and other
galaxies exhibit a gradual rise in surface brightness over a larger
range of radii. However, almost all of the galaxies in the sample do
exhibit a centrally peaked surface brightness profile. NGC 3998 shows
the sharpest peak in surface brightness, where the [\ion{N}{2}]
surface brightness falls by about a factor of 58 between $r=0$ and
$r=$ 0\farcs28. Conversely, NGC 4278, NGC 3245, and IC 989 show the
flattest profiles, where the [\ion{N}{2}] surface brightness decreases
by a factor of about 3--4 within the inner 0\farcs25. NGC 2911 does
not exhibit a central peak in the emission-line surface brightness,
although the error bars are large and may mask a weak rise in surface
brightness at smaller radii. In general, both type 1 and 2 LINERs have
central peaks in emission-line surface brightness.

\begin{figure}
\begin{center}
\figurenum{4}
\epsscale{1.0}
\plotone{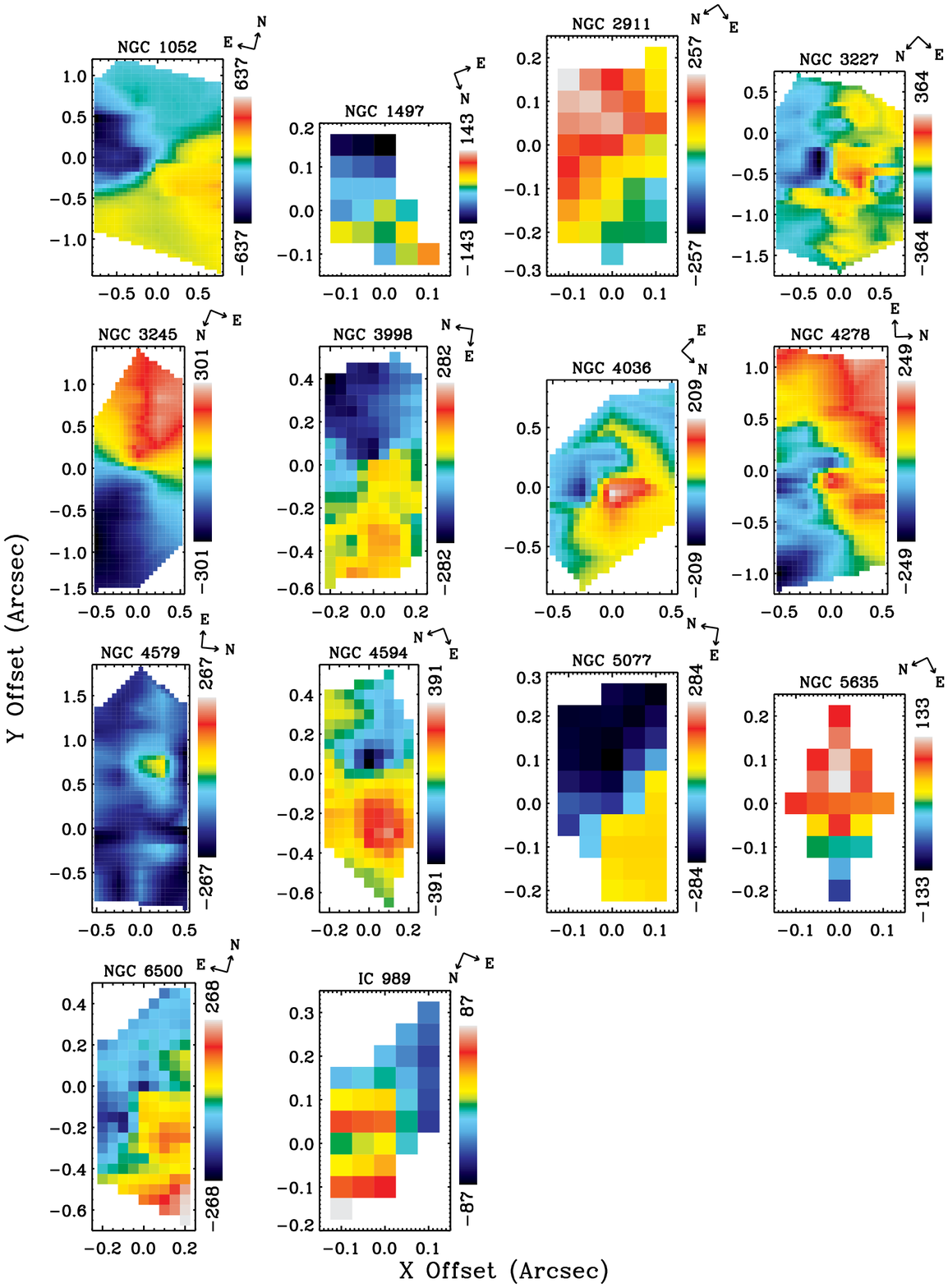}
\caption{Smoothed radial-velocity fields in the slit frame generated
by combining the multislit velocity measurements for each galaxy. For
the galaxies having multiple velocity components, we show the velocity
component plotted in black in Figures \ref{fig:ngc1052},
\ref{fig:ngc3227}, \ref{fig:ngc4278}, and \ref{fig:ngc4579}. The
minimum and maximum velocities for each galaxy relative to the
systemic velocity are given next to the key. For all of the galaxies,
green corresponds to the systemic velocity. Since the galaxies from
program GO-7403 were observed with slits separated by 0\farcs{25}, we
interpolated over the spacing to determine the velocities at those
locations. \label{fig:radvelmap}}
\end{center}
\end{figure}

Radial-velocity maps of each galaxy are shown in Figure
\ref{fig:radvelmap}. For galaxies having multiple velocity components,
we show the radial-velocity map for the velocity component plotted in
black in Figures \ref{fig:ngc1052}, \ref{fig:ngc3227},
\ref{fig:ngc4278}, and \ref{fig:ngc4579}. In these multiple-slit
velocity curves for NGC 1052, NGC 3227, NGC 4278, and NGC 4579, we
attempt to label what appears to be coherent kinematic systems by
using black and grey colors. However, we cannot definitively say that
all black points represent a single component and all grey points
represent a separate component. This ambiguity in the identification
of coherent velocity components affects the radial-velocity maps in
Figure \ref{fig:radvelmap} and may cause the velocity field we show to
be more chaotic than it is in reality. While the multiple-slit
velocity curves presented in Figures
\ref{fig:ngc1052}--\ref{fig:ic989} contain the same information, the
velocity maps show more clearly whether the gas is dominated by
rotational motion or chaotic motions.

Below we discuss the velocity fields for each galaxy in the sample,
with the exception of NGC 3245 and NGC 3998.

\subsection{NGC 1052}
\label{subsec:indivdes_ngc1052}

Overall the gas does not exhibit a regular rotation curve, and the
presence of strong outflows can clearly be seen in the STIS data. The
multiple velocity components identified in our spectra are spatially
coincident with the structure seen in the narrow-band H$\alpha +
$[\ion{N}{2}] emission-line image from \cite{Pogge_2000}, who detected
an ionization cone at a position angle of $96^{\circ}$. Two
kinematically distinct components are seen in four of the off-nucleus
slit positions and correspond to the location of jet-like features
seen in the WFPC2 F658N image. The velocity splittings are as large as
$\sim 400-600$ km s$^{-1}$. The jet-like feature emerging from the
nucleus toward the east is blueshifted by $\sim 560$ km s$^{-1}$
relative to the galaxy's systemic velocity of 1474 km s$^{-1}$, and
the western jet is redshifted by $\sim 240$ km s$^{-1}$. In the
eastern jet, in slit position 7, the line widths are extremely large,
with a velocity dispersion of 600 km s$^{-1}$.

Previous ground-based spectroscopic work by \cite{Sugai_2005} has also
found two components to the emission-line profiles, where one
component is due to disk rotation at radii greater than
$\sim1$\arcsec\ with an axis of rotation at a position angle of
153$^{\circ}$. The second component is attributed to a bipolar
outflow, seen at a radius of $\sim$1\arcsec\ from the nucleus, whose
axis is at a position angle of $\sim 103^{\circ}$. The blueshifted
side of the outflow has a higher velocity of at least $-1200$ km
s$^{-1}$ relative to the systemic velocity, while the redshifted side
has a velocity of $800$ km s$^{-1}$ relative to the systemic velocity
\citep{Sugai_2005}. Additionally, \cite{Davies_Illingworth_1986} and
\cite{Schulz_2003} found evidence for outflows on large scales from
ground-based spectra.

The fit to the spectrum from the nuclear region of the central slit
was improved by the addition of a broad H$\alpha$ component, as
expected since this is a well-known type 1.9 LINER; the broad
H$\alpha$ emission has been shown to be polarized, indicating that the
broad-line region is seen in reflected light \citep{Barth_1999}.

Although the velocity map of NGC 1052 and the central slit velocity
curve superficially resemble a rotational velocity field, the velocity
structure shown in Figure \ref{fig:radvelmap} indicates that the NLR
is actually dominated by an outflow. The color-coded velocity map
shows only one of the two velocity components, where the component we
plot shows redshifted and blueshifted sides that correspond to the
jet-like features seen in the WFPC2 image.

\subsection{NGC 1497}
\label{subsec:indivdes_ngc1497}

Due to the poor S/N of the spectra, especially from the first slit
position, only a few rows in each exposure could be fit. Consequently,
it is difficult to determine whether the gas is rotating. However, a
velocity gradient of 150 km s$^{-1}$ is seen across 0\farcs15 in the
central slit position, and this could be consistent with rotation,
although the velocity gradient is not symmetric about the location of
the peak of the narrow-line region flux.

\subsection{NGC 2911}
\label{subsec:indivdes_ngc2911}

Only a small number of velocity measurements could be made at each
slit position due to the low S/N of the spectra. However, the velocity
map suggests that the gas is dominated by rotational motion. The
central slit position exhibits a gradient of 270 km s$^{-1}$ across
the innermost 0\farcs3. With higher quality spectra, this galaxy could
be a good candidate for dynamical modeling. The flux-weighted average
velocity measured in a 0\farcs3 square aperture is 3285 km s$^{-1}$,
while the systemic velocity taken from RC3 is 3167 km s$^{-1}$. Other
optical measurements of the systemic velocity given by the NASA
Extragalactic Database (NED) span a large range of velocities, $\sim
3030 - 3250$ km s$^{-1}$. The inconsistencies between the various
systemic velocity measurements may be the result of dust obscuration
of the galaxy nucleus and one side of the emission-line disk, as a
dust lane is visible in the WFPC2 image.

\subsection{NGC 3227}
\label{subsec:indivdes_ngc3227}

The velocity field is very complex and chaotic. Two separate kinematic
components are seen in all seven slit positions. At some locations,
the velocity differences between the multiple velocity components are
as large as $\sim 600$ km s$^{-1}$. Our kinematic mapping of NGC 3227
shows that the NLR contains components with low velocities and low
velocity dispersions, as well as components with high velocities and
high velocity dispersions. In particular, in the eastern direction in
slit positions 1 and 2, one of the components has a velocity of $\sim
300$ km s$^{-1}$ relative to the systemic velocity and a velocity
dispersion of $\sim 400$ km s$^{-1}$. In slit position 7, in the
northern direction, one component has a velocity of $\sim 300-350$ km
s$^{-1}$ relative to the systemic velocity, and a velocity dispersion
of $\sim 400$ km s$^{-1}$. Similar multicomponent structures have been
seen in other Seyferts as well, including NGC 4151 \citep{Kaiser_2000,
Nelson_2000}. 

An additional H$\alpha$ broad component was fit to spectra in all but
the outermost two slit positions. The brightness of the H$\alpha$
broad component in the central slit position, as well as in the
adjacent slit positions 2 and 3, appears to be consistent with a
bright point source that has been smeared by the STIS point-spread
function (PSF). However, the flux of the broad component found at a Y
Offset of $-$0\farcs3 to 0\farcs05 in slit position 5, and the broad
component flux at a Y Offset of $-$0\farcs25 to 0\farcs15 in slit
position 6, appear to be in excess of that expected from the wings of
a point source. It is possible that NGC 3227 exhibits scattered
broad-line region light located in this region to the northwest of the
nucleus.

Previous studies of the ionized gas in NGC 3227 have also found
chaotic velocity fields, as well as asymmetries and double-peaked
profiles of the narrow H$\alpha$, [\ion{S}{2}], [\ion{N}{2}], and
[\ion{O}{3}] emission lines. These asymmetries were attributed to the
superposition of several physically and kinematically distinct
components, where the broader component was attributed to outflowing
gas \citep{Arribas_1994, Gonzales-Delgado_1997,
Garcia-Lorenzo_2001}. Although strong outflows are seen in the ionized
gas, the molecular gas in NGC 3227 appears to be in regular rotation
with a major axis at a position angle of 140$^{\circ}$
\citep{Schinnerer_2000, Reunanen_2002, Davies_2006,
Hicks_Malkan_2007}. Our data show that one of the two velocity
components, given by the black points in Figure \ref{fig:ngc3227} and
shown in Figure \ref{fig:radvelmap}, exhibits similar overall features
as the molecular gas velocity field presented in
\cite{Hicks_Malkan_2007}. The redshifted and blueshifted ionized gas
in this kinematic component roughly corresponds spatially to the
redshifted and blueshifted portions of the molecular gas. While one of
the [\ion{N}{2}] velocity components appears to trace the general
features of the molecular gas, the ionized gas is still highly
disturbed and is not participating in simple disk-like rotation.

\subsection{NGC 4036}
\label{subsec:indivdes_ngc4036}

The radial-velocity map shows the gas is rotating, although this
motion is difficult to recognize from the multiple slit velocity
curves alone. Velocities measured from the two slit positions to the
southwest of the nucleus are blueshifted by as much as 200 km
s$^{-1}$, while velocities from slit position 2 are redshifted by 150
km s$^{-1}$ relative to the systemic velocity of 1397 km s$^{-1}$. The
central slit position exhibits a velocity gradient of $\sim 300$ km
s$^{-1}$ across the inner 0\farcs2. \cite{Cinzano_1999} examined the
[\ion{O}{2}] gas kinematics and found that NGC 4036 contained gas that
was clearly not undergoing circular motion on scales less than
10\arcsec. Additionally, based upon their WFPC2 narrow-band H$\alpha +
$[\ion{N}{2}] image, \cite{Pogge_2000} noted the existence of a
complicated filamentary and clumpy structure.

We did not require a H$\alpha$ broad component in the fit to the
spectra in any of the slit positions. However, through ground-based
measurements, \cite{Ho_1997b} determined that NGC 4036 contained a
weak, very low-amplitude, broad H$\alpha$ component. This disagreement
may be the result of variability, but another possibility is that the
apparent broad H$\alpha$ component in the ground-based data arises
from the unresolved high-velocity material in the inner NLR.

\subsection{NGC 4278}
\label{subsec:indivdes_ngc4278}

From the multislit velocity curves and the velocity map, the gas
appears to be in organized motion and displays some characteristics of
rotation. However, within the inner 0\farcs5, the velocity field
appears to be twisted, and may be the result of a counter-rotating
core. Another attractive possibility is that this central decoupled
structure is related to the parsec-scale radio jet seen by
\cite{Giroletti_2005}. According to \cite{Giroletti_2005}, the
double-sided jet is seen at position angles of 155$^{\circ}$ and
$-40^{\circ}$, which suggests that the central kinematic structure
seen in the velocity field may be the result of material being pushed
by the jet. The central slit position exhibits a velocity curve that
is symmetric about the central row and has a gradient of $\sim$ 400 km
s$^{-1}$ across the inner 0\farcs2. Two velocity components were
needed to fit some of the spectra in the two outermost slit positions,
where the additional component's velocity differed by $\sim 450$ km
s$^{-1}$ from the primary component. An additional broad H$\alpha$
component was fit in the inner regions of the central slit position.

\subsection{NGC 4579}
\label{subsec:indivdes_ngc4579}

Both the multislit velocity curves and the velocity map show that the
gas is not in regular rotation. A few spectra in the off-nucleus slit
position immediately adjacent to the center slit displayed two
kinematically distinct components, where the velocity separation is
$\sim 450$ km s$^{-1}$. The nuclear disk may be nearly face-on since
the primary velocity component is almost flat at $\sim 1500$ km
s$^{-1}$. In the central slit position, the [\ion{N}{2}] velocity
dispersion peaks at 440 km s$^{-1}$ at a separation of 0\farcs25 from
the central row. A broad component of H$\alpha$ was detected in the
nuclear regions of the central slit as well as in slit position 4. In
the STIS data, the nuclear spectrum reveals an asymmetric, very broad
H$\alpha$ component with a full width at zero intensity of $\sim
18000$ km s$^{-1}$ and ``shoulders'' on the red and blue sides
\citep{Barth_2001a}. In their H$\alpha + $[\ion{N}{2}] narrow-band
image, \cite{Pogge_2000} find complex, clumpy, and filamentary
emission, which they liken to a shell or a ring. Ground-based
measurements on larger scales also find irregular kinematics
\citep{Gonzales-Delgado_1996}. The flux-weighted average velocity
measured from a 1\farcs2 square aperture is 1465 km s$^{-1}$, while
the systemic velocity taken from RC3 is 1627 km s$^{-1}$. The other
optical measurements of the systemic velocity given by NED span a wide
range of velocities, $\sim$1480--1750 km s$^{-1}$.

\subsection{NGC 4594}
\label{subsec:indivdes_ngc4594}

The data for NGC 4594 show organized motion consistent with an overall
rotation pattern, but with significant irregularities. The central
slit position contains a velocity gradient of $\sim 550$ km s$^{-1}$
across 0\farcs25. Narrow-band H$\alpha + $[\ion{N}{2}] WFPC2 images
from \cite{Pogge_2000} revealed a bright H$\alpha$ core surrounded by
spiral-like wisps that extended to 4\arcsec\ east and west and
1\arcsec\ south of the nucleus. \cite{Emsellem_2000} found that a cut
through the [\ion{N}{2}] $\lambda 6583$ velocity map along the
galaxy's major axis showed the existence of a strong velocity gradient
near the center, and that the gas kinematics within 1\arcsec\ was
decoupled from the gas located in the spiral wisps.

There has been some previous debate as to whether a broad H$\alpha$
component is present in NGC 4594. \cite{Kormendy_1996} concluded that
such a component existed; on the other hand, both \cite{Ho_1997} and
\cite{Nicholson_1998} were able to reproduce the H$\alpha +
$[\ion{N}{2}] blend with only narrow components. In our analysis, an
additional broad H$\alpha$ component improved the fit to several of
the nuclear rows of the central slit position. We also tried to fit
multiple narrow components to each of the H$\alpha$, [\ion{N}{2}], and
[\ion{S}{2}] lines without a broad component, but failed to find a
satisfactory result. However, the H$\alpha + $[\ion{N}{2}] lines are
severely blended together and the S/N is low in the central rows, so
we are unable to conclude that a broad component is definitely
required.

\subsection{NGC 5077}
\label{subsec:indivdes_ngc5077}

The velocity curves and the radial-velocity map show that the gas is
likely to be in regular rotation. This is in agreement with
\cite{DeFrancesco_2008}, who used the same archival STIS dataset to
perform gas dynamical modeling and determined the black hole mass to
be $M_{\mathrm{BH}} = 6.8^{+4.3}_{-2.8}\times 10^8$ $M_\odot$. The
flux-weighted average velocity measured from a square aperture of
0\farcs3 is 2732 km s$^{-1}$, while the systemic velocity taken from
RC3 is 2832 km s$^{-1}$. Other optical measurements given by NED cover
the range $\sim 2650 - 2850$ km s$^{-1}$.

\subsection{NGC 5635}
\label{subsec:indivdes_ngc5635}

The velocity map and the velocity curve from the central slit position
are consistent with disk rotation. The central slit position shows a
velocity gradient of $\sim 200$ km s$^{-1}$ across the nucleus. The
spectra from the off-nucleus slit positions had such low S/N that only
a single velocity measurement could be made. Unfortunately, the poor
quality of the spectra precludes detailed kinematic modeling of the
gas.

\subsection{NGC 6500}
\label{subsec:indivdes_ngc6500}

The velocity field from the multiple slit positions is irregular and
not consistent with pure circular motion. Based upon the velocity map,
the gas does appear to be dominated by rotation, though with some
additional random components. This is in agreement with
\cite{DeFrancesco_2008}, who also find complex velocity curves and
departures from regularly rotating gas, but they do note a general
rotation trend from the central slit position. They suggest that the
complicated velocity curves are the result of a nuclear expanding
bubble. \cite{Gonzales-Delgado_1996} detected multiple velocity
components out to 1 kpc away from the nucleus, indicating that the
extended NLR is experiencing an outflow.

\subsection{IC 989}
\label{subsec:indivdes_ic989}

The S/N is very low and velocities could only be measured at a few
positions, so the kinematic state of the gas is not well
constrained. With the same archival STIS dataset,
\cite{DeFrancesco_2008} note that the central slit position shows
overall rotation with blueshifted gas southwest of the nucleus and
redshifted gas northeast of the nucleus. They suggest the presence of
a counter-rotating nuclear gas disk since the velocity gradient is
reversed at the center. Alternatively, this may indicate the presence
of a radial flow component.

\section{Discussion}
\label{sec:discussion}

\subsection{Density Gradients}
\label{subsec:densitygrads}

In addition to determining the velocity, velocity dispersion, and flux
from the measurement of emission lines, we measured the [\ion{S}{2}]
$\lambda 6716/$[\ion{S}{2}] $\lambda 6731$ and the [\ion{O}{1}]
$\lambda 6300/($[\ion{S}{2}] $\lambda 6716 + $[\ion{S}{2}] $\lambda
6731)$ line intensity ratios. The [\ion{S}{2}] ratio can be used to
determine the electron density because of the difference between the
critical densities of the two lines, where \emph{n}$_{\mathrm{crit}} =
1.5 \times 10^{3}$ cm$^{-3}$ for [\ion{S}{2}] $\lambda 6716$ and
\emph{n}$_{\mathrm{crit}} = 3.9 \times 10^{3}$ cm$^{-3}$ for
[\ion{S}{2}] $\lambda 6731$ (e.g., \citealt{Peterson_book}). However,
this ratio is only sensitive to density over the range $10^{2}$
cm$^{-3} \ltsim$ \emph{n}$_{\mathrm{e}}\ \ltsim\ 10^{4}$
cm$^{-3}$. The [\ion{O}{1}]/[\ion{S}{2}] ratio is also
density-dependent, but not in a unique fashion since the ratio is
influenced by the abundances, ionization level, and shape of the
ionizing spectrum. However, the [\ion{O}{1}]/[\ion{S}{2}] ratio is
sensitive to electron density over a range wider than the [\ion{S}{2}]
ratio because of the larger critical density of the [\ion{O}{1}]
$\lambda 6300$ line, where \emph{n}$_{\mathrm{crit}} = 1.8 \times
10^{6}$ cm$^{-3}$ (e.g., \citealt{Peterson_book}). Below, we show
results from photoionization calculations to illustrate the use of the
[\ion{O}{1}]/[\ion{S}{2}] ratio to probe high-density regions.

\begin{figure}
\begin{center}
\figurenum{5}
\epsscale{1.0} 
\plotone{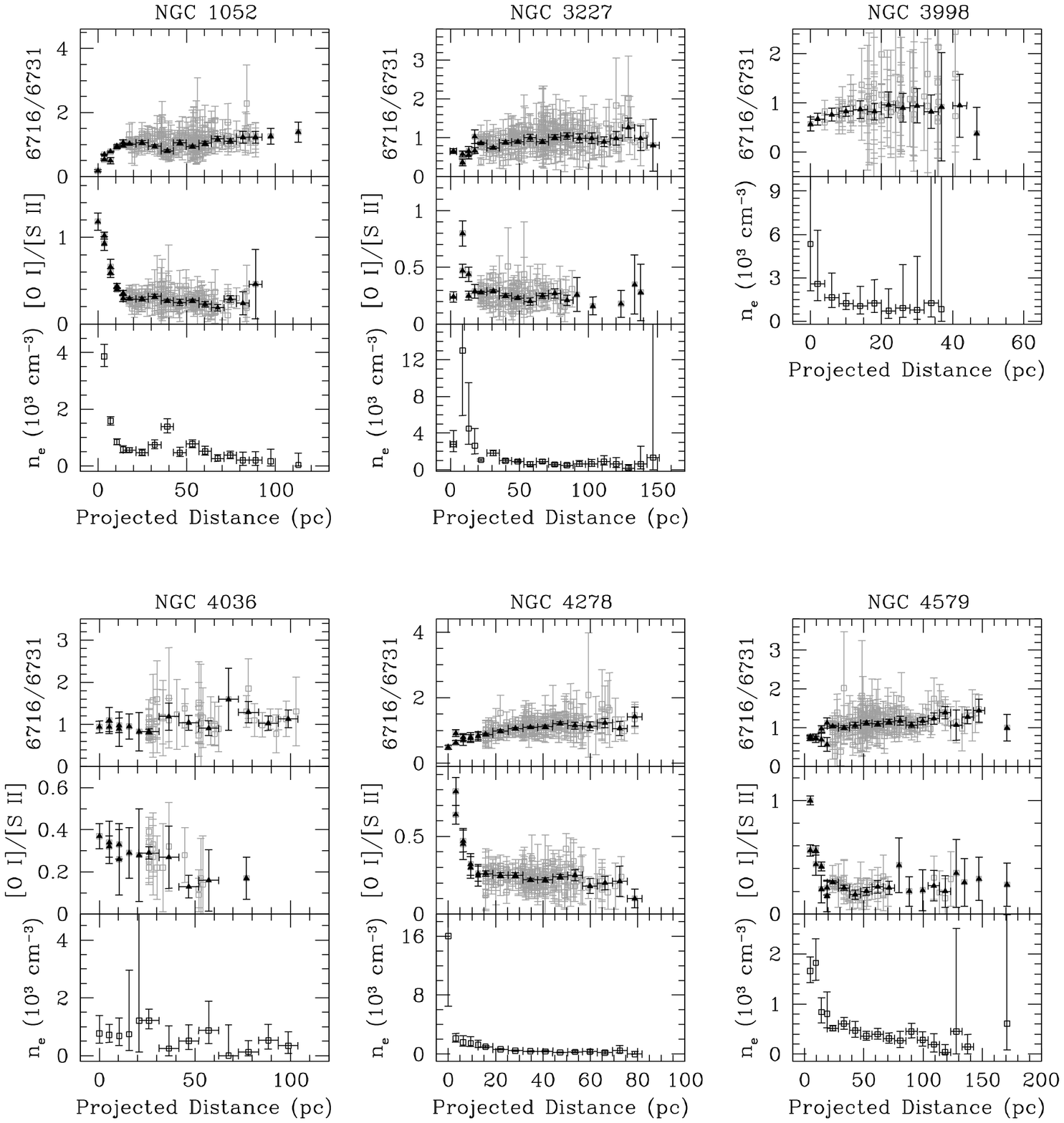}
\caption{[S II] $\lambda 6716/$[S II] $\lambda 6731$ line ratio, [O I]
$\lambda 6300/($[S II] $\lambda 6716 + $[S II] $\lambda 6731)$ line
ratio, and electron density as determined from the [S II] line ratio,
as a function of projected radial distance from the nucleus. The black
triangles are averages of the measurements in 0\farcs1 bins, and grey
squares represent the entire, unbinned set of measurements. In NGC
3998, the black triangles are averages of the ratios within a
0\farcs{05} bin. \label{fig:density}}
\end{center}
\end{figure}

Figure \ref{fig:density} contains plots of the [\ion{S}{2}] $\lambda
6716$/[\ion{S}{2}] $\lambda 6731$ ratio, the [\ion{O}{1}]/[\ion{S}{2}]
ratio, and the electron density determined from the [\ion{S}{2}] ratio
as a function of projected radial distance from the nucleus. The
[\ion{O}{1}] lines were not present in the NGC 3998 spectra since the
grating tilt was set to provide coverage of 6490--7050~\AA, so only
the [\ion{S}{2}] ratio and the electron density are plotted as a
function of distance from the nucleus. To examine the overall trends
in electron density with radius, we computed average [\ion{S}{2}]
$\lambda 6716$/$\lambda 6731$ ratios in 0\farcs1-wide radial bins (or
0\farcs05-wide bins for NGC 3998). With the radial averages of the
[\ion{S}{2}] $\lambda 6716/$[\ion{S}{2}] $\lambda 6731$ ratio, the
electron density as a function of radius was calculated using the IRAF
task {\tt temden} assuming a temperature of 10,000 K. In the instances
where velocity splitting was present, the secondary component was too
weak to produce an independent measurement of the electron density,
and the two components were summed before the [\ion{S}{2}] ratio was
calculated. The remaining eight LINERs not included in Figure
\ref{fig:density} have such large uncertainties in the density
measurements that no information on the density structure could be
extracted from the data.

Significant density gradients are seen in NGC 1052, NGC 4278, NGC
3998, and NGC 4579. Measurements of the density structure in NGC 4579
have been previously presented by \cite{Barth_2001a}. In these
galaxies, the electron density increases from $\sim 200-900$ cm$^{-3}$
in the outer regions to $\sim 1700-16,000$ cm$^{-3}$ at the center. In
contrast to these four galaxies, NGC 4036 does not exhibit any
systematic change in \emph{n}$_{\mathrm{e}}$ with respect to radial
distance from the center or a central spike in density. It is
interesting to note, however, that this galaxy did not need a broad
H$\alpha$ component to fit the central spectrum. Among the objects
with a sufficient S/N to map \emph{n}$_{\mathrm{e}}$ from the
[\ion{S}{2}] lines, all of the broad-lined AGNs show central density
peaks.

The [\ion{O}{1}]$/$[\ion{S}{2}] ratio also rises dramatically toward
the nucleus in some objects, increasing to as much as $1.2\pm0.1$ in
NGC 1052, $1.0\pm0.1$ in NGC 4579, and $0.8\pm0.1$ in NGC 4278. These
central [\ion{O}{1}]$/$[\ion{S}{2}] values are very large in
comparison to the average ratio typically found on scales of hundreds
of parsecs from ground-based spectra of LINERs. For comparison, we
used the emission-line measurements from \cite{Ho_1997} to calculate
an average [\ion{O}{1}]/[\ion{S}{2}] ratio for both type 1 and type 2
LINERs. We first removed the galaxies for which a ``low confidence''
or a ``high uncertainty'' flag was given for either the LINER
classification or the [\ion{O}{1}]$/$H$\alpha$ and
[\ion{S}{2}]$/$H$\alpha$ ratios. This left a sample of 57 galaxies
from which we calculated the average [\ion{O}{1}]$/$[\ion{S}{2}] ratio
to be $0.26$ in the ground-based 2\arcsec$\times$4\arcsec\
aperture. From the same sample of 57 galaxies, those galaxies
classified as type 1 LINERs had [\ion{O}{1}]$/$[\ion{S}{2}] ratios in
the range $0.15$--$0.54$ with an average ratio of $0.32$, while the
type 2 LINERs had [\ion{O}{1}]$/$[\ion{S}{2}] ratios of $0.13$--$0.58$
with an average ratio of $0.21$.

Although the [\ion{O}{1}]/[\ion{S}{2}] ratios at the centers of NGC
1052, NGC 4278, and NGC 4579 are well above the typical values found
in LINERs, the [\ion{O}{1}]$/$[\ion{S}{2}] ratio is not a unique
indicator of density. The ratio is additionally affected by the
abundance ratio, the ionization parameter (\emph{U}, which
characterizes the number of ionizing photons per nucleon), and the
shape of the ionizing continuum. In order to further interpret the
high values of [\ion{O}{1}]/[\ion{S}{2}], we used version 07.02.00 of
CLOUDY \citep{Ferland_1998} to generate a few grids of simple
photoionization models. In particular, we used the AGN continuum model
that takes as free parameters the temperature of the thermal big blue
bump component, the UV (2500 \AA) to X-ray (2 keV) spectral slope
$\alpha_{\mathrm{ox}}$, and the power-law slopes of the low-energy
part of the big blue bump ($\alpha_{\mathrm{uv}}$) and of the X-ray
continuum ($\alpha_{\mathrm{x}}$). We left the thermal component
temperature, $\alpha_{\mathrm{uv}}$, and $\alpha_{\mathrm{x}}$ at
their default values ($T=150,000 $ K, $\alpha_{\mathrm{uv}} = -0.5$,
$\alpha_{\mathrm{x}} = -1$). The $\alpha_{\mathrm{ox}}$ parameter was
varied over $-0.6$ to $-1.6$ in increments of $0.2$, bracketing the
typical range for low-luminosity AGNs \citep{Ho_1999}. Models were
computed for a single plane-parallel slab of constant-density gas
either at solar and twice solar metallicity, with gas density ranging
from \emph{n}$_{\rm H} = 10^2$ cm$^{-3}$ to 10$^6$ cm$^{-3}$, and
ionization parameter ranging from log \emph{U} = $-4$ to $-2$. We
calculated models that did and did not include dust grains.

\begin{figure}
\begin{center}
\figurenum{6}
\epsscale{1.0} 
\plotone{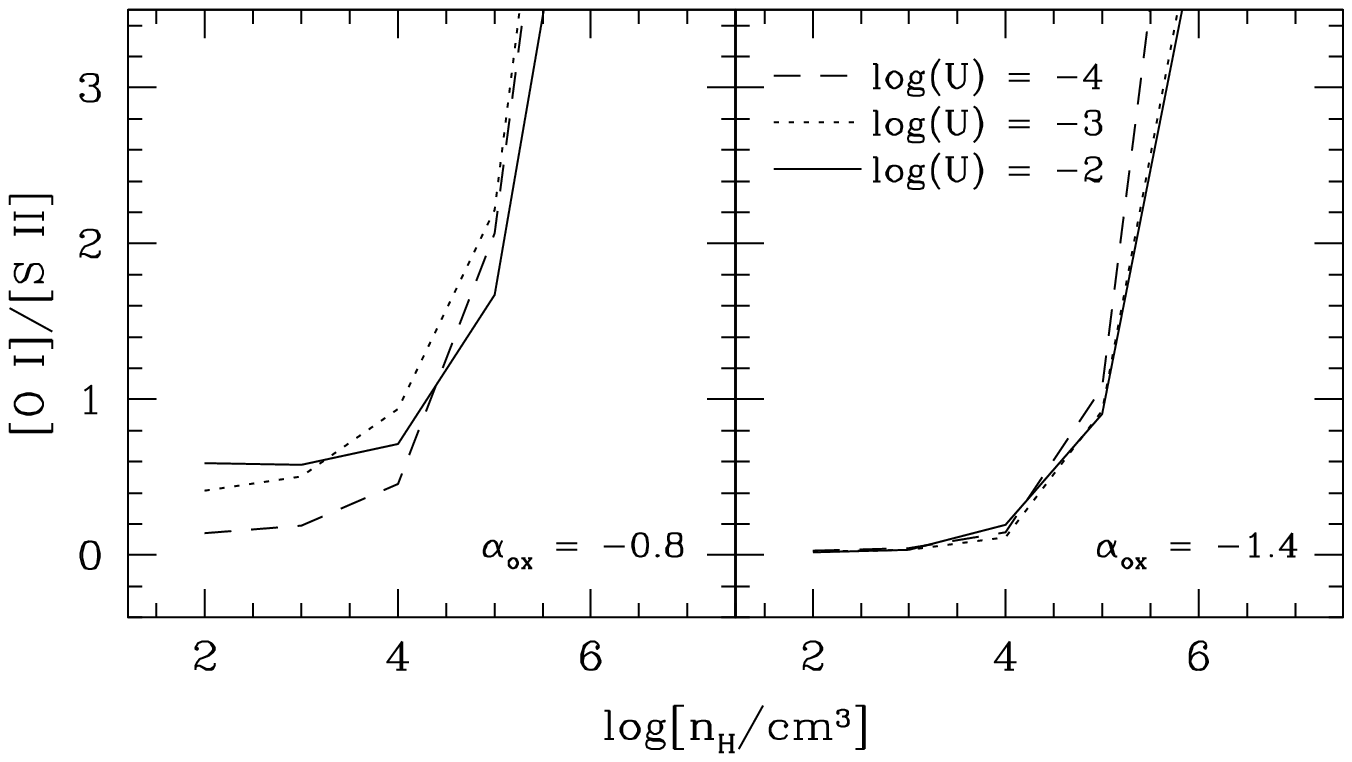}
\caption{[O I] $\lambda 6300/$([S II] $\lambda 6716$+[S II] $\lambda
6731$) as a function of density for a range of values of the
ionization parameter. These photoionization models were generated
using CLOUDY with a metallicity of two times solar in the absence of
dust and with the AGN continuum model where $\alpha_{\mathrm{ox}} =
-0.8$ and $\alpha_{\mathrm{ox}} = -1.4$. The models where solar
metallicity was used and where dust was included show qualitatively
similar results for the low ionization parameters. \label{fig:cloudy}}
\end{center}
\end{figure}

From these grids of photoionization models, we found small
[\ion{O}{1}]/[\ion{S}{2}] ratios, consistent with ground-based
results, at low values of \emph{n}$_{\rm H}$ ($10^{2}$ cm$^{-3}$ --
$10^{4}$ cm$^{-3}$). The ratio increases strongly for \emph{n}$_{\rm
H}\gtsim 10^{4}$ cm$^{-3}$. To illustrate our model results, in Figure
\ref{fig:cloudy} we plot the [\ion{O}{1}]$/$[\ion{S}{2}] ratio as a
function of gas density for the models with twice-solar metallicity
and dust-free gas, and for $\alpha_{\mathrm{ox}} = -0.8$ and
$-1.4$. For $\alpha_{\mathrm{ox}} = -0.8$, a
[\ion{O}{1}]$/$[\ion{S}{2}] ratio of $1.0$ corresponds to a density of
$\sim 2\times10^{4}$ $\mathrm{cm}^{-3}$ for \emph{U} = $10^{-2}$ and
$10^{-4}$, and a density of $\sim 1\times10^{4}$ $\mathrm{cm}^{-3}$
for \emph{U} = $10^{-3}$. For $\alpha_{\mathrm{ox}} = -1.4$, a
[\ion{O}{1}]$/$[\ion{S}{2}] ratio of $1.0$ corresponds to a density of
$\sim 1\times10^{5}$ $\mathrm{cm}^{-3}$ for \emph{U} = $10^{-2}$ and
$10^{-3}$, and a density of $\sim 5\times10^{4}$ $\mathrm{cm}^{-3}$
for \emph{U} = $10^{-4}$. The smaller [\ion{O}{1}]/[\ion{S}{2}] ratio
for $\alpha_{\mathrm{ox}} = -1.6$ is expected since the increased UV
flux relative to X-ray flux results in a less extended partially
ionized zone. The models where solar metallicity was used and where
dust was included showed qualitatively similar results to the models
presented in Figure \ref{fig:cloudy} for the low ionization
parameters, where densities greater than $\sim 10^{4}$
$\mathrm{cm}^{-3}$ were needed in order to produce an
[\ion{O}{1}]/[\ion{S}{2}] ratio of $1.0$. Thus, while the electron
density cannot be directly determined from the
[\ion{O}{1}]$/$[\ion{S}{2}] ratio, the rise in the line ratio seen in
NGC 1052, NGC 4278, and NGC 4579 gives strong evidence for a steep
inner rise in density.

The single Seyfert galaxy in the sample, NGC 3227, exhibits a density
gradient over a portion of the NLR, with \emph{n}$_{\mathrm{e}}
\approx 600$ $\mathrm{cm}^{-3}$ in the outer regions and
\emph{n}$_{\mathrm{e}} \approx 13,000$ $\mathrm{cm}^{-3}$ at a radius
of $10$ pc. The electron density then decreases at the nucleus to
$\sim 2800$ $\mathrm{cm}^{-3}$. Additionally, the
[\ion{O}{1}]$/$[\ion{S}{2}] ratio rises toward the nucleus to
$0.8\pm0.1$ at a radius of $10$ pc, then decreases to $0.4\pm0.1$ at
the center.

\begin{figure}
\begin{center}
\figurenum{7}
\epsscale{0.8}
\plotone{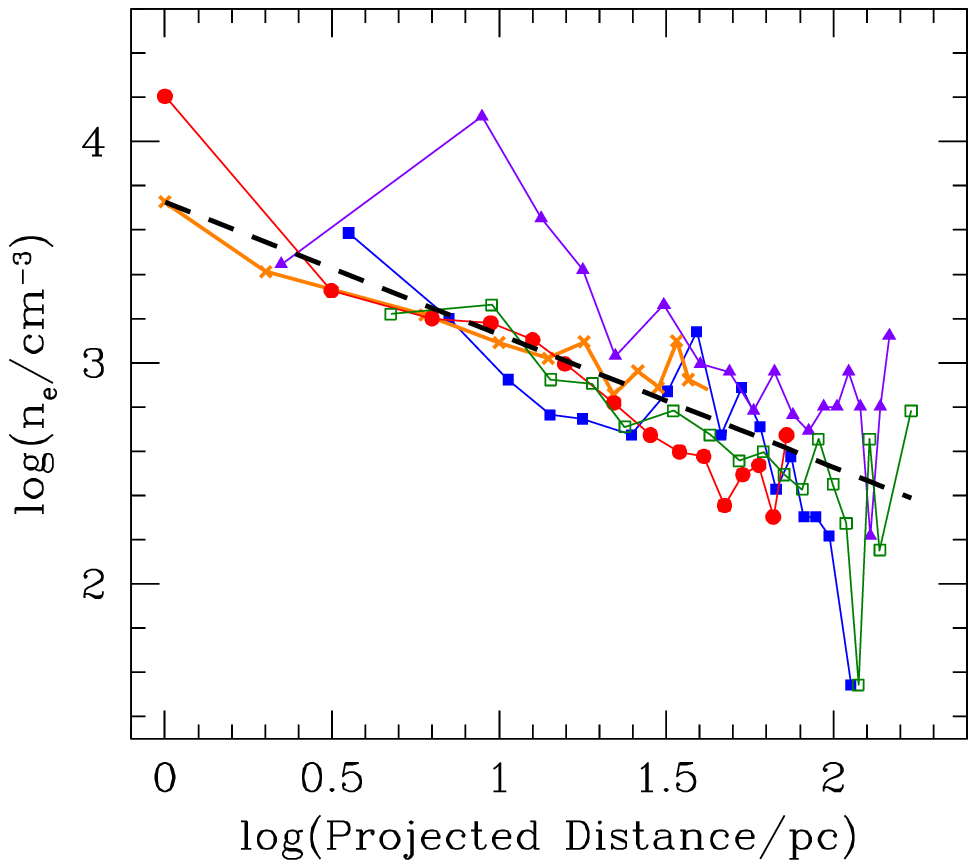}
\caption{Electron density determined from the [S II] line ratio as a
function of projected radial distance from the nucleus for the five
galaxies exhibiting a significant density gradient. Error bars have
been omitted for clarity, but are given in Figure
\ref{fig:density}. NGC 1052 (blue filled squares), NGC 3227 (purple
filled triangles), NGC 3998 (orange crosses), NGC 4278 (red filled
circles), and NGC 4579 (green open squares) show similar
electron-density gradients, where the best-fit power law, with a slope
of $-0.60\pm{0.13}$, is plotted as the dashed black
line. \label{fig:logden_logr}}
\end{center}
\end{figure}

Overall, the data from NGC 1052, NGC 3998, NGC 4278, and NGC 4579
provide clear evidence that Type 1 LINERs contain a radially
density-stratified NLR. The detection of density gradients in Type 1
LINERs is consistent with photoionization models of LINERs. It also
appears that density gradients can persist even in the presence of
strong outflows such as those seen in NGC 1052 and those seen over a
portion of the NLR in NGC 3227. Moreover, these five broad-lined AGNs
exhibit similar density structure as can be seen in Figure
\ref{fig:logden_logr}, where we show the electron density as a
function of projected radial distance from the nucleus for the five
AGNs. We fit a power law of the form $\emph{n}_{\mathrm{e}} =
\emph{n}_0(r/1~ \mathrm{pc})^{\alpha}$ to the density measurements
from each galaxy separately. We calculated the average slope and
normalization, as well as the standard deviation for each of the two
parameters, and found $\emph{n}_0 = 5370 \pm{2495}$ cm$^{-3}$ and
$\alpha = -0.60 \pm{0.13}$. Although there is a large range in the
normalization values between the five broad-lined AGNs, the density
measurements are characterized by a similar slope.

\begin{figure}
\begin{center}
\figurenum{8}
\epsscale{1.0} 
\plotone{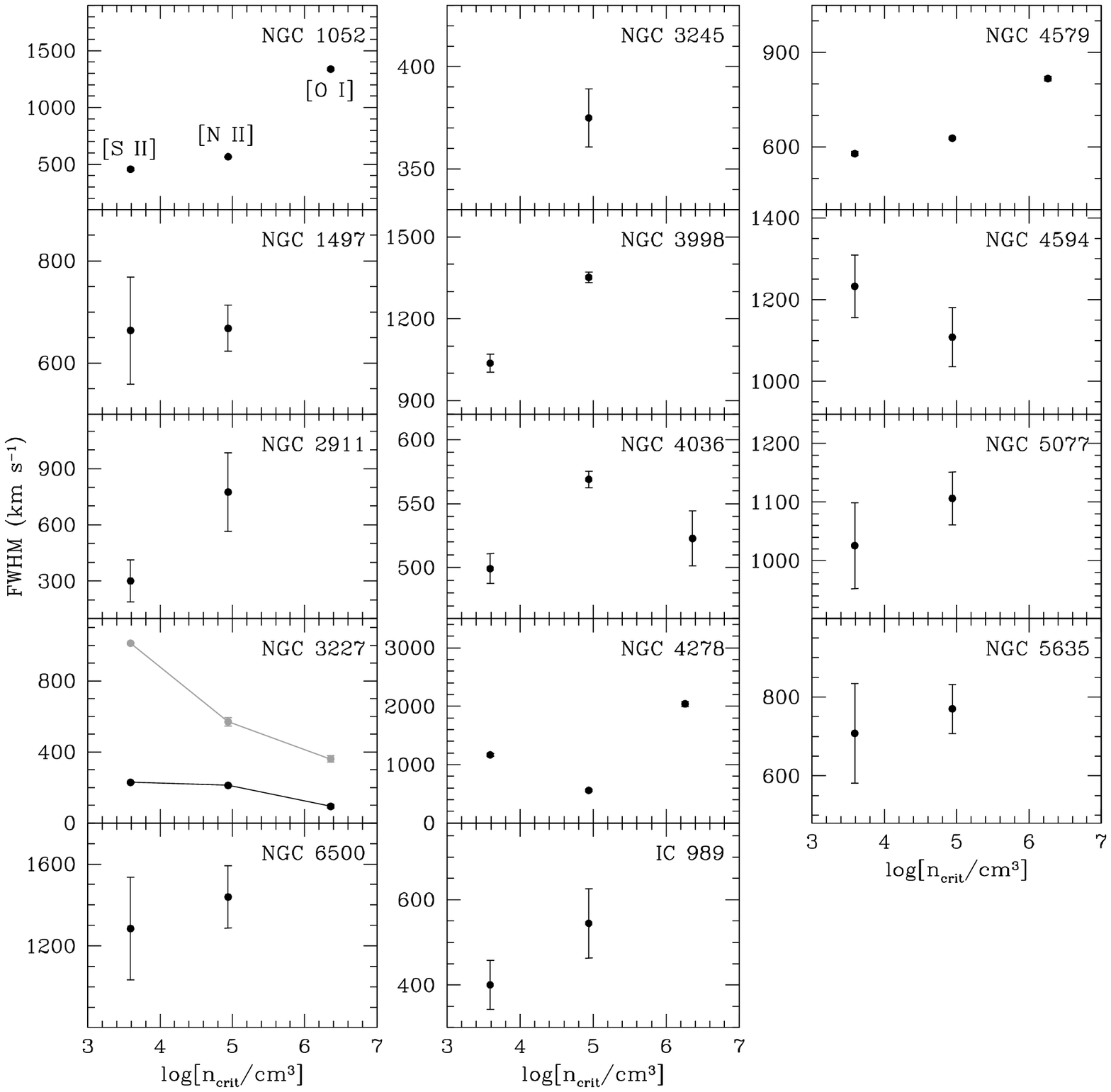}
\caption{FWHM of the [S II] $\lambda 6731$, [N II] $\lambda 6583$, and
[O I] $\lambda 6300$ lines measured from the central row of the
central slit position for each of the galaxies as a function of the
critical density, with the exception of NGC 3227 where the innermost
three rows were binned together. The correlation between line width
and \emph{n}$_{\mathrm{crit}}$ is seen in most of the LINERs in the
sample, even on the 0\farcs1 scales probed by STIS. For NGC 1052, NGC
3227, NGC 4278, and NGC 4579, the error bars are smaller than the plot
symbols. Both the low velocity dispersion (plotted in black) and high
velocity dispersion (plotted in grey) components for NGC 3227 are
shown. \label{fig:fwhmvsncrit}}
\end{center}
\end{figure}

In order to determine whether unresolved density gradients exist even
on the smallest angular scales probed by STIS, we examine the nuclear
spectrum for evidence of a correlation between line width and critical
density for collisional deexcitation
\citep[e.g.,][]{Filippenko_Halpern_1984, Filippenko_1985}. The
line-width--\emph{n}$_{\mathrm{crit}}$ relationship implies that the
NLR has radial gradients in velocity and density, where the broader
emission lines arise from gas closer to the black hole. In Figure
\ref{fig:fwhmvsncrit}, we plot the full width at half-maximum
intensity (FWHM) of the [\ion{S}{2}] $\lambda 6731$, [\ion{N}{2}]
$\lambda 6583$, and [\ion{O}{1}] $\lambda 6300$ lines as a function of
critical density.  The widths of these lines were measured from the
central row of the central slit position for all the galaxies in the
sample with the exception of NGC 3227. Due to the sharp peak in the
surface brightness at the nucleus of NGC 3227, the geometric
rectification performed during the data reduction produced anomalous
bumps in the continuum as well as an irregular [\ion{O}{1}] $\lambda
6300$ line profile. We therefore binned together the innermost three
rows from the central slit position before fitting
H$\alpha+$[\ion{N}{2}], the [\ion{S}{2}] doublet, and the [\ion{O}{1}]
$\lambda 6300$ lines to a double-Gaussian narrow-component model. In
the plot for NGC 3227 in Figure \ref{fig:fwhmvsncrit}, we show the
FWHM for each of the two Gaussian components for a single emission
line.

Of the eight objects in the sample with only [\ion{S}{2}] and
[\ion{N}{2}] measurements, seven of the galaxies show that the FWHM of
the [\ion{N}{2}] line is greater than the FWHM of the [\ion{S}{2}]
line. Although many of these width measurements have large error bars,
it does appear that the FWHM rises with increasing critical
density. Of the five galaxies with measurements of all three lines,
NGC 1052 and NGC 4579 clearly show the presence of a
line-width--\emph{n}$_{\mathrm{crit}}$ relationship, while NGC 4278
exhibits an increase in the FWHM between the [\ion{S}{2}] $\lambda
6731$ and the [\ion{O}{1}] $\lambda 6300$ line. NGC 4036 shows an
increase in the FWHM between the [\ion{S}{2}] $\lambda 6731$ and the
[\ion{N}{2}] $\lambda 6583$ lines, but a slightly smaller FWHM for the
[\ion{O}{1}] $\lambda 6300$ line. In contrast, NGC 3227 shows a
decreasing FWHM with increasing critical density for both the high
velocity dispersion and low velocity dispersion Gaussian
components. The decreasing line width with increasing critical density
indicates that the lowest density gas is actually moving the fastest,
which may be the result of the strong outflow in NGC 3227. It is also
possible, however, that this unusual trend seen in the Seyfert NGC
3227, as well as the decrease in the electron density seen in the
inner region in Figure \ref{fig:density}, may not be real and may be
due to an inadequate fit to a rather complicated nuclear
spectrum. Without additional Seyferts in the sample, or higher
luminosity AGNs similar to NGC 3227, it is impossible to determine
whether this is an anomalous result or a more common finding related
to the properties of the AGN. With the exception of the Seyfert galaxy
NGC 3227, a line-width--\emph{n}$_{\mathrm{crit}}$ relationship can be
observed even on the smallest STIS scales corresponding to $\sim 8$
pc, making the line-width--\emph{n}$_{\mathrm{crit}}$ relationship a
generic property for LINERs. We are therefore seeing unresolved
density structure, providing further evidence that the density
continues to rise even within the innermost $\sim 8$ pc in most
LINERs.

\subsection{Aperture Size and Line Width}
\label{subsec:aperatureandlinewidth}

Another question that can be addressed with our measurements concerns
the variations in emission-line width with aperture size. Such an
analysis is useful in determining the scales over which the bulge
dominates the NLR kinematics in low-luminosity AGNs. In the past, it
has been assumed that ground-based observations sample NLR gas that is
kinematically dominated by the virial motion in the bulge potential
\citep{Nelson_Whittle_1996}. However, as discussed by
\cite{Laor_2003}, low-luminosity AGNs have a more compact NLR whose
kinematics may instead be dominated by the black hole.

\cite{Rice_2006} examined narrow-line widths as a function of aperture
size using single STIS slit observations of a sample of Seyfert
galaxies. They found most of the galaxies exhibited a constant or
increasing FWHM with increasing aperture size. Our sample allows us to
perform a similar analysis on low-luminosity AGNs. We are able to
considerably improve upon the \cite{Rice_2006} analysis by using the
STIS multislit observations to study the line-width variation within
square apertures of increasing size, eliminating the biases introduced
by studying only a relatively narrow and randomly oriented portion of
the velocity field.

For the six galaxies from program GO-7403, we choose square apertures
of $0$\farcs$2$, $0$\farcs$7$, $1$\farcs$2$, and $1$\farcs$7$ on a
side, and for the eight galaxies from program GO-7354 we used square
apertures $0$\farcs$1$, $0$\farcs$3$, and $0$\farcs$5$ on a side. The
sizes of the apertures were chosen such that the width of an entire
STIS slit fell within the aperture. We then used the results of the
spectral fitting to the narrow [\ion{N}{2}] $\lambda 6583$ emission
line and summed together the Gaussian components that fell within the
various aperture bins. We characterized the profile of the summed
Gaussians in each aperture by measuring the line width at $50$\% of
the peak height, thus defining a FWHM for the summed line profile. The
FWHM was measured directly from the summed model, not from a Gaussian
fit to the summed model.

\begin{figure}
\begin{center}
\figurenum{9}
\epsscale{1.0}
\plotone{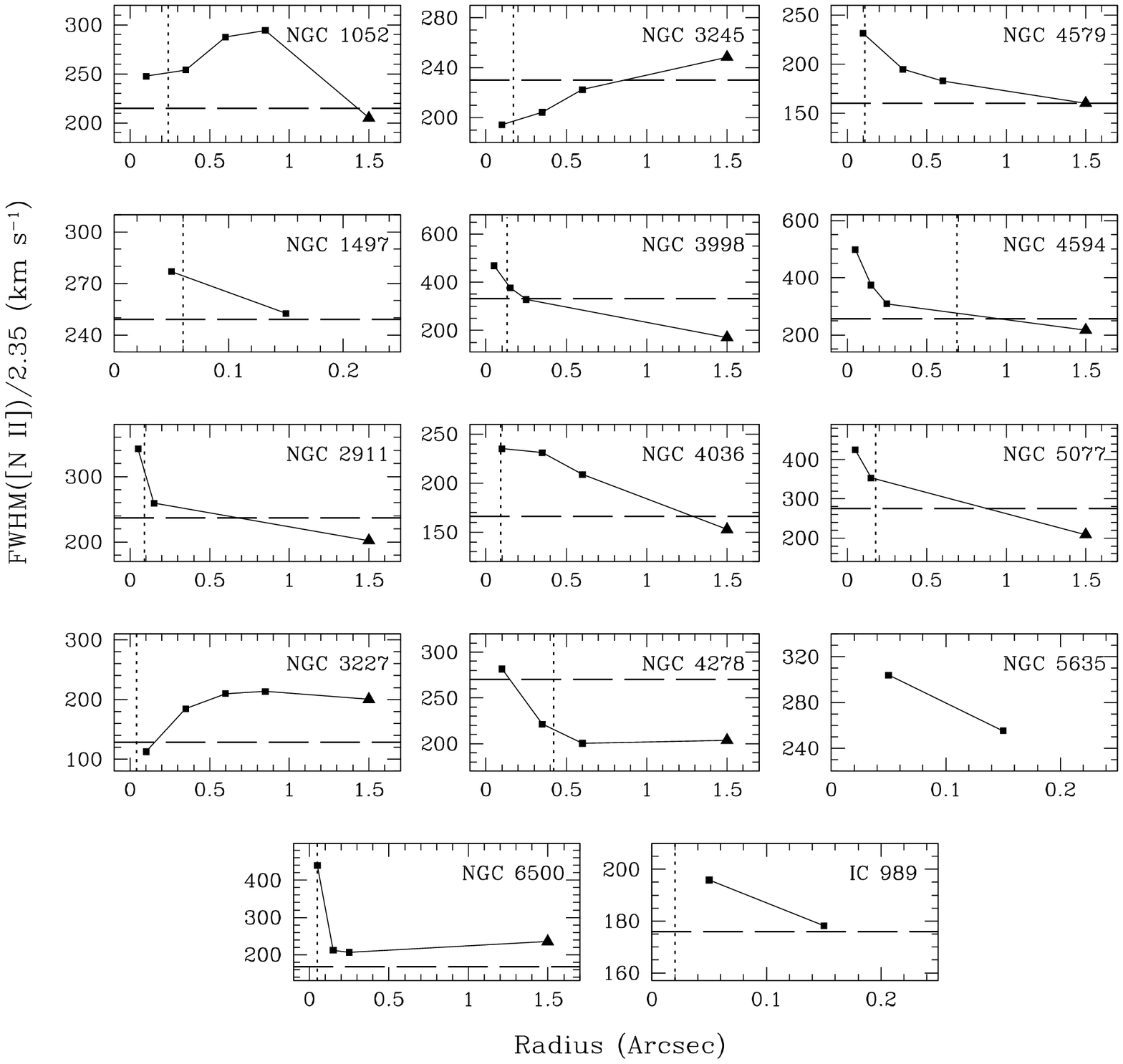}
\caption{[N II] $\lambda 6583$ line width as a function of aperture
size. Depending on how many slit positions were used to observe the
galaxy, either two, three, or four aperture bins were constructed. The
outermost point comes from a ground-based measurement \citep{Ho_1997}
using effectively a 3\arcsec\ square aperture. NGC 1497, NGC 5635, and
IC 989 were not included in the \cite{Ho_1997} sample and thus no
measurement for $1$\farcs$5$ is given. The abscissa represents half of
the square aperture size in arcseconds. The dotted vertical line is
the black hole sphere of influence and the dashed horizontal line is
the bulge stellar velocity dispersion measured from a ground-based
aperture. \label{fig:fwhm2.35_radius_stis}}
\end{center}
\end{figure}

The quantity FWHM/2.35 is plotted against half the length of the
square aperture size (i.e., the ``radius'' of the aperture) in Figure
\ref{fig:fwhm2.35_radius_stis}. We plot FWHM/2.35 in order to compare
with the bulge stellar velocity dispersion. Also included in the plots
is an additional line-width measurement for [\ion{N}{2}] $\lambda
6583$ from a ground-based $2$\arcsec$\times$$4$\arcsec aperture given
by \cite{Ho_1997} for all but three galaxies in our sample. The
$2$\arcsec$\times$$4$\arcsec aperture is taken to be an effectively
$3$\arcsec\ square aperture and is plotted at a radius of
$1$\farcs$5$. Additionally, we found ground-based measurements of the
bulge stellar velocity dispersion from the literature for $13$ of the
$14$ galaxies in the sample. These values are listed in Table
\ref{tab:galprop} and are plotted in Figure
\ref{fig:fwhm2.35_radius_stis} as the dashed horizontal lines. We
estimated the radius of the sphere of influence of the supermassive
black hole, given by ${r_g} = {GM_{\rm BH}/}\sigma_*^2$, in each of
the 13 galaxies for which a stellar velocity dispersion measurement
could be found. We used the \cite{Tremaine_2002} $M_{\rm
BH}$--$\sigma_*$ relation to estimate the black hole mass in all of
the galaxies except NGC 3227, NGC 3245, NGC 3998, NGC 4594, and NGC
5077. For these galaxies we used black hole mass measurements from the
literature (Table \ref{tab:galprop}). The sphere of influence was then
converted into an angular scale using the distances listed in Table
\ref{tab:galprop} and plotted as the dotted vertical lines in Figure
\ref{fig:fwhm2.35_radius_stis}.

For most of the galaxies in the sample, the [\ion{N}{2}] velocity
dispersion peaks within the black hole sphere of influence and then
decreases in larger apertures, approaching the bulge stellar velocity
dispersion. This implies that on scales accessible to ground-based
measurements, the width of the low-ionization lines, such as
[\ion{N}{2}], is set by the bulge and not by the black hole. The work
by \cite{Laor_2003} suggested that in low-luminosity AGNs, the NLR may
be dominated by the black hole and not the bulge. However, the
arguments put forth by \cite{Laor_2003} involved the [\ion{O}{3}]
line, which due to the higher ionization and higher critical density,
originates from a region closer to the black hole than the
[\ion{N}{2}] line. Also, our finding that, for most galaxies in the
sample, the velocity dispersion is largest within the black hole
sphere of influence supports the idea that STIS central line-width
measurements trace the black hole mass, and can therefore be used to
place upper limits on black hole masses in nearby galaxies
\citep{Sarzi_2002,Beifiori_2007}.

NGC 1052, NGC 3227, and NGC 3245, however, show a different trend,
where the FWHM increases with aperture size. Both NGC 1052 and NGC
3227 contain successively wider or roughly constant [\ion{N}{2}]
narrow lines for the increasing STIS aperture sizes, but a smaller
FWHM for the ground-based aperture. In the cases of NGC 1052 and NGC
3227, the strong radial motions create large velocity splitting
between components, causing an increase in the FWHM. The smaller line
width seen in the ground-based aperture may be the result of the
outflows decelerating at larger radii, as described in the outflow
cone models by \cite{Crenshaw_Kraemer_2000}.

An initial expectation for the behavior of line width with aperture
size for gravitationally dominated gas may be a decreasing FWHM with
increasing aperture size. This is reasonable since most of the ionized
gas flux comes from a region close to the black hole, where the gas is
moving the fastest and thus has the largest line width due to
unresolved rotation. However, the line-width variation with aperture
size in NGC 3245 clearly disagrees with the above expectation, since
the gas in NGC 3245 is known to be in near-perfect disk-like rotation
\citep{Barth_2001}. In order to gain a more complete understanding of
line-width variation with aperture size, we constructed a toy
thin-disk model for gas in circular rotation. First, we generated a
model velocity field on a grid where each model pixel was
0\farcs{00507} on a side (i.e., 10$\times$ oversampled relative to the
STIS pixel size). The velocity field is set by the black hole mass,
the disk inclination angle, and the stellar mass profile. We used a
\citet{Hernquist_1990} model to describe the stellar mass
contribution. The intrinsic line-of-sight velocity profiles were
assumed to be Gaussian before passing through the telescope optics,
and were calculated on a velocity grid with a bin size of 25.24 km
s$^{-1}$, in order to match the STIS pixel scale.

The model velocity field was then projected onto the plane of the sky,
and was synthetically ``observed'' in a manner that closely matches
the STIS observation. This synthetic observation involved summing the
intrinsic line-of-sight velocity profiles at every point in the disk
and including the blurring introduced by the nonzero STIS pixel size
and the telescope PSF. In particular, the line-of-sight velocity
profiles were weighted by the emission-line surface brightness
profile, modeled by the function $S(r) = S_0 + S_1 e^{-r/r_o}$. We
included an intrinsic velocity dispersion, similar to those observed
in other early-type galaxies \citep{vanderMarel_vandenBosch_1998}, by
setting the widths of the line-of-sight velocity profiles according to
$\sigma(r) = \sigma_0 + \sigma_1 e^{-r/r_o}$. Finally, we accounted
for the telescope PSF by using Tiny Tim \citep{Krist_Hook_2004} to
generate a 0\farcs2$\times$0\farcs2 subsection of the full STIS
PSF. The PSF was subsampled to match the model pixel grid and computed
for a monochromatic filter passband at $6600$ \AA. Each velocity slice
of the pixel grid was convolved with the PSF. These modeling steps are
a simplified version of the modeling described in detail by
\cite{Barth_2001}. For our purposes, this synthetic observation
involved placing a single STIS slit with increasing aperture lengths
at different angles measured from the gaseous disk major axis in order
to mimic the \cite{Rice_2006} analysis. We additionally placed square
apertures of increasing size over the model velocity field to
replicate our measurements previously discussed. In both instances, we
summed the line profiles that fell within the various apertures and
calculated the FWHM of the summed profile.

\begin{figure}
\begin{center}
\figurenum{10}
\epsscale{1.0}
\plotone{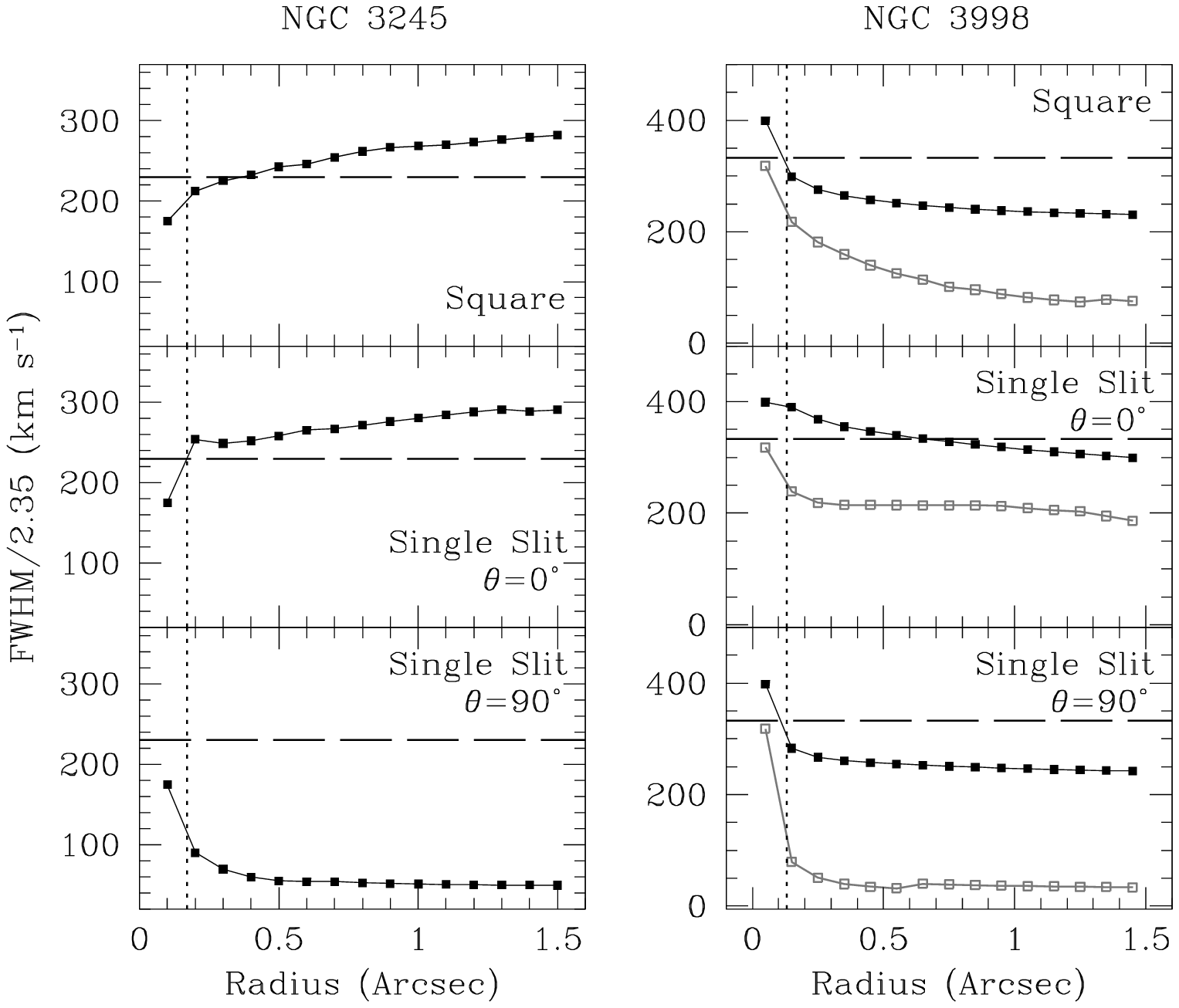}
\caption{FWHM/2.35 as a function of aperture size as determined
through the use of a circularly rotating thin-disk toy model for NGC
3245 and NGC 3998. The black hole sphere of influence for each galaxy
is plotted in the dotted vertical line and the bulge stellar velocity
dispersion for each galaxy is shown by the dashed horizontal line. The
variation in line width with aperture size in the top set of plots was
found using successively larger square apertures. For NGC 3998, the
black filled squares are the result of using a disk model with a
radially dependent intrinsic velocity dispersion, while the grey open
squares are the result of using a model with a small constant
intrinsic velocity dispersion. The bottom two sets of plots were
generated by sampling the modeled velocity field with a single slit of
width 0\farcs2 for NGC 3245 and 0\farcs1 for NGC 3998, with
progressively longer lengths. The single STIS slit was aligned along
the major axis of the gaseous disk for the middle set of plots, and
along the minor axis for the lower panels. The abscissa represents
half of the aperture size in
arcseconds. \label{fig:fwhm2.35_radius_stis_model}}
\end{center}
\end{figure}

As an example, in Figure \ref{fig:fwhm2.35_radius_stis_model} we show
the results of the toy model using the properties of NGC 3245 and NGC
3998 as the input parameters. We selected these galaxies because in
both cases the gas is known to be in circular rotation but the
behavior of line width versus aperture size is quite different for the
two objects. We used the black hole mass and inclination angle
determined from the models of \cite{Barth_2001} and
\cite{DeFrancesco_2006}, and we choose the \citet{Hernquist_1990}
bulge normalization and the scale radius such that the toy model
velocity field would match the observed velocity field as closely as
possible for the outermost velocity measurements. Similarly, we
determined the parameters of the emission-line surface brightness
profile by requiring the profile to match the observed [\ion{N}{2}]
flux distribution. Also, for NGC 3245, the parameters of the function
describing the widths of the line-of-sight velocity profiles were
given by \cite{Barth_2001}.

Note that \cite{DeFrancesco_2006} did not include any intrinsic
velocity dispersion in their model of NGC 3998. We therefore first
generated disk models using only a small constant intrinsic line width
of $\sigma_0 = 20$ km s$^{-1}$. We found an overall trend that
qualitatively matched the data, but the line widths overall were too
small, as can be seen in Figure \ref{fig:fwhm2.35_radius_stis_model},
and the line profile was double peaked in each of the apertures. Since
a double-peaked line profile is not seen in the data, we instead
included an intrinsic velocity dispersion with the parameters
$\sigma_0 = 220$ km s$^{-1}$, $\sigma_1 = 100$ km s$^{-1}$, and $r_o =
10$ pc. We chose this intrinsic velocity dispersion since it produced
a central line width more consistent with the data and a single-peaked
line profile in each of the apertures. Such values for the intrinsic
velocity dispersion are not unreasonable given previous results of
disk modeling by \cite{vanderMarel_vandenBosch_1998} and
\cite{Barth_2001}. Moreover, we stress that we are employing a
simplified version of the disk modeling; a complete treatment,
including a more appropriate emission-line surface brightness profile
and a more realistic model of how the light passes through the
telescope optics, would change the values of the intrinsic velocity
dispersion parameters needed to reproduce the observed line widths.

The top panels for both NGC 3245 and NGC 3998 in Figure
\ref{fig:fwhm2.35_radius_stis_model} are the results from using
progressively larger square apertures.  Since we adopted a simplified
disk model, the models in Figure \ref{fig:fwhm2.35_radius_stis_model}
do not exactly numerically match the observational data.  However,
with our model, we are able to reproduce the qualitative behavior of
line-width variation with aperture size. The model shows that the line
width increases with increasing aperture size for NGC 3245, and the
FWHM decreases with increasing aperture size for NGC 3998. The
additional plots in each column are the results of using a single STIS
slit, at an orientation that is aligned along the major or minor axis
of the gaseous disk. Clearly, the orientation of the single slit not
only affects the numerical values of the line widths, but also the
general behavior of the line-width variation with aperture size, as
can be seen in NGC 3245. In both cases, the disk models always
produced a monotonic trend with line width, either an increasing or
decreasing line width with increasing aperture size.

These models demonstrate how either an increasing or decreasing line
width with increasing aperture size can be seen for gravitationally
dominated gas. NGC 3245 contains a massive bulge that produces large
circular velocities at distances far from the nucleus. While this is
also true for NGC 3998, NGC 3245 has a flatter emission-line surface
brightness profile with relatively more flux at larger radii than in
NGC 3998, as can be seen in Figure
\ref{fig:nii2sb_radial}. Additionally, the gaseous disk in NGC 3245 is
highly inclined at \emph{i} = $63^{\circ}$, and the disk in NGC 3998
is less inclined with \emph{i} = $30^{\circ}$. The combination of the
flatter surface brightness profile, the highly inclined disk, and the
massive bulge of NGC 3245 results in the contribution of a significant
amount of gas with large circular velocities further away from the
nucleus. Consequently, the line width increases with increasing
aperture size.

Therefore, based upon the results of the toy model and the analysis of
the observational data, the overall behavior of line width with
aperture size is not a unique indicator of the kinematic state of the
gas. Outflowing gas can give an increasing line width with aperture
size, as seen in NGC 1052 and NGC 3227, but decelerating outflows
could produce a decreasing trend in line width with aperture size as
discussed by \cite{Rice_2006}. Similarly, rotation can generate either
a decreasing or increasing line width with increasing aperture
size. Thus, while the \cite{Rice_2006} results are useful for
identifying which aperture sizes yield similar stellar and gas line
widths, our results show that measurements such as these are not
necessarily useful as an indicator of whether the line widths are
primarily virial in origin.

Finally, many previous studies have examined the possible role of
outflows and nongravitational motions to explain the scatter in the
$\sigma_g$--$\sigma_{\star}$ relationship. However, our modeling and
data have shown that, even for galaxies whose NLR gas is in overall
rotation and dominated by virial motion, the gas line widths can
differ significantly from the stellar velocity dispersion, as seen in
NGC 3245 and NGC 3998. We find that the gaseous disk inclination, the
surface brightness distribution, and the aperture size and shape are
important factors that also most likely contribute to the scatter seen
in the $\sigma_g$--$\sigma_{\star}$ relationship.

\section{Conclusions}
\label{sec:conclusions}

We have used STIS data to measure detailed velocity, velocity
dispersion, flux, and line-intensity ratio information within $\sim
100$ pc from the nucleus of 14 nearby galaxies. All of the galaxies
harbor low-luminosity AGNs, 13 of which are classified as LINERs and
one of which is a Seyfert. We see a large diversity in the velocity
fields of the 14 galaxies, where simple disk rotation is not a common
state. Only a few galaxies have definite disk rotation, while some
others have possible disk rotation but the emission-line S/N is too
low to clearly determine the kinematic state of the gas. Some velocity
fields are dominated by irregular motions, some show an overall
rotation but with large random components, and others clearly exhibit
outflows. Dynamical modeling of NGC 2911 and NGC 4594 to determine
black hole masses may be possible and worthwhile, although chaotic and
disorganized motion will limit the accuracy of mass measurements.

Through the [\ion{S}{2}] $\lambda 6716$/[\ion{S}{2}] $\lambda 6731$
diagnostic, we were able to detect significant electron-density
gradients in four (NGC 1052, NGC 3998, NGC 4278, and NGC 4579) of the
five LINERs having a sufficient S/N. Moreover, the electron-density
gradients in these objects share a similar shape that can be described
by a power law with a slope of $-0.60 \pm {0.13}$. A similar increase
near the nucleus was seen in the [\ion{O}{1}]/[\ion{S}{2}] line ratio
for the first three of these galaxies. These results demonstrate that
density gradients within the inner $\sim 20$ pc are common in type 1
LINERs. This is consistent with expectations from photoionization
models of LINERs and from ground-based detections of correlations
between line width and \emph{n}$_{\mathrm{crit}}$ in LINERs.  We find
that the density gradients persist in NGC 1052 and over a portion of
the NLR in NGC 3227, even in the presence of strong outflows that
clearly have a large impact on the NLR gas as seen through the
disorganized velocity fields with multiple velocity components. We
also see evidence for correlations between line width and
\emph{n}$_{\mathrm{crit}}$ in most LINERs even on the 0\farcs1 scales
probed by STIS.

Measurements of the STIS spectra additionally provided a means to
study the [\ion{N}{2}] $\lambda 6583$ narrow-line width variation with
spectrograph aperture size. For most galaxies in our sample, the
emission-line velocity dispersion peaks within the black hole sphere
of influence, then decreases in larger apertures, approaching the
bulge stellar velocity dispersion. This trend is consistent with, but
does not uniquely imply, virial motion. The increasing line width with
aperture size seen in NGC 1052 and NGC 3227 is due to the outflows
that dominate the NLR kinematics. In NGC 3245, the increase is the
result of the combination of the galaxy's fairly flat emission-line
surface brightness profile, the galaxy's massive bulge, and the highly
inclined gaseous disk. NGC 3245 demonstrates that even gas in pure
disk rotation can exhibit an increasing line width as a function of
aperture size.

While work by \cite{Boroson_2005} and \cite{Greene_Ho_2005} provide
evidence for a connection between the Eddington ratio and kinematic
disturbance in the NLR, which may reflect an underlying trend of
outflow-dominated NLRs being more predominant at high AGN
luminosities, our sample of galaxies is too small and does not span a
sufficiently wide luminosity range to arrive at such a
conclusion. Since such a relationship would be helpful in
understanding AGN feedback and the effects on host-galaxy properties,
further investigation, ideally with IFU data on a much larger sample
of AGNs spanning a wide range of luminosities, is needed.

\section{Acknowledgments}
\label{sec:acknowledgments}

This material is based upon work supported by the National Science
Foundation under Grants AST--0548198 and AST--0607485. Additional
funding was provided by NASA grant GO-7403 from the Space Telescope
Science Institute (STScI), which is operated by the Association of
Universities for Research in Astronomy, Inc., under NASA contract
NAS5-26555. This research has made use of the NASA/IPAC Extragalactic
Database (NED) which is operated by the Jet Propulsion Laboratory,
California Institute of Technology, under contract with NASA. The data
presented in this paper were obtained from the Multimission Archive at
STScI (MAST); support for MAST for non-\emph{HST} data is provided by
the NASA Office of Space Science via grant NAG5-7584 and by other
grants and contracts.

\clearpage

\begin{deluxetable}{ccccccccc}
\tabletypesize{\scriptsize} 
\tablewidth{0pt} 
\tablecaption{Properties of Galaxies in the Sample \label{tab:galprop}} 
\tablehead{
\colhead{Galaxy} & 
\colhead{Morphological} & 
\colhead{$cz$} & 
\colhead{$D$} & 
\colhead{Spectroscopic} &
\colhead{$\sigma_{\star}$} & 
\colhead{$M_{\mathrm{BH}}$} &
\colhead{log $L$(H$\alpha$)} &
\colhead{References} \\ 
\colhead{} &
\colhead{Type} & 
\colhead{(km s$^{-1}$)} & 
\colhead{(Mpc)} &
\colhead{Classification} & 
\colhead{(km s$^{-1}$)} &
\colhead{($M_{\sun}$)} &
\colhead{(ergs s$^{-1}$)} & 
\colhead{for $\sigma_{\star}$} 
}

\startdata

NGC $1052$ & E & $1474$ & $14.7$ & L$1.9$ & $215$ & $1.8\times10^8$ & 
$39.28$ & $1$ \\

NGC $1497$ & S0 & $6170$ & $80.8$ & L$2^{*}$ & $249$ & $3.3\times10^8$ &
$39.05^{\dagger}$ & $2$ \\

NGC $2911$ & S0 pec & $3167$ & $45.1$ & L$2$ & $237$ & $2.7\times10^8$ & 
$39.44$ & $3$ \\

NGC $3227$ & SAB pec & $1145$ & $18.3$ & S$1.5$ & $128$ & 
$1.5\times10^7$\tablenotemark{a} & $40.26^{\dagger}$ & $1$ \\

NGC $3245$ & S0 & $1348$ & $20.3$ & T$2$: & $232$ & 
$2.0\times10^8$\tablenotemark{b} & $39.51$ & $4$ \\

NGC $3998$ & S0 & $1066$ & $16.4$ & L$1.9$ & $333$ & 
$2.7\times10^8$\tablenotemark{c} & $39.76$ & $1$ \\

NGC $4036$ & S0 & $1397$ & $21.5$ & L$1.9$ & $166$ & $6.7\times10^7$ & 
$39.23$ & $5$ \\

NGC $4278$ & E & $628$ & $13.1$ & L$1.9$ & $270$ & $4.5\times10^8$ & 
$39.43$ & $6$ \\

NGC $4579$ & SABb & $1627$ & $16.9$ & S$1.9/$L$1.9$ & $160$ & 
$5.5\times10^7$ & $39.44$ & $1$ \\

NGC $4594$ & Sa & $1082$ & $19.4$ & L$2$ & $257$ & 
$1.0\times10^9$\tablenotemark{d} & $39.67$ & $7$ \\

NGC $5077$ & E & $2832$ & $44.4$ & L$1.9$ & $275$ & 
$6.8\times10^8$\tablenotemark{e} & $39.75$ & $8$ \\

NGC $5635$ & S pec & $4352$ & $64.3$ & L$2^{*}$ & \nodata & \nodata & 
$38.91^{\dagger}$ & \nodata \\

NGC $6500$ & Sab & $2975$ & $44.6$ & L$2$ & $168$ & $6.7\times10^7$ & 
$40.42$ & $9$ \\

IC $989$ & E & $7556$ & $110.2$ & L$2^{*}$ & $176$ & $8.1\times10^7$ & 
$39.47^{\dagger}$ & $10$ \\

\enddata

\tablecomments{Morphological types and distances are taken from the
NASA Extragalactic Database (NED), where the distances are derived
using the \cite{Mould_2000} Virgo + Great Attractor + Shapley
Supercluster Infall velocity field model. Recession velocities come
from the optical line measurement of the Third Reference Catalogue of
Bright Galaxies (RC3) \citep{RC3_catalog}. Spectroscopic
classifications come from \cite{Ho_1997}, where S = Seyfert nucleus, L
= LINER, T = Transition object, and the number next the class
specifies the type. Type 2 objects have no detectable broad lines,
type 1.9 objects show a broad H$\alpha$ component, and type 1.5
objects exhibit broad H$\alpha$ and H$\beta$ lines. An uncertain
spectroscopic classification is given by ``:''. The spectral
classifications marked with an asterisk were based upon whether a
broad H$\alpha$ component was seen in the data from program GO-7354
and were not taken from \cite{Ho_1997}. Black hole masses, unless
marked, were estimated using the \cite{Tremaine_2002} $M_{\rm
BH}-\sigma_*$ relation. The black hole masses for NGC 3227, NGC 3245,
NGC 3998, NGC 4594, and NGC 5077 were determined through dynamical
measurements. H$\alpha$ luminosities were calculated using the
distances listed in column (4) and the H$\alpha$ flux taken from
\cite{Ho_1997} and \cite{Ho_2003}, except for those objects marked
with a cross: NGC 1497, NGC 3227, NGC 5635, and IC 989. The H$\alpha$
luminosities for these four objects were calculated from the flux
measured within the largest aperture that could be constructed given
the number of STIS slit positions for the galaxy and the number of
rows from which spectra could be extracted.}

\tablenotetext{a}{\citealt{Davies_2006}.}
\tablenotetext{b}{\citealt{Barth_2001}.}
\tablenotetext{c}{\citealt{DeFrancesco_2006}.}
\tablenotetext{d}{\citealt{Kormendy_1996}.}
\tablenotetext{e}{\citealt{DeFrancesco_2008}.}

\tablerefs{(1) \citealt{Nelson_Whittle_1995}; (2)
  \citealt{Wegner_2003}; (3) \citealt{DiNella_1995}; (4)
  \citealt{Tonry_1981}; (5) \citealt{Fisher_1997}; (6)
  \citealt{Bender_1994}; (7) \citealt{VanderMarel_1994}; (8)
  \citealt{Faber_1989}; (9) \citealt{Terlevich_1990}; (10)
  \citealt{White_1983}.}

\end{deluxetable}

\begin{deluxetable}{ccccccccc}
\tabletypesize{\scriptsize} 
\tablewidth{0pt} 
\tablecaption{Summary of Observations \label{tab:obsparams}} 
\tablehead{ 
\colhead{Galaxy} & 
\colhead{Program} &
\colhead{UT Date} &
\colhead{Aperture} & 
\colhead{Binning} & 
\colhead{Slit} &
\colhead{Number} & 
\colhead{Separation} &
\colhead{Exposure} \\ 
\colhead{} &
\colhead{GO ID} & 
\colhead{} &
\colhead{$'' \times ''$} & 
\colhead{} & 
\colhead{Orientation} &
\colhead{of} & 
\colhead{(Arcsec)} &
\colhead{Time} \\
\colhead{} &
\colhead{} &
\colhead{} &
\colhead{} &
\colhead{} &
\colhead{( $^\circ$ E of N)} &
\colhead{Positions} &
\colhead{} &
\colhead{(s)}
}

\startdata

NGC $1052$ & $7403$ & $1999$--$01$--$02$ & $52\times0.2$ & $1\times1$ &
$13.8$ & $7$ & $0.25$ & $2160$ \\

NGC $1497$ & $7354$ & $1997$--$11$--$13$ & $52\times0.1$ & $2\times1$ &
$159.3$ & $3$ & $0.10$ & $440$ \\

NGC $2911$ & $7354$ & $1998$--$12$--$03$ & $52\times0.1$ & $2\times1$ &
$237.1$ & $3$ & $0.10$ & $480$ \\

NGC $3227$ & $7403$ &$1999$--$01$--$31$ & $52\times0.2$ & $1\times1$ &
$222.5$ & $7$ & $0.25$ & $1840$ \\

NGC $3245$ & $7403$ & $1999$--$02$--$02$ & $52\times0.2$ & $1\times1$ &
$202.4$ & $5$ & $0.25$ & $3150$ \\

NGC $3998$ & $7354$ & $1997$--$11$--$01$ & $52\times0.1$ & $2\times1$ &
$278.3$ & $5$ & $0.10$ & $330$ \\

NGC $4036$ & $7403$ & $1999$--$03$--$25$ & $52\times0.2$ & $1\times1$ &
$133.7$ & $5$ & $0.25$ & $3360$ \\

NGC $4278$ & $7403$ & $2000$--$05$--$11$ & $52\times0.2$ & $1\times1$ &
$88.0$ & $5$ & $0.25$ & $3040$ \\

NGC $4579$ & $7403$ & $1999$--$04$--$21$ & $52\times0.2$ & $1\times1$ &
$95.4$ & $5$ & $0.25$ & $3100$ \\

NGC $4594$ & $7354$ & $1999$--$02$--$05$ & $52\times0.1$ & $2\times1$ &
$249.6$ & $5$ & $0.10$ & $280$ \\

NGC $5077$ & $7354$ & $1998$--$03$--$12$ & $52\times0.1$ & $2\times1$ &
$279.1$ & $3$ & $0.10$ & $400$ \\

NGC $5635$ & $7354$ & $1999$--$02$--$02$ & $52\times0.1$ & $2\times1$ &
$243.6$ & $3$ & $0.10$ & $480$ \\

NGC $6500$ & $7354$ & $1998$--$11$--$03$ & $52\times0.1$ & $2\times1$ &
$14.8$ & $5$ & $0.10$ & $250$ \\

IC $989$ & $7354$ & $1998$--$03$--$13$ & $52\times0.1$ & $2\times1$ &
$203.1$ & $3$ & $0.10$ & $490$ \\

\enddata

\tablecomments{The exposure time column gives the average of each slit
position's exposure time for each galaxy.}

\end{deluxetable}

\clearpage

\clearpage

\end{document}